# Wave physics as a choreographic notation for partner dance

Fernando Ramiro-Manzano[1*]

[1] *Nanomaterials for Optoelectronics, Photonics and Energy, Instituto de Tecnología Química, Universitat Politècnica de València - Consejo Superior de Investigaciones Científicas (UPV-CSIC), Avd. de los Naranjos s/n, 46022, Valencia, Spain.*

*e-mail: ferraman@fis.upv.es

**Analytical models provide interpretable descriptions of dance motion, with particular insight into its underlying physical structure. Their compactness, however, requires a focus on selected physical principles. From frequency to partner-dance synchrony, the richness of wave physics provides a natural background for describing dance motion. Here, analytical wave and oscillation models are used to reconstruct partner-dance motion and explore its underlying structure. The framework is applied to Bachata Sensual, a dance style with the wave as a leitmotif. The analysis is based on motion projections of spine-related functional groups and their constituents. Five fundamental sequences and one composite, performed by three couples (Phase I), are studied. A range of wave-related phenomena is revealed, from features emerging in the reconstructed motion, such as reflection and resonant phase shifts, to underlying effects, including constructive and destructive time-dependent modal interference. Remarkably, this combination of spine-related motion, partner dance and the analytical framework reveals coupled modes with a harmonic-like frequency ratio close to 3:1. As an acoustic analogy, these modes, when mapped to audible frequencies, form musical dyads. Overall, this wave-physics formalism can be viewed as a form of choreographic motion description or notation, highlighting connections between partner-dance motion and harmonic phenomena in nature.**

Dance provides a rich domain for the scientific study of human movement[1,2,3]. Its analysis encompasses diverse approaches, including experimental quantitative analyses, analytical models, and numerical and data-driven methods. Experimental quantitative approaches have characterized dance movement in terms of kinetic variables[4], coordination and muscle activation[5], movement variability[6], and movement characteristics associated with movement quality and evaluation[7]. Analytical models based on equations have provided compact representations of body motion, yielding tractable and interpretable descriptions with insight into its underlying physical structure. These models have addressed fundamental questions in dance, including rotation, elevation, balance and stability[8,9,10,11]. Related formulations are well established in the study of locomotion, particularly gait[12], which shares relevant mechanical features with dance stepping. However, such models are based on a selected set of physical principles and may therefore capture only part of the richness of dance. Notably, Laws and co-workers[1] showed that physical models can provide a remarkably rich analysis, with perceptual effects that appear to challenge physical intuition, such as turning after elevation or jumps that resemble floating[10,11]. Computational approaches based on numerical methods, probabilistic models and machine learning have proven powerful for analysing, and generating dance motion[13,14,15,16]. These approaches have been applied to the analysis of intrapersonal synchronization[17], hierarchical movement recognition[18] and also to movement representations informed by dance theories such as Laban Movement Analysis[19,20]. Their applications span domains including video games[14], humanoid robotics[21], and human learning and training[19,20]. Such methods can accommodate complex and high-dimensional data, enabling detailed analyses and highly realistic motion synthesis[13,14,15,16]. While broad in scope and highly successful, the internal representations arising from these approaches are not always straightforward to interpret in physical terms.

Across experimental, analytical and computational studies, wave-related and oscillatory concepts have also been used to describe dance movement, spanning rhythmic organization, propagation and synchrony. In particular, wave-propagation features have been associated with sequential movement across the upper extremities[22]. Related motion patterns across pelvic and trunk segments have been identified through combined electromyographic and kinematic analyses, with features consistent with ancestral locomotor organization[23]. At the neurophysiological level, rhythmic motor organization has been described in terms of central pattern generators represented as coupled segmental oscillators, in which intersegmental phase lags can generate travelling waves along the spinal cord[24]. Spinal motor activity has also been reformulated in terms of rotational neural dynamics[25]. In mechanical descriptions, resonant excitation has been considered in dance turning[1,18], while pendular and oscillatory mechanics are well established in models of locomotion, particularly gait[12], which shares relevant mechanical features with dance stepping, including coupling[21]. Spectral and wave-based methods have also been used to characterize dance motion through principal spatial directions and periodicities[29]. Cross-wavelet analysis has further extended this perspective to multidimensional movement interaction and syncrony[26,27].

When more than one performer is involved, dance movement encompasses interpersonal interaction, including visual-motor entrainment[27], and synchronous imitation and coordination[28]. In partner dance, interaction expands the range of possible motions, enabling movements that would be difficult or unattainable for a single performer[1]. Music- and partner-driven synchronization has been quantitatively characterized in full-body movement patterns[29]. Haptically mediated coupling in dance stepping has been described through mechanical models[21]. Partner-dance choreography has also been represented through discrete movement vocabularies and finite-state models[30], and interpersonal synchrony has been analysed to characterize coordination using cross-wavelet methods[26,29]. Partnered motion has been predicted[15] and generated[16,31], while virtual environments have supported partner-dance learning and evaluation[20].

The present study focuses on the analytical reconstruction of partner-dance motion within a wave-physics framework. The wave is a dance paradigm associated with I. Duncan[32], which establishes a connection with nature. In fact, wave physics is a fundamental field whose principles extend across harmonic phenomena in nature. These range from familiar mechanical waves to wave-particle duality and gravitational waves, with the underlying principles widely applied across science and engineering, such as Biomechanics. In this study, wave physics is applied, through analytical models, to the motion of spine-related functional groups and their constituents. Key motion features are captured through projections, facilitating compact analytical modelling without constraining the rich underlying three-dimensional motion of the performance. Although dancers may exploit body kinetics and kinematics, this work does not aim to determine biomechanical quantities in absolute terms. Nor is the primary objective to characterize fundamental wave-related parameters through numerical methods or direct data observation (apart from the resonant-phase analysis in sequence 4, providing an additional consistency check). Rather, the main objective is to reconstruct wave-related phenomena through analytical models, thereby revealing underlying physical structure and thus supporting analogies within a wave-physics framework. Here, the concept of creating illusions pointed out by K. Laws[1] remains relevant, but the approach shifts from defying physical intuition towards imitating the physics of harmonic nature.

We consider a recreational partner dance in which motion and interaction are continuously shaped in response to the evolving music. The leader guides the unfolding choreography. In response to the interaction, the follower shapes the resulting motion, potentially drawing on proprioceptive (kinesthetic) information[33]. The dance style selected is Bachata Sensual, a branch of Bachata. Bachata was inscribed on UNESCO's Representative List of the Intangible Cultural Heritage of Humanity and is described as a "couple's dance characterized by a sensual hip movement and simple eight-step structure."[34]. In Bachata Sensual, In Bachata Sensual, wave-like motions constitute a central leitmotif (see Supplementary Information S3). The study begins by examining fundamental wave propagation in a single dancer and in a synchronized pair and then focuses on unidirectional driving and bidirectional coupling. This framework reveals a range of wave-related phenomena and related analogies. These concepts, distributed across five fundamental sequences and one composite sequence, can be interpreted as a transcription of dance using a wave-physics-based choreographic notation. Based on experiments with six dancers (three couples, each dancer with more than 7 years of dance experience in Bachata Sensual), this work presents Phase I of the study. A subsequent phase is foreseen, aimed at extending the framework to a larger pool (around 40 individuals, including experts and novices).

## Results and discussion

The experiment consisted of 3D tracking of anatomical landmarks using spherical markers, recorded from anterior and posterior views (see Extended Data Table 1 and Extended Data Figs. 1 and 2). For clarity, while the manuscript mostly presents representative examples from one dancer or one couple, Extended Data Figures 1-9 include all three interpretations. In each dance sequence, analogies with Nature are proposed. For clarity, specific dance (or musical) terms are set in italics.

## I. Fundamental sequences

### Wave propagation, reflection, polarization and interference (sequences 1 and 2)

The study starts with a full-body wave motion, or simply referred to as *wave* or *body roll*, which defines sequence 1. Here, a single dancer (see Fig. 1, Extended Data Figure 1 and Supplementary Video S1) shows sequential anteroposterior displacements from the head through the trunk to the pelvis and along the lower limbs down to the ankles, reproducing in fact, a downward-propagating wave. After several repetitions, the dancer extends the sequence possibilities by reversing the direction of the sequence (often referred to as a *counterwave* or *reverse body roll*). This reversal can be interpreted as a wave reflection initiated as the wave reaches its maximum amplitude around the pelvis. An analogy of this motion could be a wave hitting a cliff face. In this dance sequence, the wave essence lies in the anterior body direction; however, similar wave-like motion could be performed by lateral displacements, in this case by a synchronous dance couple, defining sequence 2 (see Extended Data Fig. 2 and Supplementary Video S2). These two perpendicular motions correspond to orthogonal transversal wave polarizations. This is analogous to the undulations and oscillations of swimming dolphins and sharks, respectively. A transversal propagation wave can be defined as:

$$u_i = A_i \hat{u}\, cos(\omega t - 2\pi z_i/\lambda - \delta_0) = A_i \hat{u}\, cos(\omega t - \delta_i), \text{ with } \delta_i = 2\pi z_i/\lambda - \delta_0 \quad (1)$$

where $u_i$, $\hat{u}$, $A_i$, $\omega$, $t$, $z_i$, $\lambda$ and $\delta_0$ *and* $\delta_i$ denote the displacement of the *i* landmark, the unit polarization vector, the amplitude of the segment, the angular frequency, time, the position along the propagation direction, the wavelength, initial and spatially dependent phase, respectively (for axes conventions, see Supplementary Information S6). Differences of $\delta_i$ allow to characterize a dimensionless wavelength fraction $\overleftrightarrow{\lambda}_j = \Delta\delta_j/2\pi$. The fitting results (see Extended Data Figs. 1 and 2 and Supplementary Table 2) indicate that the dancers reproduces up to roughly one third or one fourth of a fraction of a wavelength when considering the Whole Body - No Upper Limbs ($\overrightarrow{\lambda}_{WB-NUL} \approx \lambda/3$) or the more uniform wave segmental emulation at Shoulder Girdle-ThoracoLumboPelvic complex (SGTLP, $\overrightarrow{\lambda}_{SGTLP} \approx \lambda/4$).

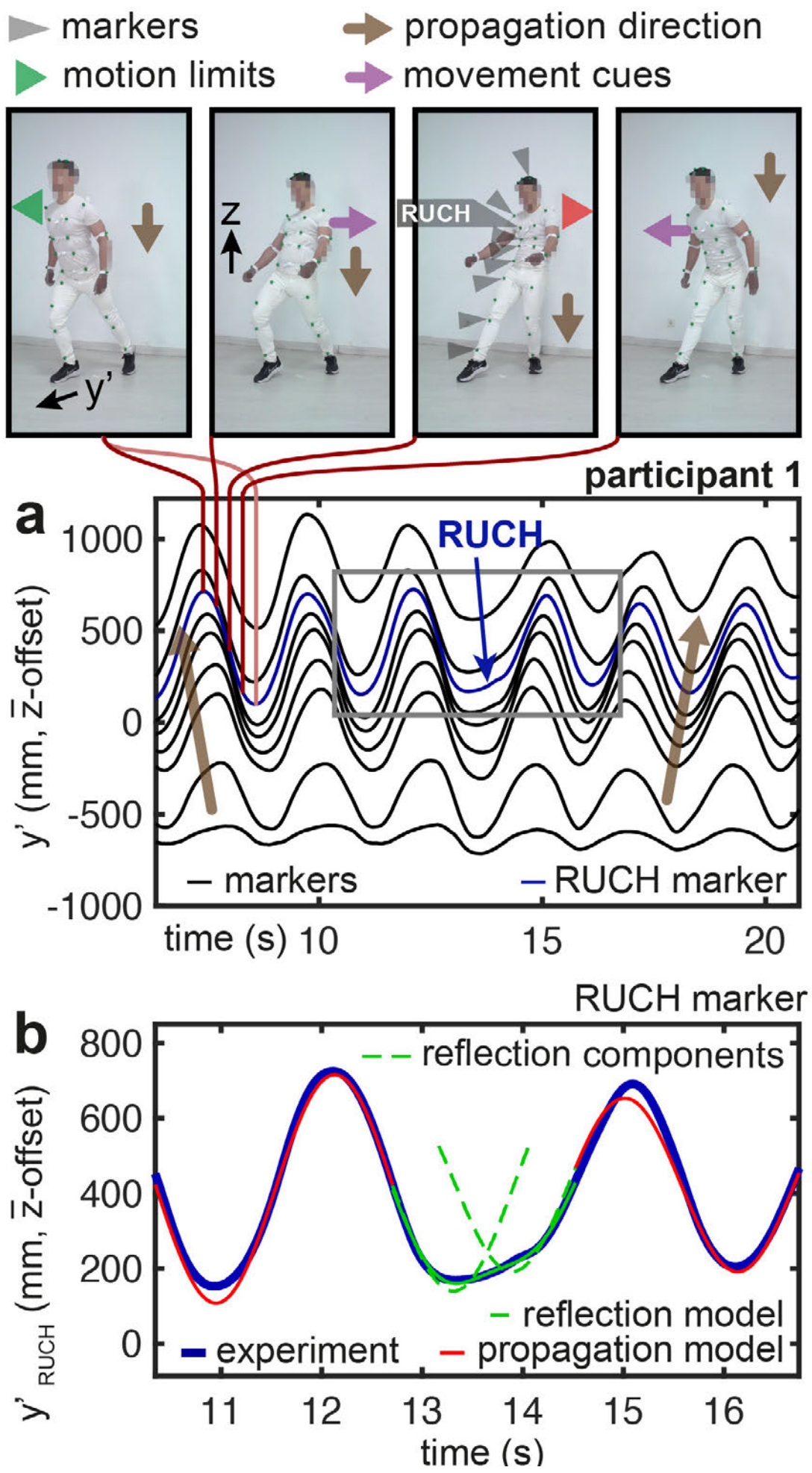

**Fig. 1. Propagation and reflection of waves. a**, Waterfall plot of marker projections. **b**, Reflection analysis of RUCH marker. Axes depicted in still images indicate their orientation.

Two synchronous dancers can be viewed as an analogy to the physical concept of coherence. This interpretation is conceptual, because it does not imply wave interference. Nonetheless, for body segments farther from the pelvis, the change in propagation direction is often accompanied by a brief pause, producing a plateau-like interval. By analogy with interference, this behaviour may be interpreted as the coexistence of counter-propagating components, with a weighted superposition that matches the data closely (see Fig. 1b). In addition, some markers show a small peak with no phase difference, suggesting a possible standing-wave feature (see Extended Data Fig. 3).

**Resonator response: amplitude and phase shifts (sequences 3 and 4)**

A wave reflected back and forth could undergo amplification or frequency-selective attenuation. This phenomenon, known as resonance, is one of the most significant oscillatory phenomena[35]. Examples of resonant behaviours include wind-driven tree motion, sound from musical instruments, and even the human voice. Resonant features are illustrated in sequence 3 with a concatenation of multiple *hip launches* spanning a spectrum of motion frequencies (see Fig. 2a,b, Extended Data Fig. 4a-f and Supplementary Video 3). Variations in motion frequency can be explored because a single hip launch is usually performed at different angular speeds. This sequence can be interpreted through a resonator model in the linear regime. Here we consider the SGTLP complex: the leader's circular drive applied at the follower's shoulder girdle results in an amplified or filtered pelvic response. These regions are represented by the virtual mid-position inferior scapular point (MBAK, extracted from left and right back markers, LBAK and RBAK, respectively, see Extended Data Table 1) and the virtual mid-posterior superior iliac spine point (MPSI, extracted from LPSI and RPSI markers), respectively. Fig. 2a,b presents the follower's normalized amplitude and phase-shift spectra, respectively.

The corresponding model is described by the differential equation[35, 36]:

$$\ddot{u} + \gamma\dot{u} + {\omega_0}^2 u = Ecos(\omega t), \text{ then} \quad (2)$$

$$\frac{A_{MPSI}}{A_{MBAK}} = \frac{\zeta}{\sqrt{({\omega_0}^2-\omega^2)^2+\gamma^2\omega^2}}, \quad \delta_{MPSI} - \delta_{\mathrm{MBAK}} = \arctan\frac{\gamma\omega}{{\omega_0}^2-\omega^2} \quad (3)$$

Where $E = \zeta A_{MBAK}$ is the amplitude of the excitation and $\zeta$, $\omega_0$, $\omega$ and $\gamma$ denotes a constant of proportionality, natural angular frequency, angular frequency of the excitation and damping/losses coefficient, respectively. Beyond amplification or filtering of the leader's drive, the relative pelvis motion shows a characteristic delay. Whereas an ideal low-loss resonator peaks at the natural frequency ($\omega_0$) with a 90° phase shift, the fitted data show the maximum amplitude shifted to a lower frequency ($\omega_{Amax}$), consistent with significant damping[36] (see methods). In fact, similar phase delays could also be observed in the broadly employed *still basic* steps of Bachata dance (both in traditional Dominican and that of Bachata Sensual, see Supplementary Information S3) defining the sequence 4 (see Fig. 2c-d, Extended data Fig. 4g-i and Supplementary Video 4). In this sequence, and continuing with the SGTLP complex, the leader induces lateral trunk displacement synchronized with the beats or counts (counts 1-4 and 5-7 define symmetric motion across the first and second bars, respectively). The follower then responds by maximizing vertical pelvic excursion. Inspired in the symmetric and antisymmetric mode analysis, we use $x'_{MSHO} = \frac{x'_{RSHO}+x'_{LSHO}}{2}$ and $z'_{DPSI} = \frac{z'_{RSHO}-z'_{LSHO}}{2}$ that denotes the virtual mid-shoulder point (MSHO) and the half-difference in vertical position of the posterior superior iliac spines (DPSI), respectively. The resulting delays (see the shift of the shaded areas under the curves) are around on-resonant phase ($\delta$=90º, Fig. 2d) in addition to an on-time optional pelvis dissociation (roughly $\delta$=0º, additional peaks/troughs $\hat{4}$ and $\hat{8}$). This suggests an effective resonant motion in response to the leader's drive.

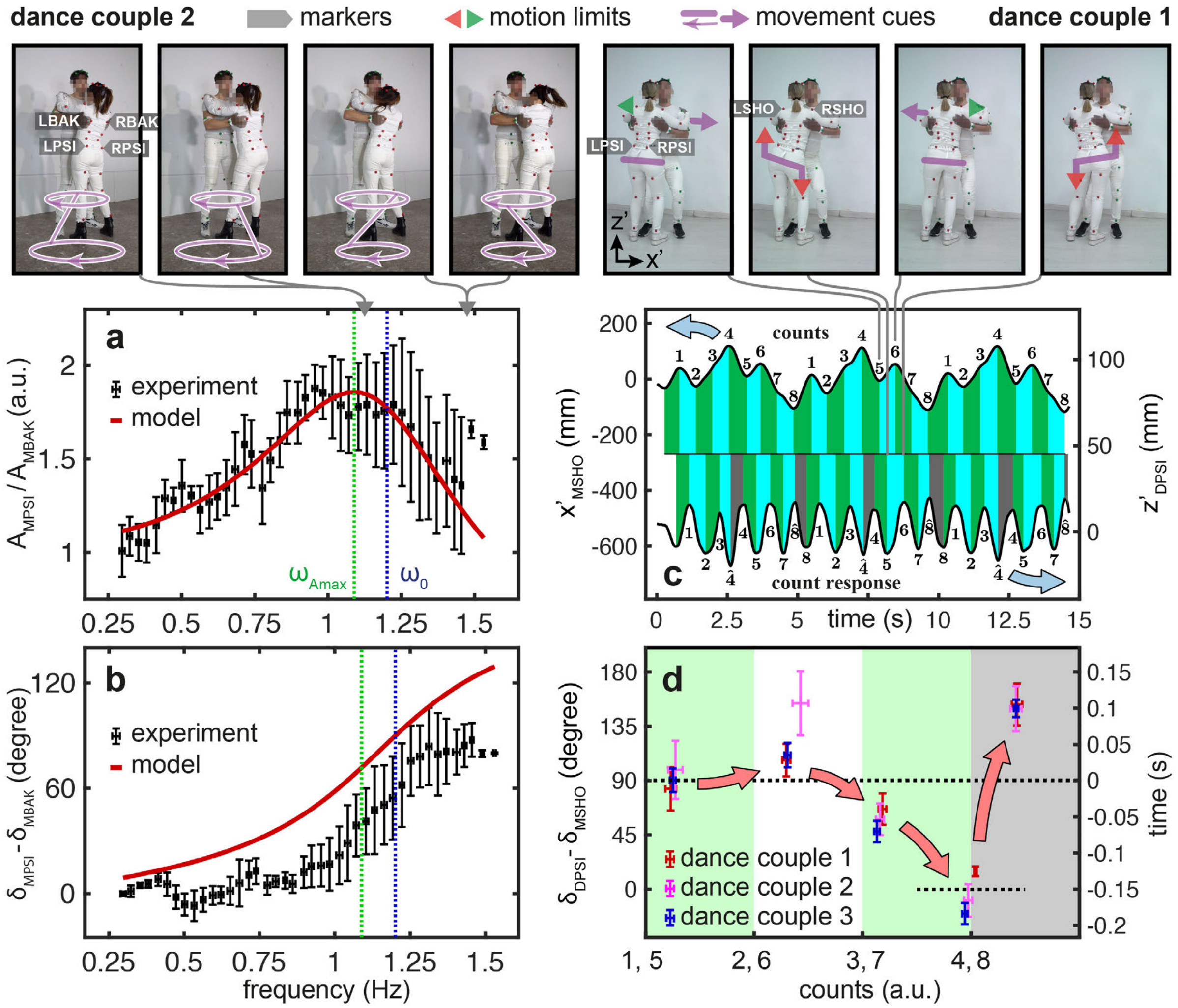

**Fig. 2. Resonator, amplitudes and phase shifts. a**, Normalized amplitude (MPSI/MBAK, mid-positions virtual markers) and b, phase shift spectra. **c,** Lateral drive (top, mid-position MSHO; with dance-count labels 1-8) and half the difference in elevation (bottom, DPSI; response labelled with the same count numbers, including decorative $\hat{4}$ and $\hat{8}$). **d,** Extracted phases versus count number for the three dance couples.

## Undamped coupled oscillators and harmonic-like modal relations (sequence 5)

In Sequence 3, the frequency shift, and in Sequence 4, the leader's motion marking the step timing, are consistent with a directional excitation scheme, in which the superior region of the SGTLP acts as an excitation source for the inferior region that is associated with a resonant response. However, both regions may also play active roles through bidirectional coupling, forming a coupled oscillator system. In nature, similar behaviour can be seen in the swaying of tree branches (supposing a linear motion). Sequence 5, termed the *V-wave*, illustrates this concept in Fig. 3 (see also Supplementary Video 5). The leader induces a lateral-superior displacement (I), then a pendular pelvic motion is performed (II-III), followed by another lateral-superior displacement (IV). This sequence is repeated symmetrically, resembling a sustained undamped oscillation over the analysed interval. Consequently, excitation and damping can be neglected in the model, and the motion is predominantly set by the initial conditions, which in dance correspond to the leader's *movement preparation*. The model of coupled equations of motion[35] (Fig. 3b) is the following:

$$\ddot{u}_1 + \omega_{0,1}^2 u_1 + \beta_{12} u_2 = 0 \qquad (4)$$

$$\ddot{u}_2 + \beta_{21} u_1 + \omega_{0,2}^2 u_2 = 0 \qquad (5)$$

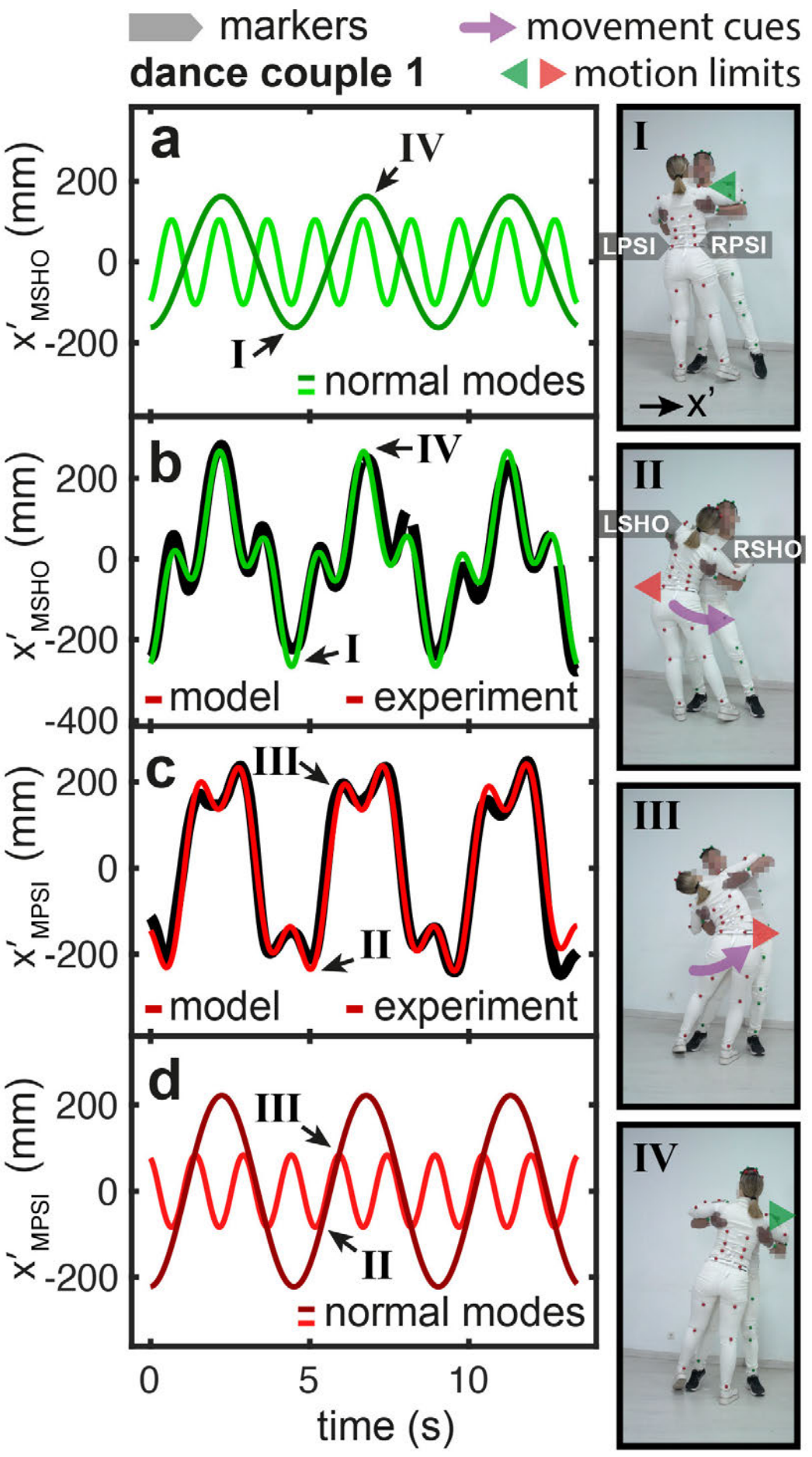

**Fig. 3.** Undamped coupled oscillators. **a/d** normal modes revealed by the **b/c** mid-position MSHO/MPSI motion projections.

where $u_1$ and $u_2$ are the displacements, $\omega_{0,1}^2$ and $\omega_{0,2}^2$ are the square of natural frequency, and $\beta_{12}$ and $\beta_{21}$ are the coupling coefficients[35,36,37] of the superior/inferior region (subindex 1/2) of SGTLP (observed at MSHO/MPSI, Fig.3 as well as their sides LSHO/LPSI and RSHO/RPSI, Extended Data Figs. 5-7). Treating the system as an eigenvalue problem yields two eigenvalues, corresponding to the normal-mode frequencies $\omega_1$ and $\omega_2$ (see Supplementary Information S7.2). Even more, the motion symmetry could be interpreted as a boundary problem, in which an integer number of waveforms or discrete modes can be obtained. Within this picture, the frequencies of these modes are related by rational numbers ($\mathcal{J}$, see Supplementary Table 1E-G). Across left, right and mid positions, all three participants show a ratio remarkably close to $\mathcal{J}$=3:1. Although no nonlinear process is involved, this modal relation resembles that between a fundamental and its third harmonic. In musical terms, this corresponds to a *dyad* of a *perfect fifth* extended by an *octave* or *compound perfect fifth*[38]. As supporting material, the signals from the dancers' performance are pitch shifted, for generating audible waveforms (see Supplementary Video 7).

## II. Combined sequence.

### Coupled oscillators: modal amplification, attenuation and time-dependent interference (sequence 6)

Thus far, we have described fundamental sequences; sequence 6 combines two distinct motions within a continuous dance performance (see Fig. 4 and Supplementary Video 6). It starts (sequence 6a, see also Extended Data Fig. 8) with the leader inducing a transverse-plane rotation, driven primarily by the motion of the left shoulder girdle. Next, the movement enters a descent phase driven by knee flexion, producing downward pelvic excursion into a squat posture, or *chair pose*, followed by an ascent motion. Because the displacement primarily affects the left side, we use the trajectories of LSHO and LPSI for a conceptual ellipsoidal approximation (Fig. 4a). Both the motion projections (Fig. 4b-e) and the azimuthal angles about the ellipsoid centre are consistent with a coupled-oscillator model (see Supplementary Figs. 2-4). The leader then induces a head rotation, or *head roll* (in sequence 6b, see also Extended Data Fig. 9), with the trajectory of the Top Back Head landmark (TBHD) providing another conceptual ellipsoidal approximation (Fig. 4f). Finally, the sequence ends with a smooth termination of the motion in an upright posture. Here the projected curves of TBHD and MSHO are also consistent with a coupled oscillatory behaviour. For analysing sequences 6a and 6b, the coupled system includes dissipative/active parameters:

$$\ddot{u}_1 + \gamma_{11}\dot{u}_1 + \gamma_{12}\dot{u}_2 + \omega_{0,1}^2 u_1 + \beta_{12}u_2 = 0 \qquad (6)$$

$$\ddot{u}_2 + \gamma_{21}\dot{u}_1 + \gamma_{22}\dot{u}_2 + \beta_{21}u_1 + \omega_{0,2}^2 u_2 = 0 \qquad (7)$$

The resulting eigenmodes of sequence 6a show both attenuation and reinforcement. Indeed, the amplified component could act as a transition to the initial conditions of the sequence 6b. In the latter, although both modes show attenuation, the sequence starts with time-domain constructive interference and ends with destructive one, allowing a rapid and smooth onset and termination of the motion. An analogous effect of such interference in Nature could be observed in the alternating periods of strong waves and almost no waves approaching a shore or cliff.

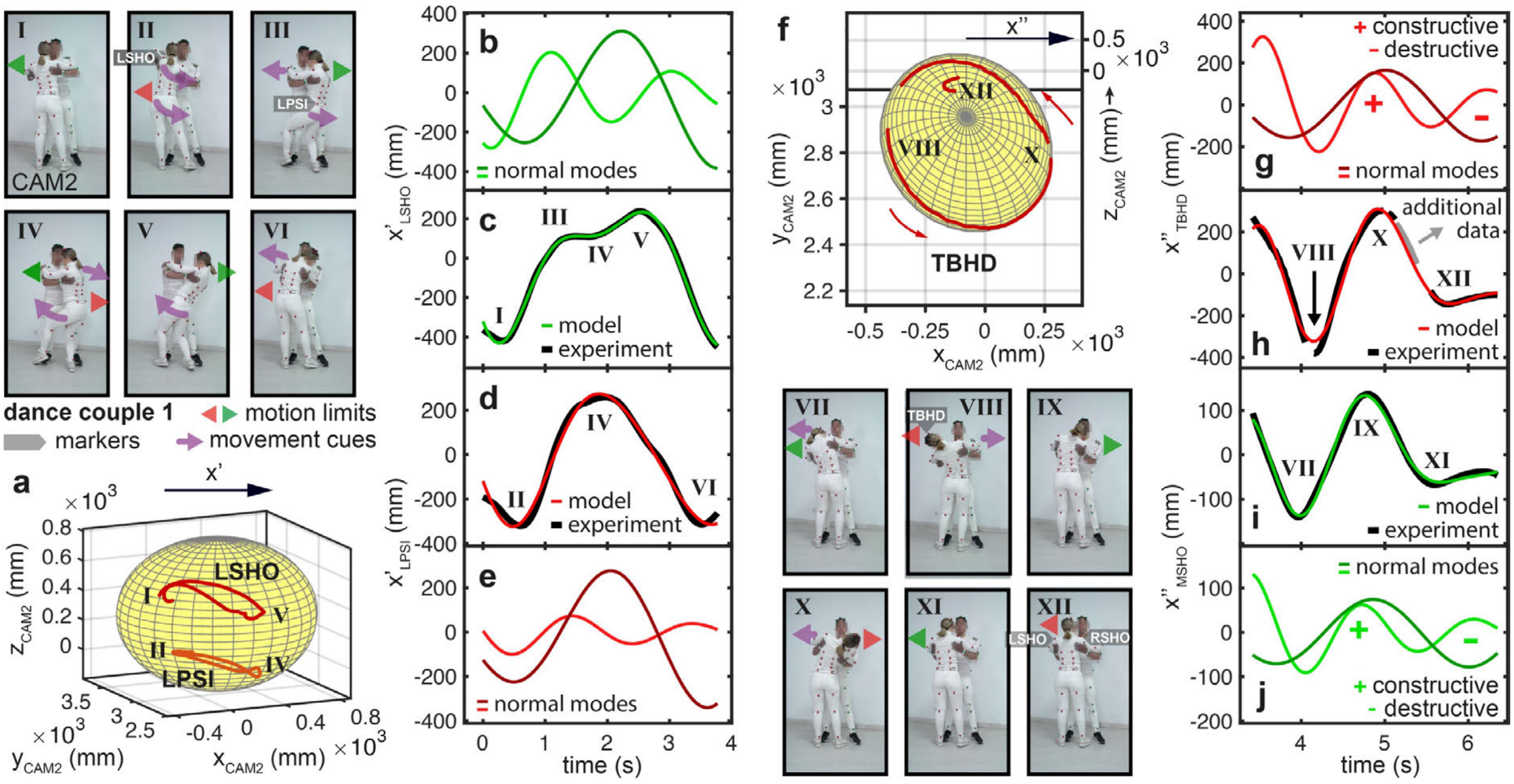

**Fig. 4. Combined sequence of coupled oscillators.** **a**, LSHO/LPSI motion on an ellipsoidal surface. **b,e**, modal decomposition of the corresponding projections in **c,d**. **f**, TBHD motion on an ellipsoid; **h,i**, projections of TBHD and the mid-position marker MSHO; **g,j**, corresponding modal decomposition. '+'/'**-**' in **g,j** indicate constructive/destructive time-dependent interference. Grey curve in **h**: additional reconstructed data not used for fitting.

In the coupled-oscillator scenario, dancers may exploit the intrinsic oscillatory behaviour of different body regions through the resulting modal features, supporting wave-like fluidity in the motion. By modulating coupling and damping in and between body regions[24,39], these modal features may be reshaped (see Supplementary Information S7.4) without being rigidly determined by intrinsic body parameters, allowing, for example, adaptation to rhythm. While the full set of system parameters (Eqs. 6 and 7) may admit multiple equivalent parametrizations consistent with the fitted trajectories, the resulting eigenvalues provide a compact and robust description of the underlying modal structure. The initial state, understood in dance terms as *preparation*, determines the relative modal contributions and thus shapes the observed amplitudes and phase relations, enabling diverse responses from the same underlying modal structure. By analogy, the same cliff may face calm waves or a storm. Because the fit does not define absolute reference points in the trajectory (including at onset), the initial state cannot be uniquely prescribed and is therefore treated as a set of free parameters. The coexistence of stable modal characteristics across variable system realizations may suggest an underlying structure alongside variability in dance[6].

## Conclusion

In conclusion, as L. Brillouin wrote, "waves always behave in a similar way”[40], in this work, the analysis of Bachata Sensual partner dance reveals a range of wave-related phenomena. It is based on motion projections of spine-related functional groups and their constituents across five individual sequences as well as in a combined sequence. The reconstructed wave-related phenomena include propagation, reflection, polarization, resonant amplitude and phase shifts and coupled oscillations. The underlying physical features revealed by the analytical models include interference, wavelength proportions, a possible standing-wave feature, resonant damping, modal amplification and attenuation, and constructive and destructive time-dependent interference. In this sense, the wave-physics framework can act as a form of choreographic motion description or notation, highlighting connections between partner-dance motion and harmonic phenomena in nature. Remarkably, this combination of spine-related functional groups, partner-dance motion and analytical access to underlying physical structure also revealed, under certain conditions, coupled modes emerge with a frequency ratio remarkably close to 3:1, revealing a harmonic-like modal relation. When mapped into the audible range, these modes can be heard as musical dyads, providing a direct acoustic analogy between bodily motion and harmonic structure. By modulating coupling and damping, dancers may shape the modal structure, characterized by eigenvalues that, within certain limits, are not rigidly constrained by body morphology. This suggests an underlying modal structure alongside variability in movement realization. Within this picture, the initial conditions of the interpretation shape the realised response through the relative modal contributions. In Phase II, this study is foreseen to be extended to a larger participant sample of approximately 20 couples. Building on the present findings, the work could also be extended to evaluate other symmetric movement sequences constrained by boundaries, and to explore different musical *dyads*. A complementary analysis of the vertical component of motion, and associated potential energy variations, may reveal further analogies involving a two- to three-level energetic structure (shoulder girdle-trunk, pelvis, and head). Drawing on the pedagogical use of analogy[41], this study may offer an educational bridge between dance and wave and oscillatory physics. More broadly, because the wave is considered a paradigm in dance, this approach could be applied to other dance styles. In addition, potential links between these wave-based descriptors and neuronal or physiological activity in individual dancers or couples could be explored[42]. As for now, the open-ended question to lovers of dance is: what harmonic splendours of Nature do you wish to interpret? let’s dance.

# Methods

## Rhythm

With the exception of the *hip-launch* sequence (Fig. 2a,b and Extended Data Figure 4), a metronome set to 100 beats per minute (bpm) was used as a timing reference, similar to the rhythmic guidance provided by music. However, timing was not strictly enforced, preserving the dancer's interpretive freedom. This is most evident in sequence 6b (*head roll*), where the end timing clearly varies across interpreters.

## Motion Tracking, algorithms and data reconstruction

**Motion tracking.** The motion was recorded using identical cameras (CAM1 and CAM2): Sony Alpha 7SII, with Canon EF 18-55mm f/3.5-5.6 lens. The resolution was set to 1920 × 1080 pixels in progressive mode, with a frame rate of 119.88 frames per second. The cameras were positioned at about 3 meters of distance from the dancers, forming an angle of about 37º. Lens distortion was corrected by analysing a square-lattice reference using the image processing toolbox of MATLAB. Twenty-two spheres for anterior view and 22 for posterior view markers (25 mm in diameter) were attached to the leader (green) and follower (red) so as to define anatomical landmark for tracking (see Extended Data Table 1 and Extended Data Fig.1g and 2g). For sequence 6b, additional head landmarks were used for 3D–3D reconstruction.

**Automatic algorithm.** The automatic algorithm assigns each pixel a three-component RGB colour vector. By calculating the distance in RGB colour space between a marker colour and each image pixel, colour proximity is obtained and used to identify possible tracking points. Each cluster of pixels that fulfils certain colour proximity and size is identified as a marker. Next, each marker centre is linked with the nearest marker from the previous frame, thus constructing the motion trajectory. Then the 3D paths from the dual or stereoscopic information are built by employing the image processing toolbox from MATLAB. The raw 3D trajectories of the tracked points were obtained in a camera-based global Cartesian coordinate system defined with respect to either CAM1 or CAM2 as ($x_{CAM1}$, $y_{CAM1}$, $z_{CAM1}$) or ($x_{CAM2}$, $y_{CAM2}$, $z_{CAM2}$), respectively. For analysis, we additionally used relative projection axes, denoted by primes, whose indicative directions are shown in Figs. 1 to 4. For each sequence, the origin of the relative axes was placed at the centre of the corresponding motion, introducing a constant offset between the raw 3D coordinates and the theoretical model; this offset can be estimated during fitting.

**Manual verification and computer-based correction.** Markers were occasionally partially occluded, which caused the automated tracking to localize them imprecisely or fail to detect the marker. In these cases, we manually pre-selected the marker region and applied a circle-detection procedure based on Random Sample Consensus (RANSAC), with optional filtering, and subsequent fit refinement (see Supplementary Information S9). All traces were manually verified, and when partial occlusions were present the marker position was identified or corrected using the procedure described above. In some frames, the marker was nearly or completely occluded and was therefore considered unavailable.

**2D-3D reconstruction.** In order to reconstruct trajectories for occluded markers, we leveraged the fact that the motion is often approximately planar and perpendicular to one of the cameras, which typically provides higher marker visibility. We therefore implemented a simple reconstruction procedure that maximizes the number of usable marker samples by fitting the undistorted, single-camera trajectories to their corresponding 3D projection, using an additive and multiplicative biases. This approach was adopted to mitigate pronounced occlusions affecting some of the 3D motion tracking data (see Supplementary Figs. 1 and 5). The deviations are typically on the order of millimetres and are reported in Supplementary Table 2, indicated by a superscript "2D-3D" (see Supplementary Figs. 1 and 5).

**3D-3D reconstruction.** The rigidity of the head allowed the use of additional markers and reference points. This enabled the application of the orthogonal Procrustes method[43] (see Supplementary Information S10). As a result, a large fraction of the TBHD marker data in sequence 6b was reconstructed (see Supplementary Fig. 5). The deviations are reported in Supplementary Table 2, indicated by a superscript "3D-3D".

## Sequence analysis methods.

Most analyses were performed by fitting analytical models to the experimental data. Supplementary Table 1 reports the key extracted parameters, and Supplementary Table 2 summarizes fit quality using the root mean square deviation (RMSD), the experimental data range (EDR), and a normalized goodness-of-fit metric RMSD/EDR.

**Propagating wave in anteroposterior displacements, sequence 1.** Extended Data Fig. 1 shows waterfall plots for participants 1-3, for both downward- and upward-propagating waves. Each curve in the plot is represented by $y_i' + \bar{z}_i - \bar{y}_i'$, where $y_i'/\bar{y}_i'$ is the projection/average projection of the motion in the $y'$ axis; and $\bar{z}_i$ is the average $z$ coordinate for an $i$ tracked marker. Consequently, the projected data $y_i'$ are vertically offset and ordered according to the average height of each marker. To evaluate the full range of wave propagation, we considered markers from the whole body without upper limbs (WB-NUL). Because anteroposterior motion can be assumed to be equivalent on both sides of the body, the most visible markers were selected from a predominantly right-side sequence, with the head marker represented by LFHD (see Supplementary Video 1). The sequence from top to bottom is therefore LFHD, RSHO, RUCH, REOB, RICH, RASI, RTHI, RTIB and RANK (see Extended Data Table 1 and Extended Data Fig. 1). Only a single dancer is shown here, as this sequence serves as the starting point of the manuscript before introducing coupled dancing. These motions can be performed synchronously, symmetrically, or individually; the individual performance additionally allows reliable tracking of most target markers.

**Propagating wave in mediolateral displacements. Sequence 2.** Extended Data Fig. 2 shows the same waterfall plots as Extended Data Fig. 1, but for mediolateral body displacements, traced along the orthogonal direction $x'$. In this case, mid-position markers, mostly virtual, were used. The virtual marker positions were computed by averaging the corresponding left and right marker coordinates. The sequence from top to bottom is therefore TBHD, MSHO, MBAK, MIDO, MLUM, MPSI, MTHI, MKNE and MANK (see Extended Data Table 1).

**Propagating waves. Extracted parameters.** Extended Data Figs. 1 and 2 show the experimental curves (black) together with the model fits (red curves) obtained using Eq. 1. The fitted parameters are reported in Supplementary Tables 1A and 1B, respectively. The frequencies ($\omega^{dir}/2\pi$, with the direction $dir$ = $D$ or $U$ denoting downward- and upward-propagating waves) are constrained to be the same across all body segments in the fit. This shared frequency enables the extraction of phase differences. In these sequences, we focus on WB-NUL ($\Delta\delta_{WB-NUL}^{dir}$) and SGTLP ($\Delta\delta_{SGTLP}^{dir}$), see Supplementary Information S11. In addition, mean amplitudes are reported for each configuration ($\bar{A}_{WB-NUL}^{dir}$, $\bar{A}_{SGTLP}^{dir}$). Supplementary Tables 2A and 2B report fit qualities via the maximum and minimum deviations across the t ${y'}_i^{dir}$ or ${x'}_i^{dir}$ curves (where $i$ denotes the marker).

**Reflections in sequences 1 and 2.** The markers selected to analyse reflection were RUCH and MBACK (see Extended Data Fig.3) because they are located at similar heights in the anterior and posterior views and are not affected by occlusion. The fit was constructed by extending the previously fitted downward- and upward-propagating responses into the intermediate region and combining them in a superposition analogous to wave reflection, but with weighted contributions. The fit region is assumed to be bounded by the points where each component crosses its mean level. Phase and frequency were constrained using the previous sinusoidal fit, whereas amplitudes were treated as free parameters constrained between the overall maximum and overall minimum across the two sinusoidal fits. Each contribution was weighted by a logistic envelope,

$$u^L = \frac{1}{1+e^{\pm g(t-t_0)}},$$

where $g$ is the growth rate, $t$ is time and $t_0$ defines the transition onset (see Supplementary Table 1A,B). The $\pm$ sign indicates that the downward- and upward-propagating sinusoidal responses were weighted using the two opposite logistic branches. This yields a weighted superposition in which each contribution is modulated and gated by its corresponding logistic envelope.

**Spectral response in sequence 3.** In order to obtain the spectra of experimental amplitude and phase, large amount of data was analysed by rotating the projection axis $x'$ around $z'$ (performing steps of one degree). This procedure is analogous to retrieving the real part of a complex wave amplitude ($x(t) + i\,y(t)$). As a result, 180 datasets MSHO and MPSI positions have been evaluated in 1° increments. Here, MSHO was analysed as a representative marker of the shoulder girdle because it shows the largest amplitude and is mostly visible in the stereoscopic recordings. The experimental normalized-amplitude data were used to fit the model defined by Eqs. 3, and the theoretical phase values were computed from the fitted parameters. As the damping value is significant, the spectral position ($\omega_{Amax}$) of the amplitude peak $A_{max}(\omega_{Amax})$ is shifted by a factor[36] of

$$\frac{\omega_{Amax}}{\omega_0} = \sqrt{1 - \frac{\gamma^2}{2\omega_0{}^2}}\,.$$

Consequently, the phase delay at $\omega_{Amax}$ is different from that at $\omega_0$ . Supplementary Table 1C lists the fitting parameters, including the extrapolated amplitude to zero frequency $A(0)$ and the ratio $A_{max}/A(0)$ (see Supplementary Information S12).

**Phase delay in sequence 4.** The data analysis focuses on the peak and trough positions of the MSHO and DPSI curves, extracted using Gaussian fits. For MSHO, these peaks and troughs mark transitions between dance counts, 1-4 and the symmetric 5-8, or more generally a change in the leg supporting the dancer's centre of mass. Although a subtle inflection is visible, no prominent peak or trough occurs at counts 3 and 7, where the centre of mass basically continues its displacement until the end of the bar, at counts 4 and 8. Accordingly, the timing of counts 3 and 7 was assumed to occur at the temporal midpoint between the preceding and subsequent count. The DPSI signal defines the responses over counts 1-4 and the symmetric 5-8, including the additional on-beat decorative $\hat{4}$ and $\hat{8}$ pelvic motions. Consequently, their symmetric phases ($\Delta\delta_{1,5}$ , $\Delta\delta_{2,6}$ , $\Delta\delta_{3,7}$ , $\Delta\delta_{4,8}$ and $\Delta\delta_{\hat{4},\hat{8}}$ were grouped, see Supplementary Table 1D). Note that both $\Delta\delta_{\hat{4},\hat{8}}$ and $\Delta\delta_{4,8}$ are computed with respect to the leader's timing at counts 4 and 8.

**Mode analysis in coupled oscillators.** Projections of experimental data from sequences 5 and 6 (as well the angular variable) were modelled as a coupled-oscillator system using a general two-degree-of-freedom, linear, second-order formulation. The equations of motion are written in reduced-mass form as $\boldsymbol{\ddot{u}} + \boldsymbol{C\dot{u}} + \boldsymbol{Ku} = 0$, where $\boldsymbol{C}$ and $\boldsymbol{K}$ capture damping and stiffness coupling terms, respectively (Supplementary Eqs. S1-S7). For modal analysis we use the standard first-order state-space form $\boldsymbol{\dot{w}} = \boldsymbol{Aw}$, whose eigenvalues occur as two complex-conjugate pairs $\lambda_n = -\kappa_n \pm j\omega_n$, defining the modal frequencies $\omega_n$ and damping factors $\kappa_n$ (Supplementary Eqs. S9-S13). The displacement of each component can then be written as a superposition of two damped sinusoids, with mode-specific amplitudes $A_{mn}$ and phase offsets $\delta_{mn}$ determined by the eigenvectors and the initial state (Supplementary Eqs. S14-S20).

Individual entries of $\boldsymbol{C}$ and $\boldsymbol{K}$ can correspond to different configurations that yield comparable fits, whereas the modal quantities $\omega_n$ and $\kappa_n$, derived from the fitted model, remain robust across equivalent system representations (particularly in the case of $\omega_n$, and for $\kappa_n$ within appropriate observation windows, see Supplementary Eqs. S21-S34). Moreover, the modal parameters can absorb uncertainties arising from the estimation of the projection-plane position or from deviations from ideal viewing alignment in the 2D-3D reconstruction.

**Sustained undamped oscillatory motion and musical *dyad* formation in sequence 5.** Because marker occlusions occurred during this sequence, we applied a 2D-3D reconstruction. This procedure is illustrated in Supplementary Fig. 1, and Supplementary Table 2D reports normalized deviations (RMSD/EDR) below 1%. Here, left-, right- and virtual mid-position markers were analysed to compare measurements and minimise the impact of missing data. The dancers' motion (see Supplementary Video 5) resembles a pendulum-like system, here consisting of two coupled entities oscillating laterally. The corresponding coupled-oscillator model is described by Eqs. (4) and (5). In this case, neither direct nor cross-damping terms are included, so that $\boldsymbol{C}=\boldsymbol{0}_{2x2}$ . The model solution used in the fitting procedure was obtained by numerically solving the system of equations with an Ordinary Differential Equation (ODE) solver in MATLAB. This approach yielded good agreement with the experimental data while requiring fewer parameters than in sequence 6. Although the lower-frequency component resembles a fundamental, naturally entrained to the 1-8 count phrase, the accompanying 3:1 *ternary* component, is unusual within a *binary* musical bar. We interpret this behaviour as arising from two bodies interacting within a bounded system. We suggest that dancers' kinesthetic coupling, grounded in proprioceptive sensing[33] and sensorimotor responses, contributes to the observed harmonic-like content.

**Ellipsoid fitting process.** Because the data do not fully constrain the ellipsoid along its axial directions, the fit may converge to different parameter configurations. In particular, increasing one or more semi-axes ($s_1$, $s_2$ or $s_3$) can reduce the coordinate contribution to the residual without improving geometric fidelity. For this reason, we performed a 3600-step rotational sweep to minimize the ellipsoid residual

$$\varepsilon_{\text{ellipsoid}} = \sqrt{\left(\frac{x-c_x}{s_1}\right)^2 + \left(\frac{y-c_y}{s_2}\right)^2 + \left(\frac{z-c_z}{s_3}\right)^2} - 1 ,$$

where $(s_1, s_2, s_3)$ are the semi-axis lengths oriented at an azimuthal angle $\theta$, and $(c_x, c_y, c_z)$ denote the centre coordinates. We then retained a statistical mode (which is robust to outliers) across all solutions by computing the mode of the semi-axes and the RMSD (see Supplementary Information S8). Although the goodness-of-fit metrics indicate reasonable agreement (especially for Sequence 6b, see Supplementary Tables 1H, 1K, 2E and 2F), the fitted ellipsoids should be regarded as conceptual geometric approximations of intrinsically 3D complex motions, and their parameters interpreted as qualitative rather than quantitative estimates. Here, the fitting uses stereoscopic marker data for sequence 6a and 3D-3D reconstruction for sequence 6b.

**Coupled oscillators of sequence 6a.** For sequence 6a, we performed a coupled-oscillator analysis using (a) the horizontal components of motion, expressed as azimuthal angles $\theta$ about the ellipsoid centre, and (b) the motion projected onto an $x'$ axis obtained by rotating the traces to maximize the experimental data range. The resulting angular and projected trajectories were fitted with multiplicative and additive biases (Supplementary Figs. 2-4), and fit quality is reported as $x^{\theta\text{-}3D}$ in Supplementary Table 2E. The close agreement between these trajectories, including the extracted normal modes, supports the reliability of the analysis despite sensitivity in estimating the ellipsoid centre and the approximation inherent in using a spherical coordinate representation. Consequently, in Fig. 4c-d and Supplementary Video 6, we show only projections onto the $x'$ axis, whose parameters are more robust than those derived from the ellipsoid-based approach. In this case, direct and cross-damping coefficients were included in the fitting procedure, yielding normal modes with both decay and growth terms. In this three-dimensional motion, the CAM2 axes were used as the global reference frame for plotting.

**Coupled oscillators in sequence 6b.** To mitigate missing data due to occlusion of the components that define the MSHO trajectories, we applied a 2D-3D reconstruction. In the case of dance couples 1 and 3, after the rotational motions of sequence 6a, we continued to use the global axes of CAM2. For dance couple 2, we switched to CAM1, as its coordinate system appeared better aligned with the motion and resulted in fewer marker occlusions during 2D-3D reconstruction. For the TBHD marker, both 2D-3D and 3D-3D reconstructions were computed (see Supplementary Fig. 5). In both cases, the analysed data correspond to the motion projected onto the $x''$ axis. For the 3D-3D reconstruction, we incorporated auxiliary visual landmarks to provide additional spatial information. In the case of dance couple 1: (a) a stud earring tracked at the earlobe, (b) the intersection between a high-contrast hair strand and the elastic band, assumed fixed relative to the head, and (c) the upper ear point, defined as the highest point of the ear at the superior helix. Landmark (a) was tracked using a RANSAC-based circle fit (see Supplementary Information S9). For landmark (b), we fitted a line to the hair strand and detected its intersection with the tonal transition corresponding to the elastic band. For landmark (c), we applied two RANSAC-based partial circle fits to the upper and lower ear contours and used the line connecting the circle centres to determine the upper intersection point, which was defined as the upper ear point. Dance couples 2 and 3 wore an additional headband carrying three markers (placed below the LHBD, THBD and RHBD markers), enabling reconstruction of the full TBHD trajectory. In the case of dance couple 3, an additional visual landmark was used, corresponding to the intersection between a hoop earring (approximately 10 mm in radius) and the earlobe. A strategy similar to that in (b) was followed. Here, the earring and the earlobe defined two colour ranges. Depending on the size of the selected region of interest and subsequent upsampling, either a circular arc or a line was used to represent the earring portion that intersects with the earlobe. To validate the 3D reconstruction procedure, for all three dance couples we compared 2D-3D and 3D-3D deviations in $x''$ (see Supplementary Table 2F). Visually, Supplementary Fig. 5 illustrates the 2D-3D and 3D-3D solutions, showing close overlap between the two approaches. In the case of 3D-3D reconstruction of dance couple 1, fewer time instants were available at which stereoscopically tracked markers coincided with additional landmarks. As a result, this dataset provides fewer independent estimates for reconstructing the trajectory and fewer samples for computing deviation statistics. Despite these limitations, the 3D-3D reconstruction (evaluated along the $x''$ axis) yielded RMSD/EDR $\approx$ 0.15%, the lowest among all participants. Although the 2D-3D reconstruction provided reasonable agreement (average RMSD/EDR $\approx$ 1.34%), the 3D-3D approach resulted in smaller deviations (average RMSD/EDR $\approx$ 0.29%) than the 2D-3D method. This can be explained by the predominantly non-planar nature of TBHD motion. Accordingly, the 3D–3D reconstruction was selected for the analysis of sequence 6b, as it further enabled the reconstruction of longer TBHD trajectories. For dance couple 1, a portion of the 2D-3D reconstructed trajectory is shown in grey (Fig. 4h) to illustrate gap filling; however, it was excluded from the fitting to maintain methodological consistency across dance couples.

## Extended Data and Supplementary Information

Extended Data Table 1 and Extended Data Figs. 1-9 are available in this document. Supplementary Information is also available. Supplementary Figs. 1-5 are included in this document. Supplementary Videos 1-7 can be accessed at the following URL:

https://osf.io/2vkqg/overview?view_only=e3693b66672d4c2786e2bea143fc5edf

## Ethics declarations

### Competing interests

The author declares no competing interests.

### Ethical approval

This study involving human participants was approved by the Ethics Committee of The Spanish National Research Council (CSIC) being assessed under the internal registration codes 165/2023 (internal project: "Analogías entre la expresión corporal en la danza y la física ondulatoria - modificación") and 160/2025 (FECYT project: PAREIDOLIA. PARamEtrización e IDentificación de fenómenos Ondulatorios en el baiLe y sus analogÍAs con la naturaleza). All procedures were conducted in accordance with the Declaration of Helsinki. Written informed consent was obtained from all participants. Participants also provided explicit consent for the recording and use of their image data for scientific dissemination and public communication, including scientific publications and preprints. To protect participant identity, faces and any identifiable features were pixelated; additionally, hands were also masked. Special attention was paid to participants' welfare, and they were covered by accident insurance during the recording sessions.

## Acknowledgments

The author gratefully acknowledges J. Escalona and J. Cordero for their helpful comments and discussions on dance, F. Cantos-Prieto, R. Fenollosa, F. Meseguer and E. Navarro-Moratalla on physics, A. Dominguez on biomechanics, A. Lluch on mathematics and music, A. Cervera and L. Salguero on arts in general, J. Ros for technical support and P. Atienzar and D. Fernández for general scientific considerations. The author expresses profound gratitude to the Ethics Committee of CSIC and gratefully acknowledges the support of the Spanish Foundation for Science and Technology (FECYT) and the Spanish Ministry of Science, Innovation and Universities for funding the project "PAREIDOLIA. PARamEtrización e IDentificación de fenómenos Ondulatorios en el baiLe y sus analogÍAs con la naturaleza" (reference: FCT-23-19135). In addition, financial support by the Spanish Ministry of Science and Innovation (CEX2021-001230-S grant funded by MCIN/AEI/10.13039/501100011033) is gratefully acknowledged.

**Extended Data Table 1. Marker descriptions, placements, virtual markers and analysis sets.**

| Name | View | Description | Placement region | Mid/ diff. | SGTLP | SGCH | WB-NUL |
|---|---|---|---|---|---|---|---|
| TFHD[1] | Anterior | Top Front Head | High frontal region, midline | TFHD | - | X | X |
| TBHD[1] | Posterior | Top Back Head | Galea aponeurotica, high region, midline | TBHD | - | X | X |
| LFHD/RFHD | Anterior | Left/Right Forehead | High frontal, above left/ right temporal region | - | - | X | X |
| LBHD/RBHD | Posterior | Left/Right Back Head | Galea aponeurotica, high region, left/right | - | - | X | X |
| LSHO/RSHO | Both | Left/Right Shoulder | Left/right acromio-clavicular joint | MSHO | X | X | X |
| LBAK/RBAK | Posterior | Left/Right Back | Left/right inferior scapular region | MBAK | X | X | X |
| LUCH/RUCH[2] | Anterior | Left Right Upper Chest | Left/right upper pectoral region | - | X | - | X |
| LIDO/RIDO[2] | Posterior | Left/Right Inferior Dorsal | Left/right inferior dorsal region | MIDO | X | - | X |
| LEOB/REOB[2] | Anterior | Left/Right External Oblique | Left/right superior external oblique region | - | X | - | X |
| LLUM/RLUM[2] | Posterior | Left/Right Lumbar | Left/right lumbar region | MLUM | X | - | X |
| LICH/RICH[2] | Anterior | Left Right Inferior Chest | Left/right costal margin | - | X | - | X |
| LPSI/RPSI | Posterior | Left/Right PSIS | Left/right posterior superior iliac spine | MPSI/ DPSI | X | - | X |
| LASI/RASI | Anterior | Left/Right ASIS | Left/right anterior superior iliac spine | - | X | - | X |
| LTHI/RTHI | Both | Left/Right Thigh | Left/right upper thigh | MTHI | - | - | X |
| LKNE/RKNE | Posterior | Left/Right Knee | Flexion-extension axis of left/right knee | MKNE | - | - | X |
| LTIB/RTIB | Anterior | Left/Right Tibia | Left/right upper tibia | - | - | - | X |
| LANK/RANK | Both | Left/Right Ankle | Left/right ankle region | MANK | - | - | X |
| LELB/RELB | Both | Left/Right Elbow | Left/right lateral epicondyle | - | - | - | - |
| LWR/RWR[1] | Both | Left/Right Wrist | Left/right dorsal wrist | - | - | - | - |

Markers are classified by recording orientation as anterior, posterior or present in both views (see Extended Data Figs. 1g and 2g). The Mid/Diff. column indicates mid-position markers or half-difference quantities: markers with the prefix M are virtual and computed as the average (half the sum) of the left and right markers, whereas markers with the prefix D represent half the difference in vertical position between them. The analysis sets are **SGTLP**, the Shoulder Girdle-ThoracoLumboPelvic complex; **SGCH**, the Shoulder Girdle-Cervical-Head complex; and **WB-NUL**, Whole Body No Upper Limbs. For illustrative purposes, all real markers are considered to define **WB**, Whole Body. Most real markers follow the Plug-in Gait naming convention, except that superscript 1 denotes intermediate marker positions and superscript 2 denotes additional trunk markers placed as left-right pairs.

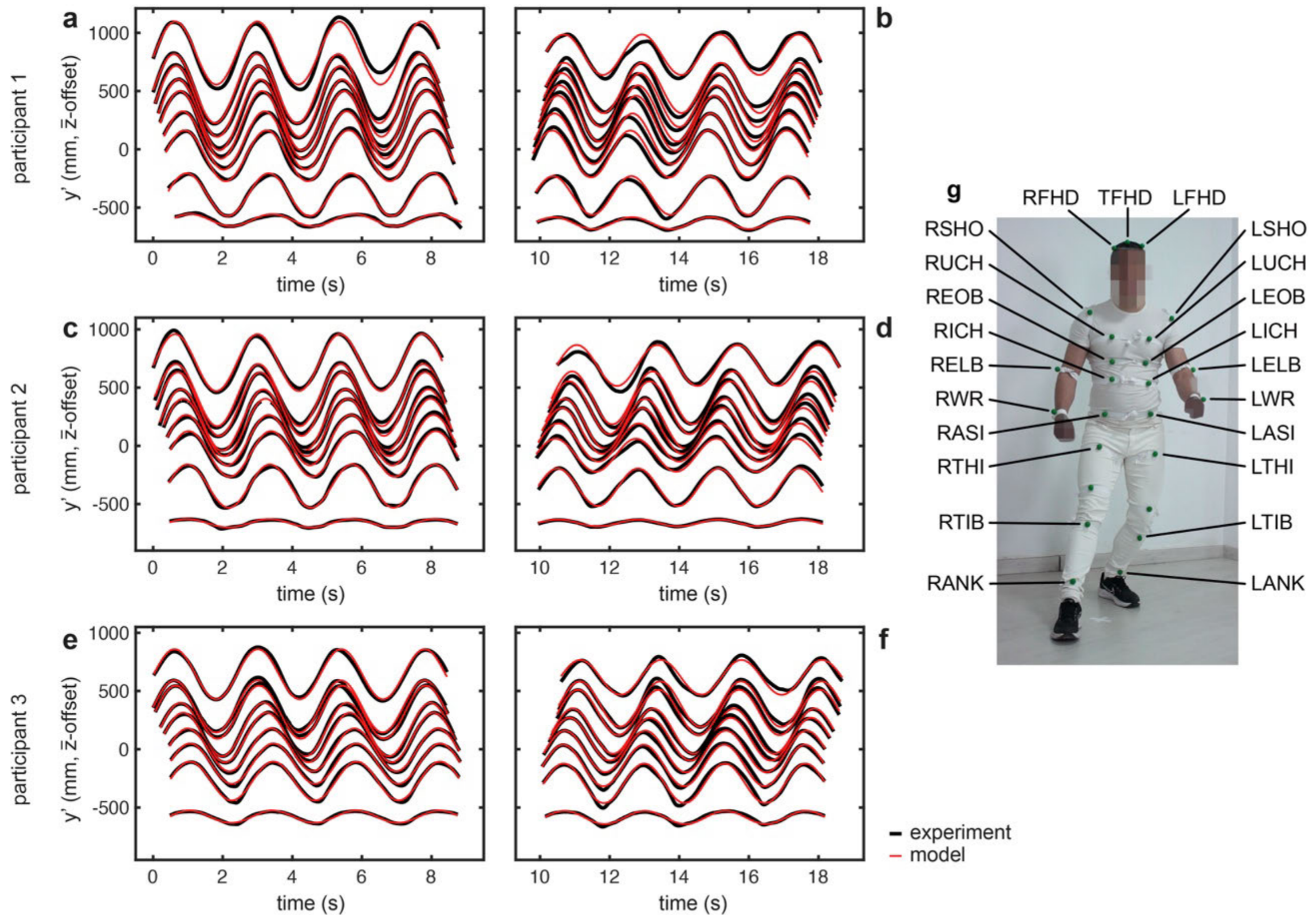

**Extended Data Fig.1. Waterfall representation of downward- and upward-propagating waves for anteroposterior body displacements (sequence 1).** Panels **a**, **c** and **e** and panels **b**, **d** and **f** correspond to the downward- and upward-propagating waves, respectively. Panel **g** shows the anterior view with marker locations labelled by their acronyms (see Extended Data Table 1 for acronym definitions). Panels **a-b**, **c-d** and **e-f** correspond to the participants 1, 2 and 3, respectively. Black lines depict the experimental data, and red lines show the model fits. The vertical offset corresponds to their respective mean z position, given that the data are mean-subtracted. From top to bottom, the waterfall displays the sequence of markers defined for the anterior view: LFHD, RSHO, RUCH, REOB, RICH, RASI, RTHI, RTIB, and RANK (see Extended Table 1). These selected markers are almost always visible to both cameras (see Supplementary Video 1). With the exception of LFHD, all selected markers are right-side markers. We assumed that right- or left-side markers yield equivalent anteroposterior motion projections.

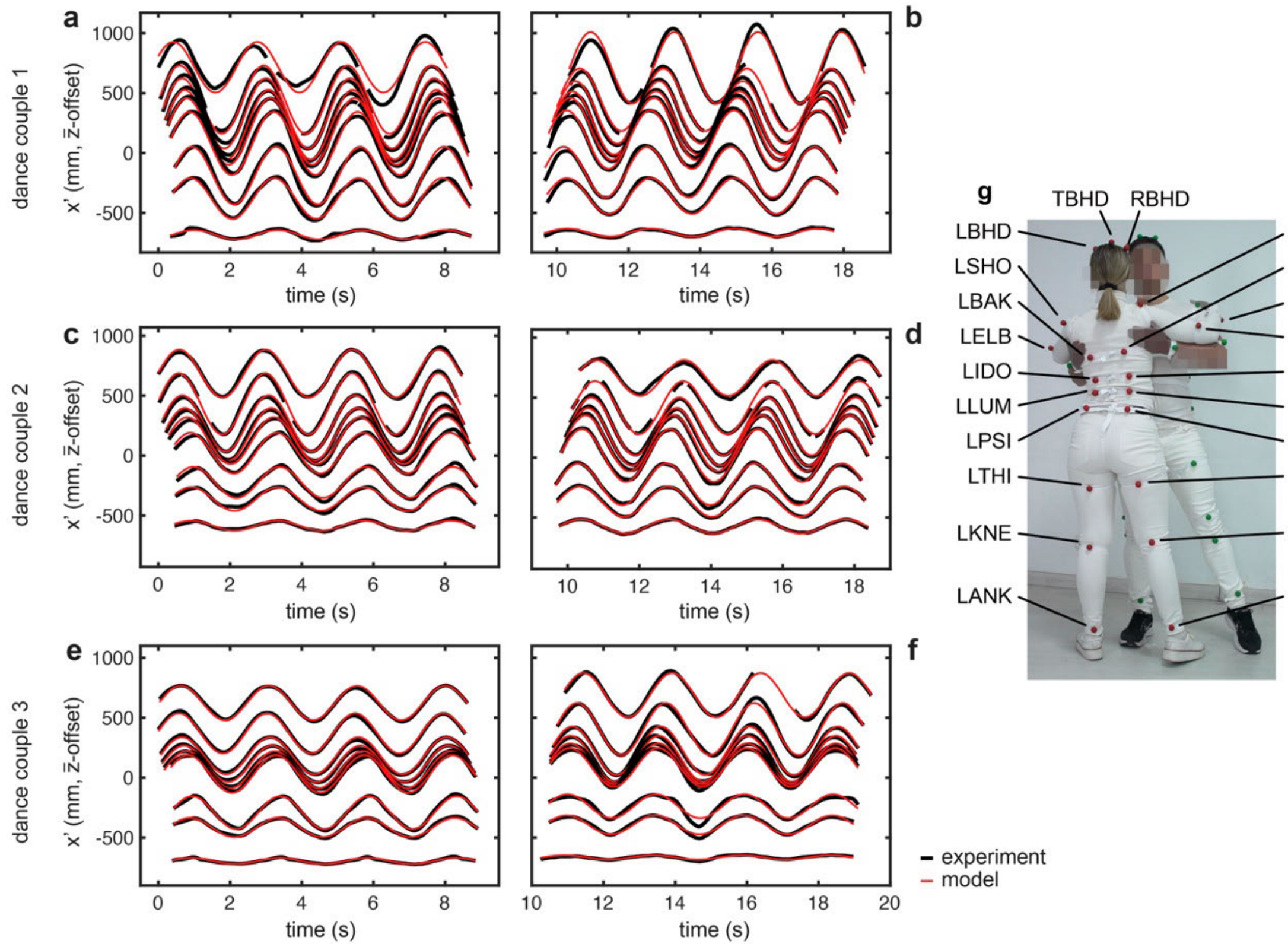

**Extended Data Fig.2. Waterfall representation of downward- and upward- propagating waves for mediolateral body displacements (sequence 2).** Panels **a**, **c** and **e** and panels **b**, **d** and **f** correspond to the downward- and upward-propagating waves, respectively. Panel g shows the posterior view with marker locations labelled by their acronyms (see Extended Data Table 1 for acronym definitions). **a-b**, **c-d** and **e-f** correspond to the performance of the dance couples 1, 2 and 3, respectively. Black lines depict the experimental data, and red lines show the model fits. The vertical offset corresponds to their respective mean z position, given that the data are mean-subtracted. The marker sequence in the waterfall corresponds to the mid-position markers defined for the posterior view, which are mostly virtual and start with the letter M to indicate the midpoint between the corresponding left and right markers, which start with L and R, respectively. From top to bottom: TBHD, MSHO, MBAK, MIDO, MLUM, MPSI, MTHI, MKNE and MANK (see Extended Data Table 1, Supplementary Table 1, and Supplementary Video 2).

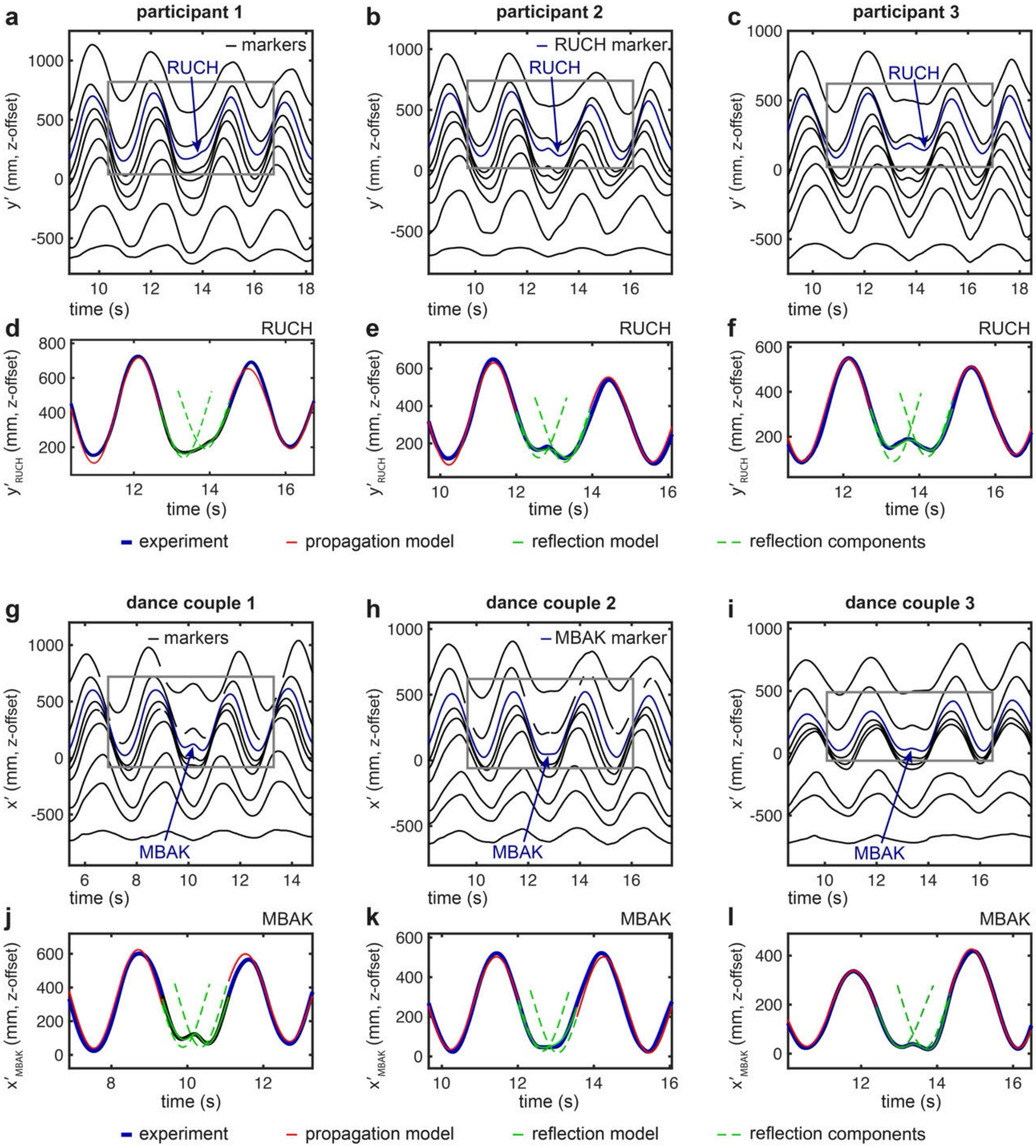

**Extended Data Fig.3. Reflection between the downward- and upward-propagating waves (sequence 1 and 2).** Panels **a-f** and **g-l** correspond to anteroposterior (sequence 1) and mediolateral (sequence 2) body displacements, respectively. Panels **a-c** and **g-i** show the curves interpreted as downward- and upward-propagating waves, where the intermediate region is considered the reflection zone, for participants 1-3 and dance couples 1-3, respectively. Here, the RUCH and MBACK trajectories are highlighted in dark blue. These curves are shown in zoomed panels **d-f** and **j-l** (corresponding to the grey rectagle region of panels **a-c** and **g-i**, respectively). In these panels, dark-blue curves indicate only the data segments used to fit the sinusoidal response (red curves) from downward- and upward-propagating waves (see Extended Data Figs. 1 and 2). The reflection response (solid green) is modelled as a reflectance-weighted superposition of extensions of the sinusoidal fit components, with amplitudes allowed to vary within prescribed bounds. The reflection fit region is assumed to be bounded by the points where each component crosses its mean level.

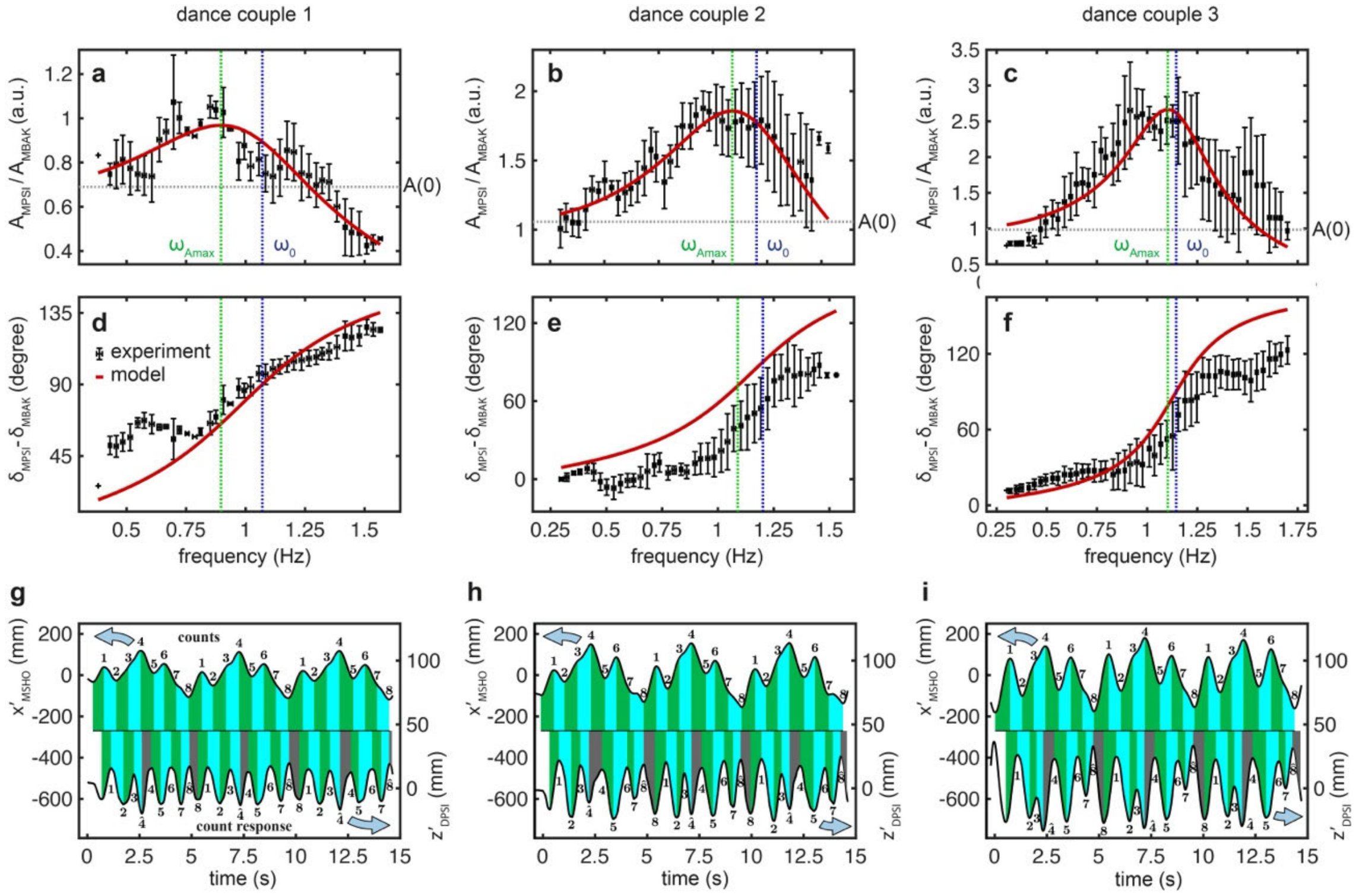

**Extended Data Fig. 4. Resonance spectra (sequence 3) and phase shifts resembling a resonant response (sequence 4).** Panels **a-c** and **d-f** show the normalized amplitude (MPSI/MBACK) and phase-difference spectra, respectively, for sequence 3 (see Supplementary Video 3). Panels **g-i** show pelvis half elevation, quantified by DPSI, relative to the lateral drive from the MSHO motion during sequence 4 (see Supplementary Video 4). Panels **a**,**d**,**g**; **b**,**e**,**h**; and **c**,**f**,**i** correspond to participants 1, 2 and 3, respectively. In panels **a-f**, black points show the experimental data and red lines show the model fits. Blue, green and grey dotted lines indicate the natural frequency ($\omega_0$), the frequency at which the amplitude is maximal ($\omega_{Amax}$), and the extrapolated amplitude to zero frequency $A(0)$, respectively. In panels **g-i**, MSHO traces are plotted at the top, with dance-count labels 1-8, and DPSI response at the bottom, labelled with the same count numbers, including decorative $\hat{4}$ and $\hat{8}$. The relative displacement of the top and bottom colour-band boundaries highlights the phase shifts.

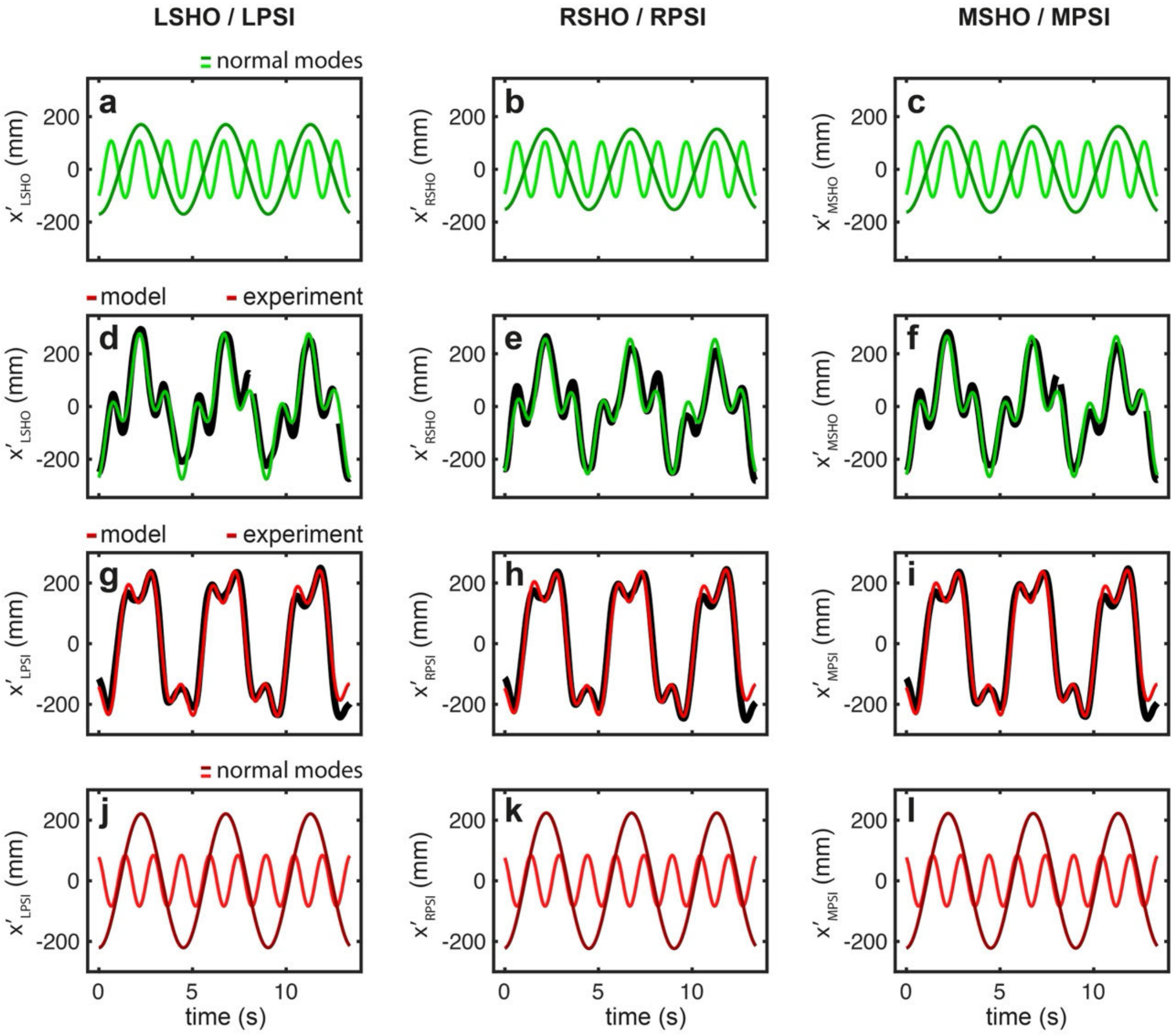

**Extended Data Fig.5. Coupled undamped oscillators, dance couple 1 (sequence 5).** Coupled oscillatory and modal responses of the upper and lower parts of the SGTLP complex. Left (**a**, **d**, **g** and **j**), centre (**b**, **e**, **h** and **k**) and right (**c**, **f**, **i** and **l**) panel columns, correspond to the left, right and virtual markers, respectively. Specifically, LSHO and LPSI are shown in panels **a,d** and **g,j**, respectively, RSHO and RPSI in panels **b,e** and **h,k**, and MSHO and MPSI in panels **c,f** and **i,l** (see Supplementary Video 5). Panels **d-f** and **g-l** show the coupled undamped oscillator model fits, displayed as green and red curves, to the projections of the experimental data, shown as black curves, for markers LSHO, RSHO and MSHO and for markers LPSI, RPSI and MPSI, respectively. Panels **a-c** and **j-l** display the normal modes corresponding to panels **d-f** and **g-l**, respectively.

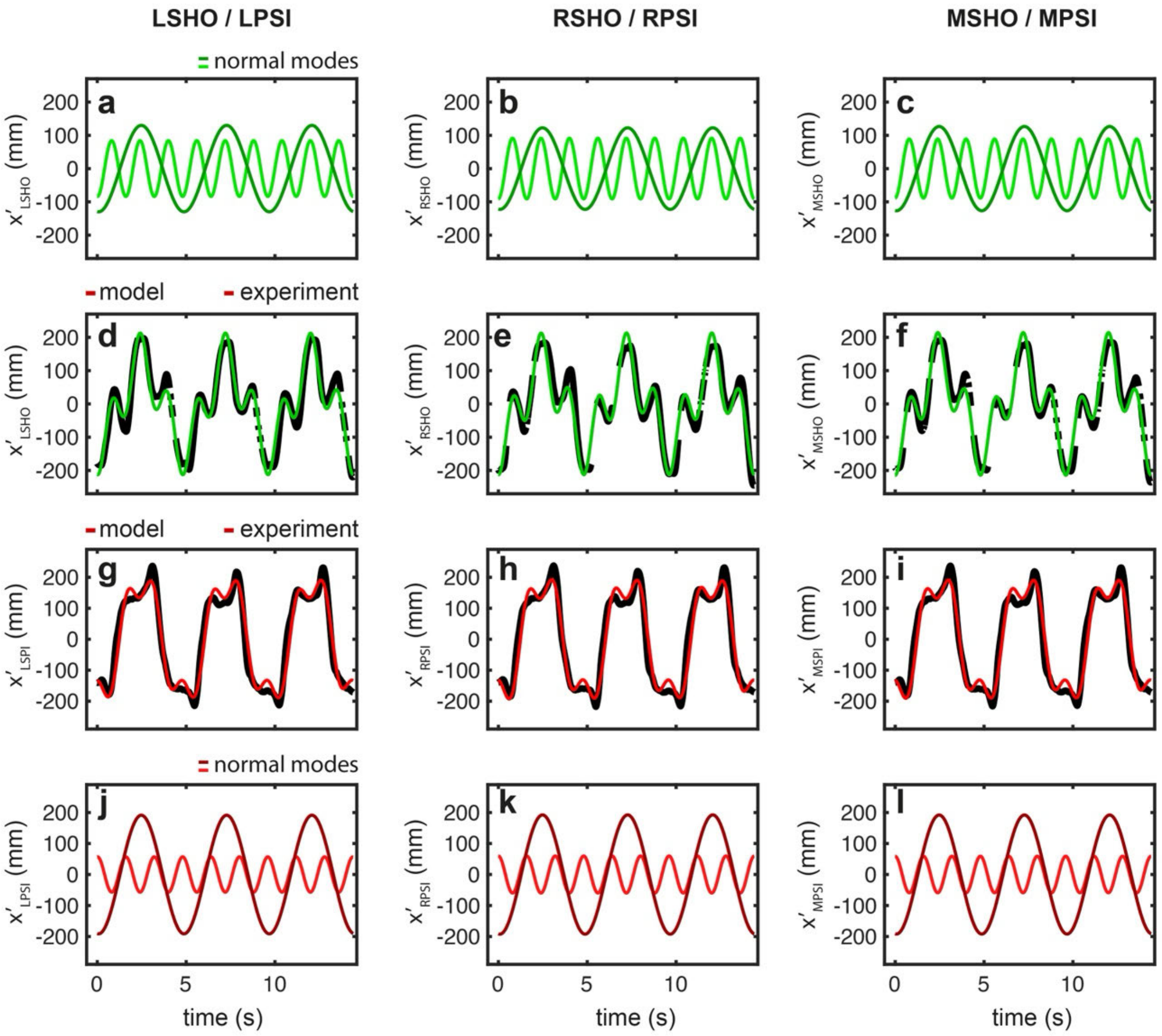

**Extended Data Fig. 6. Coupled undamped oscillators, dance couple 2 (sequence 5).** Coupled oscillatory and modal responses of the upper and lower parts of the SGTLP complex for dance couple 2. The figure follows the same layout and conventions as Extended Data Fig. 5.

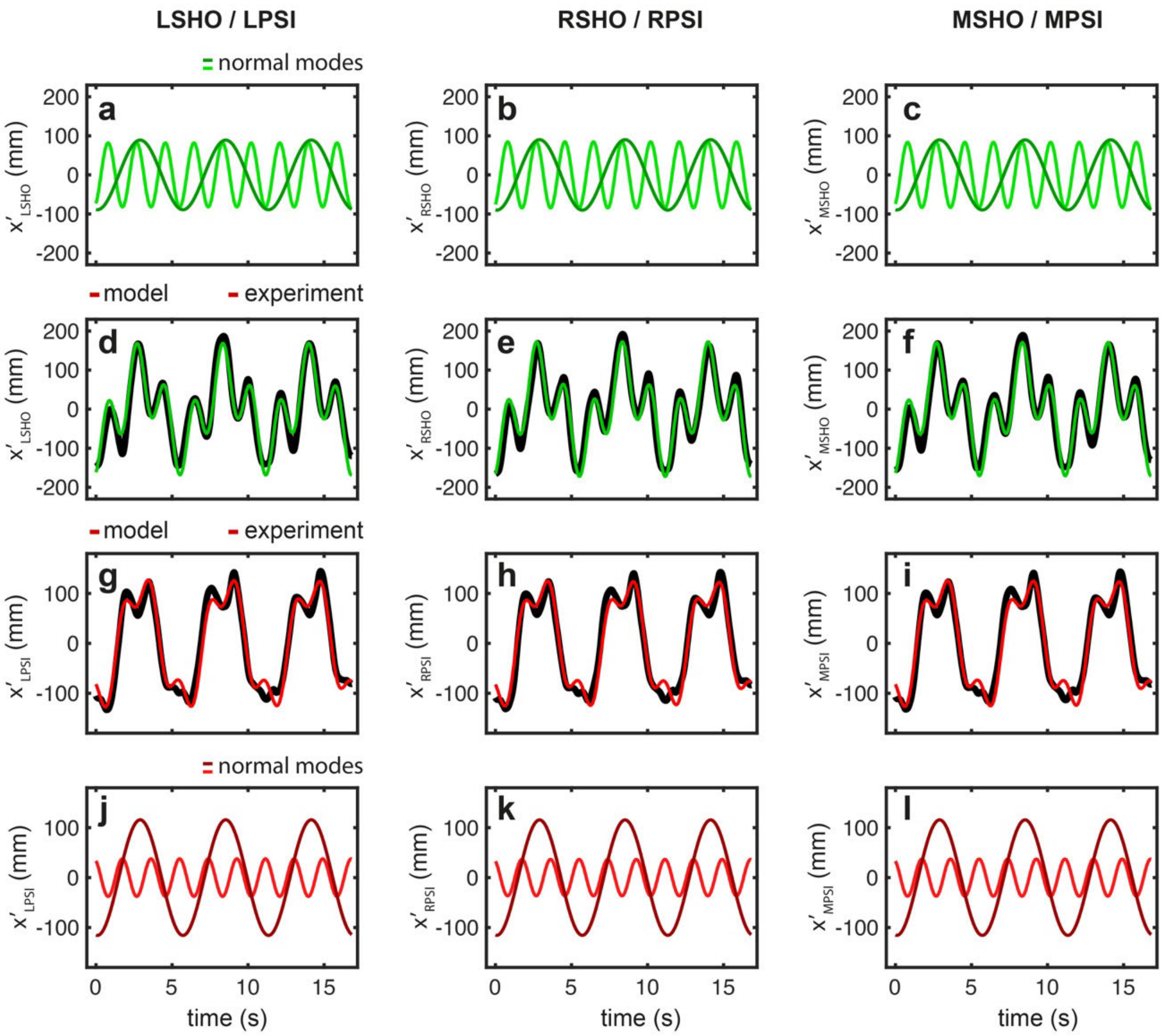

**Extended Data Fig. 7. Coupled undamped oscillators, dance couple 3 (sequence 5).** Coupled oscillatory and modal responses of the upper and lower parts of the SGTLP complex for dance couple 3. The figure follows the same layout and conventions as Extended Data Fig. 5.

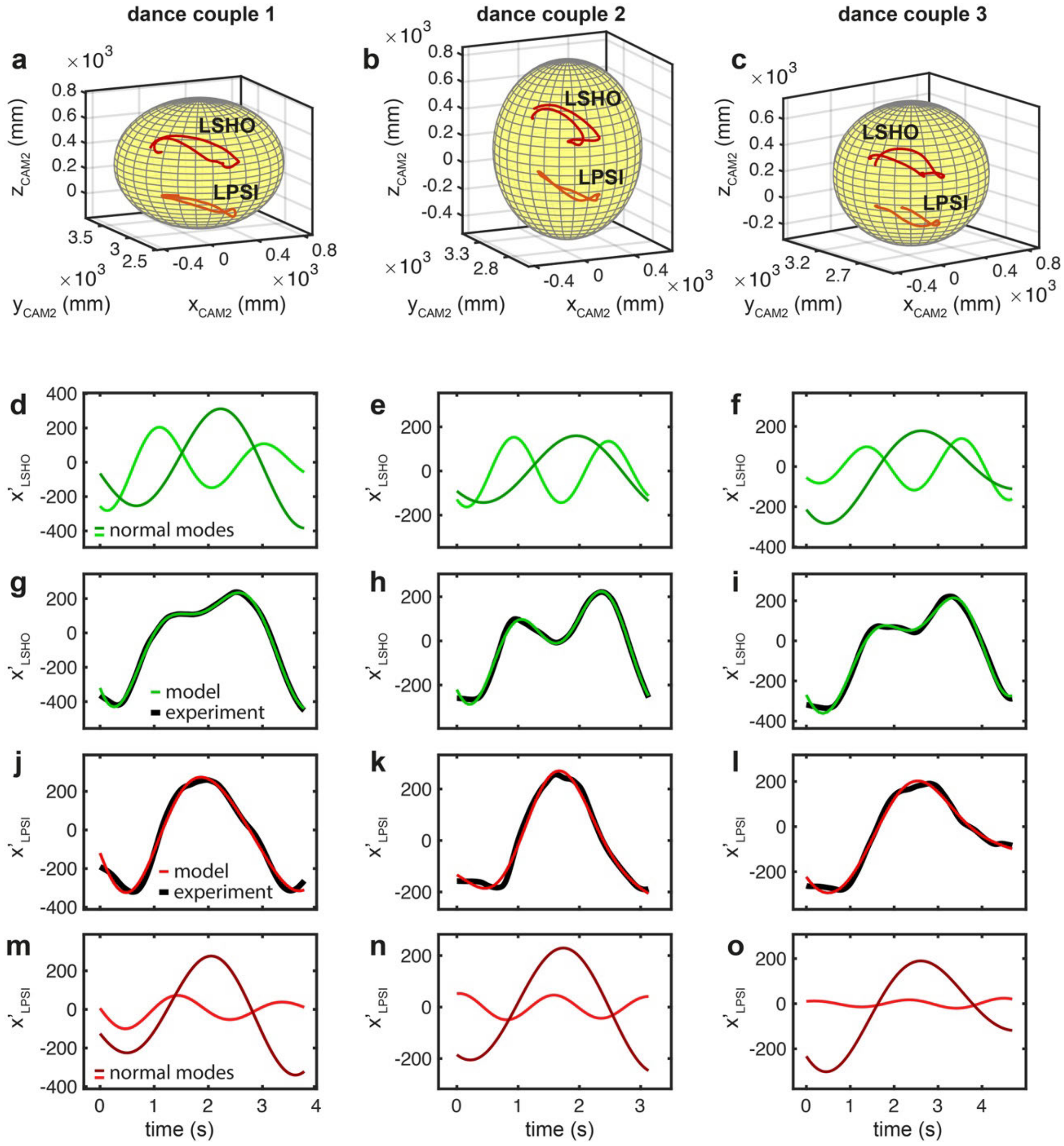

**Extended Data Fig. 8. Ellipsoids and coupled oscillators for SGTLP complex (sequence 6a).** The left (**a,d,g,j,m**), centre (**b,e,h,k,n**) and right (**c,f,i,l,o**) columns correspond to dance couples 1 (see Supplementary Video 6), 2 and 3, respectively. Panels **a-c** show ellipsoids fitted to the experimental trajectories of the LSHO (green) and LPSI (red) markers. Panels **d-o** show the coupled-oscillator response. Panels **g-i** and **j-l** show the model fits (green and red curves, respectively) to the projections of the experimental data (black curves) for LSHO and LPSI, respectively. Panels **d-f** and **m-o** display the normal modes corresponding to panels **g-i** and **j-l**, respectively.

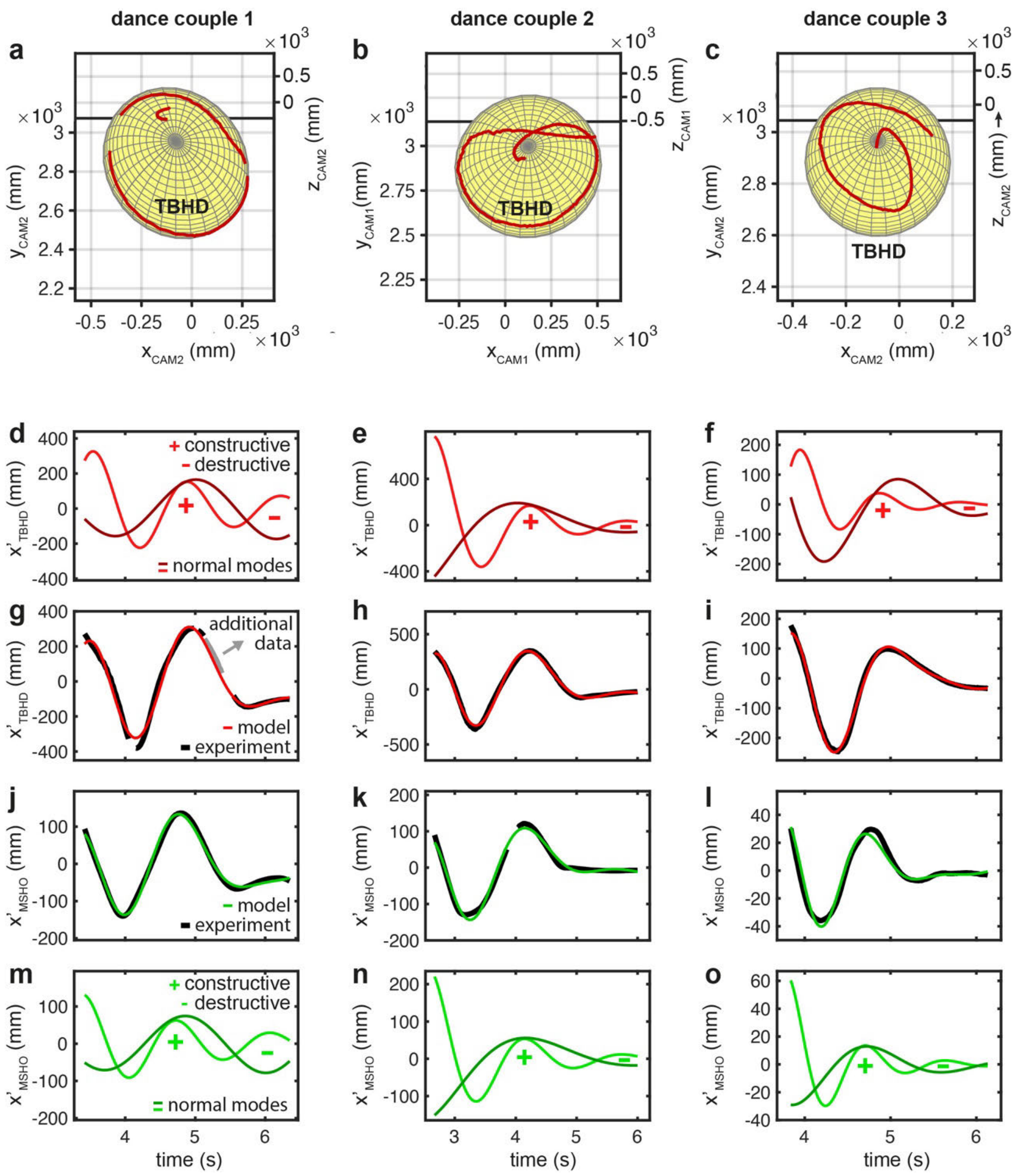

**Extended Data Fig. 9. Ellipsoids and coupled oscillators for the SGCH complex (sequence 6b)**. The left column (**a,d,g,j,m**), centre column (**b,e,h,k,n**) and right column (**c,f,i,l,o**) correspond to dance couples 1 (see Supplementary Video 6), 2 and 3, respectively. Panels **a-c** show ellipsoids fitted to the experimental trajectories of the TBHD marker. After the rotations in Sequence 6a, the motion of dance couples 1 and 3 becomes more aligned with CAM2, while dance couple 2 is more aligned with CAM1, resulting in fewer marker occlusions. Panels **d-o** show the coupled-oscillator response. Panels **g-i** and **j-l** show the model fits (red and green curves, respectively) to the projections of the experimental data (black curves) for TBHD and the mid-position MSHO, respectively. The grey portion of the curve in panel **g** shows illustrative reconstructed data that were not used for fitting. Panels **d-f** and **m-o** show the normal modes corresponding to panels **g-i** and **j-l**, respectively. '+' and '-' annotations indicate constructive and destructive time-dependent interference, respectively.

SUPPLEMENTARY INFORMATION

# Wave physics as a choreographic notation for partner dance

Fernando Ramiro-Manzano[1*]

[1]*Nanomaterials for Optoelectronics, Photonics and Energy, Instituto de Tecnología Química, Universitat Politècnica de València - Consejo Superior de Investigaciones Científicas (UPV-CSIC), Avd. de los Naranjos s/n, 46022, Valencia, Spain.*

*e-mail: ferraman@fis.upv.es

## Contents

## S1. Introduction

This Supplementary Information provides additional general aspects of the study, mathematical background and extended details.

## S2. Anatomical segments and complexes

Given the functional relevance of the spine in this study, and considering that the scapulae (shoulder girdle) constitute the leader's primary contact interface, we examined how this interface connects to pelvic motion through the Shoulder Girdle-Thoraco-Lumbo-Pelvic complex (SGTLP; see Supplementary Table 1), and how it relates to head motion through the Shoulder Girdle-Cervical-Head complex (SGCH). For the analysis of the first sequences, we used WB-NUL (Whole Body-No Upper Limbs) together with the SGTLP complex. In WB-NUL, upper-limb segments were excluded from the model. In addition, WB (Whole Body) was used to visualise the skeletal model, as shown in Supplementary Video 1.

We adopted a modified Plug-in Gait marker set[44], primarily by adding several thoracic, abdominal, and lumbar marker pairs to register additional anatomical landmarks along the spine (see Extended Data Figs. 1 and 2 and Extended Data Table 1). Here, marker pairs were selected following the leader's two-hand drive, with the aim of capturing possible angular dependencies (sequence 4) and side motion (sequence 5), and increasing robustness to possible occlusions (sequences 1 and 6a). Paired markers may also provide additional information on left-right patterns of intersegmental behaviour in sequences 3-6, which could be leveraged in future analyses. Although previous studies have reported more complex segmental behaviour (e.g., full pelvic rotation[5]), the movements examined here were primarily characterised in the frontal (coronal) and sagittal planes. These planes are considered sufficient to capture the fundamental motion required for the proposed analysis.

NOTE: Although most analyses do not include the lower limbs, they are essential for supporting and coordinating pelvic motion. A clear example is sequence 3 (*hip launch*): the motion may appear similar to a purely mechanical conical pendulum, raising questions about the voluntariness of the follower's response. However, the follower actively supports their own weight while remaining upright, and the SGTLP complex is not immobilized. Instead, the leader's guidance provides the driving input that induces a fluid, resonant response in the follower, including associated phase shifts, with the lower limbs contributing to the execution of the movement. The follower may sense the induced motion proprioceptively (kinesthetically) and interpret it through coordinated engagement of the SGTLP and lower limbs, in line with theories of sensorimotor coordination[45,33,46].

## S3. Notes on Bachata and Bachata Sensual.

Bachata, a music and dance style, was inscribed on UNESCO's Representative List of the Intangible Cultural Heritage of Humanity[34]. This performance art originated in the rural areas of the Dominican Republic in the 1960s, as a fusion of rhythmic bolero with other Afro-Antillean genres such as Son, Cha-cha-cha and Merengue, among others[34]. The instrumentation is based on guitars and decorated with rich percussion[47]. The lyrics usually express heightened romanticism and emotions[48,49]. While dancing, the couple follows riffs of two 4/4 bars, "characterized by a sensual hip movement" [34]. In the basic motion, weight, or centre-of-mass support, alternates between legs across the counts; however, from 3 to 4 and from 7 to 8 the support leg remains unchanged. Although counts 3 and 4, and 7 and 8, usually correspond to different actions (for example, counts 1-3 and 5-7 may involve stepping, whereas counts 4 and 8 serve as holding steps), there is no inversion of motion between 3 and 4 or between 7 and 8. This permits a symmetric inversion/change of the motion direction in each bar. A branch of this dance, Bachata Sensual, is characterized by rich combinations of wave-like and oscillatory movement.

Sequences 1-6 involve contact between the leader's hands and the follower's scapular region, providing a direct pathway for driving input into the SGTLP complex. In Sequence 4, due to the slight angle between the leader and the follower, the lateral sides of the leader's right knee and the follower's left knee may briefly contact; however, this is considered transient and secondary, with the hand-scapula interface as the primary drive interface. Alternative executions of the same sequences may employ upper-limb connections (e.g., hand-elbow or hand-wrist), while in other sequences a lateral knee interface may serve as the primary driver of the follower's pelvic motion. These variants are not considered here. In the present study, we therefore focus on the hand-scapula interface, which is shared across all sequences and provides a consistent driving input for analysing SGTLP motion in the follower.

Although the term 'choreographic' is often associated with pre-established dance phrases, recreational partner dance may also emerge through the real-time combination and variation of movement sequences. Sequence 6 provides one example by combining Sequences 6a and 6b. Other combinations could be formed from Sequences 1-5, 6a and 6b, or segments thereof. Moreover, in some performance settings, both the music and the sequence structure may be selected in advance. In this work, we use the term choreographic notation to denote a description and analysis of motion, without implying a fixed, pre-composed choreography.

## S4. Dance Nomenclature

Waves are a paradigm of dance[32] and are present in many styles. However, terminology for these motions may vary across disciplines and, in particular cases, may even depend on the language employed or the individual instructor. In this document, we primarily adopt Bachata Sensual terminology, while occasionally including more generic alternative terms also used within this style. For example, a *wave* and *counterwave* may also be described as a *body roll* and *reverse body roll*. Regardless of the terminology used, for each sequence (labelled 1-6) we provide a written description of the motion together with supplementary videos (Supplementary Videos 1-6, respectively).

## S5. Number of participants (Phase I)

Across the works cited here that focus on dance or movement analysis, the reported numbers of dancers or performers ($N_p$), couples or dyads ($N_d$), and, where applicable, the extended number of participants, including non-dancers, observers, judges or controls ($N_{p,ext}$), vary widely:

**- Conceptual or review contributions with no reported participants:** refs. 1 (individual and partnered dance), 2 (conceptual study of solo dancers, dance couples, and larger interaction groups), 3, 12, 26, 30, 32 and 39.

**- Dataset-based studies:** refs. 13, 14, 16 (including $N_{p,ext}$ = 18 evaluators), 18 and 31. For these studies, the number of performers represented in the datasets is not treated here as an experimental participant count unless it is explicitly reported and directly comparable. Ref. 13 is also listed below because its motion-capture dataset explicitly comprises five performers.

- $\boldsymbol{N_p = 1}$: single-participant designs in empirical studies (e.g., single-subject biomechanics studies) such as refs. 9 (based on a previous study with more participants), 8 (following preliminary trials with student dancers who showed the same general characteristics), 11, 5 (defining a context and guidelines for future investigations involving cross-subject comparisons) and 19 (computational motion analysis); a single-case plus controls design: ref. 42 ($N_p$ = 1, one case participant with $N_{p,ext}$ = 12 control participants).

- $\boldsymbol{N_p = 2}$: two-dancer study: ref. 17 (two dancers analysed individually rather than as a dyad).

- $\boldsymbol{N_p}$ **= 4**: two studies involving four dancers, refs. 15 (two dyads or dance couples, $N_d = 2$) and 7 (with $N_{p,ext} = 108$ non-expert observers). In addition, two studies (refs. 24 and 25) involve four non-human subjects.

- $\boldsymbol{N_p}$ **= 5**: refs. 21 (forming human-robot dyads) and 13 (five performers represented in the motion-capture dataset rather than five experimentally recruited participants).

- $\boldsymbol{N_p}$ **= 8**: refs. 4 and 10.

- $\boldsymbol{N_p}$ **= 26**: a larger participant group with 26 dancers and $N_{p,ext} = 12$ judges, ref. 22.

- $\boldsymbol{N_p}$ **= 35**: 35 dancers with $N_{p,ext} = 35$ non-dancers: ref. 27.

- $\boldsymbol{N_p}$ **= 40**: single-dancer analyses within human-virtual dyads: ref. 20.

- $\boldsymbol{N_p}$ **= 42**: novice dancers: ref. 28.

- $\boldsymbol{N_p}$ **= 80**: 80 dancers forming $N_d = 40$ dyads, at the higher end, enabling large-scale statistical analyses of interpersonal synchrony: ref. 29.

In this context, the present work is designed as Phase I, in which we analyse the movement sequences performed by six participants forming three dance couples ($N_d = 3$, $N_p = 6$) within the wave-physics framework. Building on the results of this initial phase, a Phase II study is foreseen to further extend the framework to approximately $N_d = 20$ dyads, or dance couples ($\sim N_p = 40$ participants), including an expert-non-expert study.

Note: In this section, dyad refers to a pair of participants; the musical term *dyad* is written in italics elsewhere in the text.

## S6. Definitions of variables and axis conventions

We use $u$ as a generic perturbation variable. Because this work draws on wave physics and related analogies, the specific meaning of $u$ depends on the context: in mechanical vibrations $u$ may denote (for example) a displacement $x$, $y$, $z$ or a rotation $\theta$; in acoustics, a pressure $p$; in audio processing, an audio waveform $a$; in electromagnetics, an electric-field component $E$; in envelope or waveguide descriptions, an amplitude $A$; in quantum mechanics, a wave function $\psi$ (state variable); and a degree of freedom $u$ in Finite Element Methods (FEM) formulations. Likewise, $M$ may refer to a mass $m$, a moment of inertia $I$, a mass density $\rho$, an inductance $L$, or the FEM mass matrix.

For our specific calculations, $u$ is expressed in the coordinates $\{x_{CAM1}, y_{\mathrm{CAM1}}, z_{\mathrm{CAM1}}\}$, $\{x_{CAM2}, y_{\mathrm{CAM2}}, z_{\mathrm{CAM2}}\}$ , $\{z\}$, $\{x', x'', y', z', z''\}$, $\{\theta\}$ or $\{a\}$, as appropriate. The subscripts CAM1 and CAM2 denote the global coordinate axes associated with each camera. The choice of this axis determines the camera used for the 2D-3D reconstruction and the visualization of the 3D figures. In this 3D representation, the motion data for sequence 6 are shown on ellipsoidal surfaces. Primes (and double primes) on the coordinates indicate that they refer to a relative axis rather than the absolute stereoscopic axis of a given camera. This choice of axis is particularly relevant in Sequence 6, where CAM1 or CAM2 absolute axes and two relative axes, $x'$ (for sequence 6a) and $x''$(for sequence 6b), are considered across participants. As $z_{CAM1}$ and $z_{CAM2}$ correspond to the same nominal axis (with different origins in the horizontal plane for each camera), they can for simplicity, be denoted as $z$ in the text and figures, except in 3D plots that correspond to a specific camera. This $z$ axis could likewise be denoted $z'$ and $z''$ for illustrative purposes, with the origin shifted to the mean position so that the plotted values remain small and centered (see Supplementary Video 4c). Notably, the DPSI quantity is invariant to whether it is expressed in absolute $z$ or relative $z'$ axis: the direction remains unchanged, and any shift in the origin cancels in the differential. However, because the other coordinate in these analyses is expressed in the relative frame ($x'$), we therefore retain the $z'$ notation in the DPSI analysis and plots for consistency. The approximate orientations of the relative axes are shown in the still images (Figs. 1-3) and 3D representations (Fig. 4). The same visualization strategy is used in the dancer video panels and 3D plots in Supplementary Videos 1-6. Note that the stereoscopic reference axes can be transformed from one camera frame to another by using the same procedure adopted for ellipsoid fitting (see section S8).

## S7. Description of the model of coupled oscillators

This section describes the coupled oscillator model, including eigenvalue extraction, modal decomposition, and the distinction between robust modal quantities and fit-variant parameters.

### S7.1 General model

We start from a 2-degrees of freedom (DOF) linear second-order model:

$$M_1\ddot{u}_1 + \gamma_{0,11}\dot{u}_1 + \gamma_{0,12}\dot{u}_2 + \Omega_{0,1}u_1 + \beta_{0,12}u_2 = 0 \quad \text{(S1)}$$

$$M_2\ddot{u}_2 + \gamma_{0,21}\dot{u}_1 + \gamma_{0,22}\dot{u}_2 + \beta_{0,21}u_1 + \Omega_{0,2}u_2 = 0 \;.\text{(S2)}$$

For convenience, this system of equations can be expressed in matrix form

$$\boldsymbol{M\ddot{u}} + \boldsymbol{C_0\dot{u}} + \boldsymbol{K_0u} = 0\;, \qquad \boldsymbol{u} = \begin{bmatrix} u_1 \\ u_2 \end{bmatrix} \quad \text{(S3)},$$

$$\boldsymbol{M} = \begin{bmatrix} \mathrm{M}_1 & 0 \\ 0 & \mathrm{M}_2 \end{bmatrix},\; \boldsymbol{C_0} = \begin{bmatrix} \gamma_{0,11} & \gamma_{0,12} \\ \gamma_{0,21} & \gamma_{0,22} \end{bmatrix},\quad \boldsymbol{K_0} = \begin{bmatrix} \Omega_{0,1} & \beta_{0,12} \\ \beta_{0,21} & \Omega_{0,2} \end{bmatrix}. \text{(S4)}$$

At the same time, we could consider the $\boldsymbol{M}$ matrix that represents either effective mass or effective moment of inertia for the selected body regions. However, these terms introduce a scale indeterminacy and cannot be independently identified from the fitting procedure. Since $M_1,\ M_2 \neq 0$ premultiplying by $\boldsymbol{M^{-1}}$ recasts the system in reduced-mass form, yielding the rescaled parameters

$$\gamma_{mn} = \frac{\gamma_{0,mn}}{M_m}\;,\; \Omega_m = \frac{\Omega_{0,m}}{M_m} = \omega_{0,m}^2 \text{ and } \beta_{mn} = \frac{\beta_{0,mn}}{M_m}\;, \text{(m, n=1,2)}\;. \quad \text{(S5)}$$

$$\text{Where we denote } \Omega_m = \omega_{0,m}^2 \text{ with } \omega_{0,m} \geq 0\;, \qquad \text{(S6)}$$

$$\boldsymbol{\ddot{u}} + \boldsymbol{C\dot{u}} + \boldsymbol{Ku} = 0, \quad \boldsymbol{C} = \boldsymbol{M^{-1}C_0} = \begin{bmatrix} \gamma_{11} & \gamma_{12} \\ \gamma_{21} & \gamma_{22} \end{bmatrix},\quad \boldsymbol{K} = \boldsymbol{M^{-1}K_0} = \begin{bmatrix} \omega_{0,1}^2 & \beta_{12} \\ \beta_{21} & \omega_{0,2}^2 \end{bmatrix}. \text{(S7)}$$

This represents the matrix form of the coupled equations of the manuscript (see Eq. 6,7 and Eq. 4,5 with $\boldsymbol{C} = \boldsymbol{0_{2x2}}$). The model parameters, together with the initial state

$$\boldsymbol{w}(0) = \begin{bmatrix} \boldsymbol{u}(0) \\ \boldsymbol{\dot{u}}(0) \end{bmatrix} \qquad \text{(S8)}$$

and an offset baseline vector $\boldsymbol{u_{offset}}$ (used to centre the relative axis on the mean motion) were fitted by numerically integrating the system with an ordinary differential equation (ODE) solver.

In summary, we employ a general second-order formulation that retains the complete description of the coupled system considered. Well-established approaches such as coupled-mode theory (CMT)[50] have demonstrated outstanding capability in capturing the essence of resonant coupling phenomena within their respective domains of applicability, typically under near-resonance or weak-damping conditions, where a compact first-order complex representation is often employed. The present framework is compatible with these formalisms while keeping all parameters in their most general form, enabling direct application to cases involving low-frequency oscillations, asymmetric damping, time-domain solutions, or transient behaviour.

### S7.2. Eigenvalues and eigenvectors

The second-order system is rewritten in first-order form to perform eigenvalue analysis[51].

$$\boldsymbol{w} = \begin{bmatrix} \boldsymbol{u} \\ \dot{\boldsymbol{u}} \end{bmatrix}, \qquad \dot{\boldsymbol{w}} = \boldsymbol{A}\boldsymbol{w}, \qquad \boldsymbol{A} = \begin{bmatrix} \boldsymbol{0}_{2\times 2} & \boldsymbol{I}_{2\times 2} \\ -\boldsymbol{K} & -\boldsymbol{C} \end{bmatrix}, \quad \text{(S9)}$$

with $two$ physical DOF, $\boldsymbol{w} \in R^4$ and $\boldsymbol{A} \in R^{4\times 4}$. Since the system matrices are real, complex eigenvalues occur in conjugate pairs. The eigenvalues ($\lambda_k$) and eigenvectors ($\boldsymbol{v}_k$) are computed as:

$$\boldsymbol{A}\boldsymbol{v}_k = \lambda_k \boldsymbol{v}_k \quad \text{for } k\text{=1,2,3,4.} \quad \text{(S10)}$$

or, indexed by $n = 1{,}2$, and assuming $\omega_n \neq 0$, the eigenvalues form two complex-conjugate pairs, given by

$$\lambda_n = -\kappa_n \pm j\,\omega_n \;. \quad \text{(S11)}$$

The same eigenvalues satisfy the quadratic problem

$$\det(\lambda^2 \boldsymbol{I} + \lambda \boldsymbol{C} + \boldsymbol{K}) = 0 \;. \text{(S12)}$$

The eigenvalues $\{\lambda_k\}_{k=1}^{4}$ are indexed according to the rule $k = 2n-1, 2n$ ($n = 1{,}2$) where $\text{Im}(\lambda_{k=2n-1}) > 0$. The modal frequencies and decay rates are defined as

$$\omega_n = \text{Im}(\lambda_{k=2n-1})\,,\; \kappa_n = -\text{Re}(\lambda_{k=2n-1}),\; n = 1{,}2\,. \qquad \text{(S13)}$$

Collect the right eigenvectors in

$$\boldsymbol{V} = [\boldsymbol{v_1}, \boldsymbol{v_1^*}, \boldsymbol{v_2}, \boldsymbol{v_2^*}] \text{ with } \boldsymbol{\Lambda} = \text{diag}(\lambda_1, \lambda_1^*, \lambda_2, \lambda_2^*) \text{ so that } \boldsymbol{A} = \boldsymbol{V}\boldsymbol{\Lambda}\,\boldsymbol{V}^{-1}\;. \quad \text{(S14)}$$

Assuming $\boldsymbol{A}$ is diagonalizable and given initial conditions (Eq. S8),

$$\boldsymbol{w}(0) = \boldsymbol{V}\boldsymbol{c} \Rightarrow \boldsymbol{c} = \boldsymbol{V}^{-1}\boldsymbol{w}(0)\,, \qquad \text{(S15)}$$

then the solution can be expressed as[52,53]

$$\boldsymbol{w}(t) = \boldsymbol{V}e^{\boldsymbol{\Lambda}\text{t}}\boldsymbol{V}^{-1}\boldsymbol{w}(0) = \boldsymbol{V}e^{\boldsymbol{\Lambda}\text{t}}\boldsymbol{c} = \textstyle\sum_{k=1}^{4} c_k\, \boldsymbol{v}_k\, e^{\lambda_k t}\,, \qquad \text{(S16)}$$

where $\boldsymbol{c} \in C^4$ are the model coefficients to be determined. Each eigenvector is written in block form $\boldsymbol{v_k} = \left[\boldsymbol{v_k^{(u)}}\; \boldsymbol{v_k^{(\dot{u})}}\right]^{\text{T}}$. Extracting the displacement block for the m-th degree of freedom ($m = 1{,}2$) yields

$$u_m(t) = \textstyle\sum_{k=1}^{4} c_k \left(\boldsymbol{v_k^{(u)}}\right)_m e^{\lambda_k t}\,, \text{(S17)}$$

where $c_{k=2n} = c_{k=2n-1}^*$ and $\boldsymbol{v}_{k=2n}^{(\boldsymbol{u})} = \left(\boldsymbol{v}_{k=2n-1}^{(\boldsymbol{u})}\right)^*$. Restricting to $k = 2n-1$ (positive imaginary part) and defining $z_{mn} = c_{k=2n-1}\left(\boldsymbol{v}_{k=2n-1}^{(\boldsymbol{u})}\right)_m$, each conjugate pair yields

$$e^{-\kappa_n t}\left(z_{mn}e^{j\omega_n t} + z_{mn}^* e^{-j\omega_n t}\right) = 2|z_{mn}|\, e^{-\kappa_n t}\cos(\omega_n t + \arg z_{mn}). \text{(S18)}$$

Summing over $n = 1{,}2$

$$u_m(t) = \textstyle\sum_{n=1}^{2} A_{mn}\, \text{e}^{-\kappa_n t}\cos(\omega_n t + \delta_{mn}) \;.\text{(S19)}$$

The mode-specific amplitudes and phase offsets are given by

$$A_{mn} = 2|z_{mn}| \qquad ,\, \delta_{mn} = \arg(z_{mn})\,. \quad \text{(S20)}$$

These quantities result from the fitted time-domain evolution of the system and its implicit initial conditions.

## S7.3. Model fitting uncertainty

While the standard errors of the fitted ODE parameters were obtained as the square roots of the diagonal elements of the parameter covariance matrix $\boldsymbol{CovB}$ (returned by MATLAB's *nlinfit*), uncertainties in the derived modal quantities, (modal frequencies, damping factors, amplitudes and phases) were estimated a posteriori from the fit residuals using local uncertainty propagation. After obtaining the best-fit parameters, the reconstructed signals were expressed using the modal representation in Eq. S19, together with the offset baseline vector $\boldsymbol{u_{offset}}$ . The residual variance was estimated from the reconstruction residuals[54] ($\varepsilon_d = u_{\exp,d} - u_d(\boldsymbol{\varphi})$) as

$$\sigma^2 \approx \frac{1}{N-P}\sum_{d=1}^{N} \varepsilon_d^2 \ . \quad \text{(S21)}$$

Here, $u_{\exp,d}$ denotes the experimentally measured signals sampled at the same time points as the model prediction $\boldsymbol{u}(t;\boldsymbol{\varphi})$, $d$ indexes the sampled data points, $N$ is the number of data samples and $P$ the number of parameters. A Jacobian matrix $\boldsymbol{J} = \partial\boldsymbol{u}/\partial\boldsymbol{\varphi}$ was computed numerically by finite differences around the parameter values, with the parameter vector collecting the coefficients into an array

$$\boldsymbol{\varphi} = \left(\omega_n, \kappa_n, \delta_{mn}, A_{mn}, \boldsymbol{u_{offset}}\right), n, m = 1{,}2 \ . \qquad \text{(S22)}$$

Under a local linear approximation of the reconstruction model, the covariance matrix $CovB$ was estimated as[55,56]

$$\boldsymbol{CovB} \approx \sigma^2(\boldsymbol{J}^{\top}\boldsymbol{J})^{-1}. \quad \text{(S23)}$$

Then, standard uncertainties were taken as the square roots of the diagonal elements of $\boldsymbol{CovB}$.

Note: All models from which we extracted uncertainties were fitted in MATLAB using *nlinfit*[55], which directly returns the parameter covariance matrix $\boldsymbol{CovB}$, except for the reflection analysis, where *lsqcurvefit* was employed to enforce explicit bounds on the amplitudes. In that case, $\boldsymbol{CovB}$ was computed a posteriori from the Jacobian returned by *lsqcurvefit* and the residual variance, following the same uncertainty propagation as above.

## S7.4. Parameter Identifiability: Robustness and parametric variability.

In Section S7.3, the modal quantities were treated as parameters of the reconstructed time-domain model when estimating uncertainties from the residuals. Accordingly, although the fitting is formulated in terms of the system matrices and initial conditions ($\boldsymbol{K}$, $\boldsymbol{C}$ and $\boldsymbol{w}(0)$), we analyze the modal parameters ($\omega_n$, $\kappa_n$, $A_{mn}$ and $\delta_{mn}$) in the same spirit here, as they directly shape the reconstructed trajectories.

Consider a single damped mode, $u(t) = A_{mn}e^{-\kappa_n t}\cos(\omega_n t + \delta_{mn})$, the parameter derivatives are

$$\frac{\partial u_m}{\partial \omega_n} = -A_{mn}\, t\, e^{-\kappa_n t}\sin(\omega_n t + \delta_{mn}) \ ,$$

$$\frac{\partial u_m}{\partial \kappa_n} = -A_{mn}\, t\, e^{-\kappa_n t}\cos(\omega_n t + \delta_{mn}) \ ,$$

$$\frac{\partial u_m}{\partial A_{mn}} = e^{-\kappa_n t}\cos(\omega_n t + \delta_{mn}) \ ,$$

$$\frac{\partial u_m}{\partial \delta_{mn}} = -A_{mn}e^{-\kappa_n t}\sin(\omega_n t + \delta_{mn}) \ . \quad \text{(S24)}$$

In a least-squares setting, the local identifiability of a parameter $\theta_f$ is captured (up to the noise variance and sampling density) by the Fisher-information matrix[57],

$$\mathcal{M} = \sum_{i=0}^{S} \left[ \left( \nabla_{\theta_f} \boldsymbol{u}(t_i) \right)^{\mathrm{T}} \boldsymbol{Y}_i^{-1} \left( \nabla_{\theta_f} \boldsymbol{u}(t_i) \right) \right], \qquad \text{(S25)}$$

where $t_i$ denotes the discrete sampling times and S+1 is the total number of time samples in the fitting window.

Assuming additive, zero-mean Gaussian measurement noise that is uncorrelated across coordinates and samples, with constant variance $\sigma^2$, such that $(\boldsymbol{Y}_i = \sigma^2 \boldsymbol{I})$, this reduces for a single parameter to

$$\mathcal{M}_{ff} = \sum_{i=0}^{S} \frac{1}{\sigma^2} \left( \frac{\partial u_m(t_i)}{\partial \theta_f} \right)^2 . \qquad \text{(S26)}$$

Although finite-sample covariances depend on the specific estimator and noise realization, the relative conditioning and identifiability of the parameters is already encoded in the structure of $\mathcal{M}_{ff}$. Consequently, inspection of the Fisher-information contributions shows that the information associated with the modal frequency $\omega_n$ is dominated by phase accumulation, since the corresponding sensitivity grows linearly in time and the squared derivative therefore carries a $t^2$ weighting in $\mathcal{M}_{ff}$. As a result, $\omega_n$ remains well-constrained. The damping rate $\kappa_n$ exhibits a similar time-weighted sensitivity, but is most robustly constrained when the observation window is comparable to the decay time, i.e. for $\kappa_n T = \mathcal{O}(1)$, and becomes progressively less well conditioned outside this regime. By contrast, the sensitivities to the amplitude $A_{mn}$ and phase offset $\delta_{mn}$ do not exhibit such time-weighted amplification and consequently provide substantially weaker information. Moreover, the modal poles $(\omega_n, \kappa_n)$ are common to all measured curves, since they are determined by the same fitted system matrices. In contrast, the amplitudes $A_{mn}$ and phases $\delta_{mn}$ are curve-specific quantities.

We now examine the fitted parameters $\boldsymbol{K}$ , $\boldsymbol{C}$ and $\boldsymbol{w}(0)$ as the quantities from which $\omega_n$, $\kappa_n$, $A_{mn}$ , $\delta_{mn}$ arise.

In the undamped limit ($\boldsymbol{C} = 0_{2\times 2}$), the characteristic equation (S12) reduces to $\det(\lambda^2 \boldsymbol{I} + \boldsymbol{K}) = 0$ which, for a two-degree-of-freedom system, can be written as $\lambda^4 + a_2\lambda^2 + a_0 = 0$. Thus, the two observed modal frequencies $\omega_1$ and $\omega_2$ uniquely determine only the two scalar quantities

$$a_2 = \omega_1^2 + \omega_2^2 = tr\,\boldsymbol{K} = \omega_{0,1}^2 + \omega_{0,2}^2 \,,$$

$$a_0 = \omega_1^2 \omega_2^2 = \det \boldsymbol{K} = \omega_{0,1}^2 \omega_{0,2}^2 - \beta_{12}\beta_{21} \,. \qquad \text{(S27)}$$

Therefore, the same modal frequencies $\omega_n$ can be reproduced by different combinations of $\omega_{0,m}^2$ and $\beta_{mn}$.

In the general damped case, the eigenvalues $\lambda_n = -\kappa_n \pm i\omega_n$ obtained from the characteristic equation (S12) yield the fourth-order polynomial

$$\lambda^4 + a_3\lambda^3 + a_2\lambda^2 + a_1\lambda + a_0 = 0 \,. \qquad \text{(S28)}$$

For the 2-DOF case considered here, the individual coefficients can be written as:

$$a_3 = \mathrm{tr}(\mathbf{C}) = \gamma_{11} + \gamma_{22} = 2\kappa_1 + 2\kappa_2 \,,$$

$$a_2 = tr(\boldsymbol{K}) + \det(\boldsymbol{C}) = \left(\omega_{0,1}^2 + \omega_{0,2}^2\right) + (\gamma_{11}\gamma_{22} - \gamma_{12}\gamma_{21}) =$$

$$(\kappa_1^2 + \omega_1^2) + (\kappa_2^2 + \omega_2^2) + 4\kappa_1\kappa_2 \,,$$

$$a_1 = \gamma_{11}\omega_{0,2}^2 + \gamma_{22}\omega_{0,1}^2 - \gamma_{12}\beta_{21} - \gamma_{21}\beta_{12} = 2\kappa_1(\kappa_2^2 + \omega_2^2) + 2\kappa_2(\kappa_1^2 + \omega_1^2) \text{ and}$$

$$a_0 = \det(\boldsymbol{K}) = \omega_{0,1}^2\omega_{0,2}^2 - \beta_{12}\beta_{21} = (\kappa_1^2 + \omega_1^2)(\kappa_2^2 + \omega_2^2) \,. \qquad \text{(S29)}$$

Therefore, the robust modal frequencies $\omega_n$, and, depending on the observation window, also the damping rates $\kappa_n$, can in general be reproduced by different combinations of the natural frequencies $\omega_{0,m}^2$ and the coupling and damping coefficients $\beta_{mn}$ and $\gamma_{mn}$.

On the other hand, one may apply a simultaneous similarity transformation of the generalized coordinates (such as mix, scale, rotation) $u = \boldsymbol{S}z$ where $z$ denotes an alternative set of generalized coordinates related to $u$ by an invertible linear transformation (e.g. mixing, scaling, or rotation). Under this transformation, the system matrices transform as

$$\boldsymbol{C}' = \boldsymbol{S}^{-1}\boldsymbol{C}\boldsymbol{S} \text{ and } \boldsymbol{K}' = \boldsymbol{S}^{-1}\boldsymbol{K}\boldsymbol{S} \,. \quad \text{(S30)}$$

Defining the block-diagonal matrix $\boldsymbol{T} = \mathrm{diag}(\boldsymbol{S}, \boldsymbol{S})$, and using $\boldsymbol{A}\boldsymbol{V} = \boldsymbol{V}\boldsymbol{\Lambda}$ (see S14), the corresponding state-space matrix transforms as $\boldsymbol{A}' = \boldsymbol{T}^{-1}\boldsymbol{A}\boldsymbol{T}$ , $\boldsymbol{V}' = \boldsymbol{T}^{-1}\boldsymbol{V}$, and therefore

$$\boldsymbol{A}'\boldsymbol{V}' = (\boldsymbol{T}^{-1}\boldsymbol{A}\boldsymbol{T})(\boldsymbol{T}^{-1}\boldsymbol{V}) = \boldsymbol{T}^{-1}\boldsymbol{A}\boldsymbol{V} = \boldsymbol{T}^{-1}\boldsymbol{V}\boldsymbol{\Lambda} = \boldsymbol{V}'\boldsymbol{\Lambda} \,. \qquad \text{(S31)}$$

The initial state transforms consistently as

$$\boldsymbol{w}'(0) = \boldsymbol{T}^{-1}\boldsymbol{w}(0) \,. \quad \text{(S32)}$$

With these transformations, the modal coefficient vector becomes

$$\boldsymbol{c}' = (\boldsymbol{V}')^{-1}\boldsymbol{w}'(0) = \boldsymbol{c} \,. \qquad \text{(S33)}$$

Equation (S33) provides a direct example of how changes in the eigenvector basis can be exactly compensated by corresponding changes in the initial state, leaving the modal coefficient vector $\boldsymbol{c}$ unchanged. For the reconstructed motion, however, the mode-specific amplitudes and phase offsets are determined through $z_{mn}$, which also depends on the eigenvector components. Consider simultaneous perturbations of the system parameters and initial state about a fitted solution. To first order

$$\Delta\boldsymbol{u} \simeq \boldsymbol{J}_{\boldsymbol{KC}}\Delta\boldsymbol{p}_{\boldsymbol{KC}} + \boldsymbol{J}_{\boldsymbol{w}(0)}\Delta\boldsymbol{w}(0) \qquad \text{(S34)}$$

where $\Delta\boldsymbol{p}_{KC}$ denotes variations of the independent parameters defining $\boldsymbol{K}$ and $\boldsymbol{C}$, while $\boldsymbol{J}_{KC} = \partial\boldsymbol{u}/\partial\boldsymbol{p}_{KC}$ and $\boldsymbol{J}_{\boldsymbol{w}(0)} = \partial\boldsymbol{u}/\partial\boldsymbol{w}(0)$ are the corresponding Jacobian matrices of the reconstructed trajectory. Thus, changes in $K$ and $C$ may be approximately compensated by changes in the initial state when $\boldsymbol{J}_{\boldsymbol{KC}}\Delta p_{\boldsymbol{KC}} + \boldsymbol{J}_{\boldsymbol{w}(0)}\Delta\boldsymbol{w}(0) \simeq 0$. As a result, distinct combinations of the triad ($\boldsymbol{K}$, $\boldsymbol{C}$, $\boldsymbol{w}(0)$) may produce reconstructed trajectories $\boldsymbol{u}(t)$ that are numerically indistinguishable within the fitting tolerance or can lead to multiple local minima in the optimization.

We finally focus on the role of the initial conditions in the fitting procedure. For fixed $\boldsymbol{K}$ and $\boldsymbol{C}$, the initial conditions determine the modal contributions and, hence the amplitudes and phase offsets, but do not affect the modal poles. In contrast to the modal poles, $A_{mn}$ and $\delta_{mn}$ are neither shared across datasets nor as robustly constrained by the full time-domain trajectory. In the hypothetical case that the initial state were fixed from a small number of early-time data points, this could in principle reduce the number of free parameters. However, this does not lead to a better-conditioned estimation of the system matrices, because the initial position and velocity would then be inferred from a very limited subset of the data and could therefore be strongly affected by a nearly instantaneous snapshot of the performance represented by those few points. Instead, the fit is designed to reproduce the global time evolution rather than to pass exactly through isolated points. For this reason, the initial conditions are treated as free fit parameters.

In conclusion, the modal poles provide a robust characterization of the observed system evolution, whereas the fitted matrices ($\boldsymbol{K}$ and $\boldsymbol{C}$) may be interpreted as effective parameters. Due to the reduced-mass formulation and the interdependence of the intrinsic frequencies $\omega_{0,m}^2$ with the coupling terms $\beta_{mn}$ and, in the damped case, with the damping coefficients $\gamma_{mn}$ , similar modal responses can arise from different underlying parameter configurations. Accordingly, the fitting procedure may not distinguish these underlying parameter configurations with the same robustness as the resulting modal quantities. Interestingly, this reduced distinguishability of the underlying parameter configurations accompanies the structural flexibility of the model, while the modal quantities remain comparatively robust.

The physical flexibility reflected in the modal description suggests that performers may exploit the intrinsic oscillatory behaviour of different body regions through the resulting modal features, supporting wave-like fluidity. Within this picture, coupling and damping reshape the modal structure, whereas the initial state plays an important role in establishing the relative modal contributions. In response to a given musical input, this flexibility may allow different motion realizations, involving variations in frequency content and decay rates, as well as amplitude and phase. Within certain limits (e.g., due to initial-condition limits, sensorimotor constraints, or maximum attainable amplitudes), this flexibility is not rigidly constrained by body morphology. The coexistence of robust modal characteristics and flexible motion realizations may suggest considerations of an underlying structure alongside variability in dance[6].

## S8. Supplementary details on ellipsoid fitting

This section describes the numerical procedure used to fit ellipsoids to the marker trajectories. As noted in the Methods, the data do not fully constrain the ellipsoid along its axial directions, such that the semi-axes may increase without improving the geometric fit. To mitigate this parametric variability, we proceeded as follows. We first initialized the ellipsoid parameters by fitting a sphere defined by

$$(x - c_x)^2 + (y - c_y)^2 + (z - c_z)^2 = {R_s}^2, \quad \text{(S35)}$$

where ($c_x$, $c_y$, $c_z$) denote the centre coordinates and $R_s$ is the radius. We then express the residual as

$$\varepsilon_{\text{sphere}} = \sqrt{(x - c_x)^2 + (y - c_y)^2 + (z - c_z)^2} - R_s\,. \qquad \text{(S36)}$$

and fit it to a vector of zeros using MATLAB *nlinfit*. The (unrotated) ellipsoid is defined by

$$\left(\frac{x-c_x}{s_1}\right)^2 + \left(\frac{y-c_y}{s_2}\right)^2 + \left(\frac{z-c_z}{s_3}\right)^2 = 1\,, \quad \text{(S37)}$$

where ($s_1$, $s_2$, $s_3$) are the semiaxis lengths. The residual minimized in the ellipsoid fit is

$$\varepsilon_{\text{ellipsoid}} = \sqrt{\left(\frac{x-c_x}{s_1}\right)^2 + \left(\frac{y-c_y}{s_2}\right)^2 + \left(\frac{z-c_z}{s_3}\right)^2} - 1\,. \qquad \text{(S38)}$$

Here we use $s_1 = s_2 = s_3 = R_s$ and the sphere centre as initial values. We then introduce homogeneous translation matrices that shift the coordinates to the centroid frame defined by the fitted sphere centre ($\boldsymbol{T_m}$), and subsequently restore them to the original reference frame ($\boldsymbol{T_M}$), together with a rotation matrix $\boldsymbol{R}$ around the $z$-axis:

$$\boldsymbol{T_m} = \begin{bmatrix} 1 & 0 & 0 & -c_x \\ 0 & 1 & 0 & -c_y \\ 0 & 0 & 1 & -c_z \\ 0 & 0 & 0 & 1 \end{bmatrix}, \boldsymbol{T_M} = \begin{bmatrix} 1 & 0 & 0 & c_x \\ 0 & 1 & 0 & c_y \\ 0 & 0 & 1 & c_z \\ 0 & 0 & 0 & 1 \end{bmatrix}, \boldsymbol{R} = \begin{bmatrix} cos\,\theta & -sin\,\theta & 0 & 0 \\ sin\,\theta & cos\,\theta & 0 & 0 \\ 0 & 0 & 1 & 0 \\ 0 & 0 & 0 & 1 \end{bmatrix}. \quad \text{(S39)}$$

For a point in homogeneous coordinates

$$\mathbf{r} = \begin{bmatrix} x \\ y \\ z \\ 1 \end{bmatrix}, \quad \text{(S40)}$$

the rotated point is obtained as

$$\boldsymbol{r'} = \boldsymbol{T_M R T_m r}\ . \quad \text{(S41)}$$

We sweep the rotation angle $\theta$ from 0° to 359.9° in 0.1° steps, performing a nonlinear least-squares fit for each $\theta$. In particular, two curves were simultaneously fitted for sequence 6a, and a single curve for sequence 6b. Solutions with one or more semi-axes exceeding $10^6$ mm (1 km) were discarded, as they correspond to a divergence of the semi-axes. Among the remaining solutions, we selected a representative configuration based on the statistical mode of several parameters. Specifically, we retained the maxima of the kernel density estimates (computed in MATLAB using *ksdensity*) of the three semi-axes and the RMSD. We then calculated a normalized distance among all solutions:

$$dist = \sqrt{\left(\frac{s_a - s_{mode,a}}{s_{iqr,a}}\right)^2 + \left(\frac{s_b - s_{mode,b}}{s_{iqr,b}}\right)^2 + \left(\frac{s_c - s_{mode,c}}{s_{iqr,c}}\right)^2 + \left(\frac{RMSD - RMSD_{mode}}{RMSD_{iqr}}\right)^2}, \quad \text{(S42)}$$

Here, $s_a \geq s_b \geq s_c$ denote the ordered semi-axis lengths. The subscript *mode* ($s_{mode,a}$ , $s_{mode,b}$ , $s_{mode,c}$ , $RMSD_{mode}$) indicates the maxima of the kernel density estimates, whereas *iqr* ($s_{iqr,a}$ , $s_{iqr,b}$ , $s_{iqr,c}$ , $RMSD_{inq}$) denotes the interquartile range (computed in MATLAB using the function *iqr*). The final solution was selected as the one minimizing this distance. In this way, the retained solution corresponds to the configuration closest to the joint mode of the parameter distributions across all local minima. The mode and interquartile range were used because of their robustness to outliers and large deviations in the data distributions. Note that the fitting metrics assume a simplified radial intersection of the data points with respect to the centre of the ellipsoid (see $r_i$ , where *i* denotes a landmark, in Supplementary Table 2E and F). This approximation is similar to the spherical coordinate representation of the trajectories shown in Supplementary Figs. 2-4.

## S9 Supplementary details on manual verification and computer-based correction

We manually pre-selected the marker region (by zooming in) and applied a circle-detection procedure based on Random Sample Consensus (RANSAC) with an approximately known radius range. Edges were detected using the Canny method implemented with the MATLAB function *edge*, applied to an upsampled image region to improve sub-pixel accuracy. Depending on contrast conditions, the image could optionally be inverted prior to edge detection. Edge detection was performed on grayscale intensity, a selected RGB channel, or the HSV saturation channel, with optional angular weighting favouring a predefined sector. The selected circle was subsequently refined by nonlinear least squares (using MATLAB *lsqnonlin*) to estimate the final centre and radius.

## S10 Supplementary details on 3D-3D reconstruction

This section describes the 3D-3D reconstruction procedure used to estimate the trajectory of an occluded target marker from simultaneously tracked auxiliary markers or position references, employing the orthogonal Procrustes solution. The method was formulated in a general setting by Schönemann[43], with related formulations for rigid 3D alignment given by Kabsch[58] and Arun et al[59].

We start from a reference time $t_0$ at which the target marker $\boldsymbol{SM_0}$ and three auxiliary markers $\boldsymbol{M_a}, \boldsymbol{M_b}, \boldsymbol{M_c}$ are all visible. We denote their 3D positions at this reference time: $\boldsymbol{SM_0^{t0}}, \boldsymbol{M_a^{t0}}, \boldsymbol{M_b^{t0}}, \boldsymbol{M_c^{t0}}$. At a later time $t$, the target marker $\boldsymbol{SM_0}$ is no longer visible, but the three auxiliary markers can still be tracked: $\boldsymbol{M_a^t}, \boldsymbol{M_b^t}, \boldsymbol{M_c^t}$. Assuming that all four markers are rigidly attached to the same body, we seek a rigid transformation, defined by a rotation matrix $\boldsymbol{R}$ and a translation vector $\boldsymbol{T}$, that best maps the reference configuration of the auxiliary markers to their configuration at time $t$. Specifically, $\boldsymbol{R}$ and $\boldsymbol{T}$ are determined by minimizing[59]

$$\sum_{k \in a,b,c} \| \boldsymbol{M_k^t} - (\boldsymbol{R}\, \boldsymbol{M_k^{t_0}} + \boldsymbol{T}) \|^2 \;. \quad \text{(S43)}$$

The rigid transformation can be obtained by centering both point sets at their centroids before estimating the optimal rotation via singular value decomposition (SVD). In practice, the solution was obtained using MATLAB's *procrustes* routine, with scaling disabled and reflections disallowed, thereby enforcing a proper rotation.

Once this transformation $(\boldsymbol{R}, \boldsymbol{T})$ has been estimated from the three auxiliary markers, the target marker $\boldsymbol{SM_0}$ is treated as an additional rigidly attached point and reconstructed as

$$\boldsymbol{SM_0^t} \approx \boldsymbol{R}\, \boldsymbol{SM_0^{t_0}} + \boldsymbol{T}. \quad \text{(S44)}$$

To minimize sensitivity to the choice of a single reference frame $t_0$ and reduce estimation errors, the reconstruction was repeated for multiple time lapses in which all four markers were simultaneously visible, and the resulting estimates were averaged.

## S11. Supplementary details on the comparison between segment sets for wave propagation

A more general WB-NUL segment set and a more specific SGTLP subset (see Extended Data Table 1) are considered in the manuscript and Supplementary Tables 1A, 1B, 2A and 2B. In these tables, data corresponding to averaged amplitudes ($\bar{A}^{dir}_{WB-NUL}$, $\bar{A}^{dir}_{SGTLP}$ with the direction *dir* = *D* or *U* denoting downward- and upward-propagating waves), phase differences ($\Delta\delta^{dir}_{WB-NUL}$, $\Delta\delta^{dir}_{SGTLP}$) as well as the associated wavelength fractions ($\overleftrightarrow{\lambda}^{dir}_{WB-NUL}$ and $\overleftrightarrow{\lambda}^{dir}_{SGTLP}$) are shown. In contrast to averaged amplitudes, their variability (in this context, corresponding to the mean of their standard deviations) as well as the remaining parameters are relative quantities and can therefore be considered independent of participants' body dimensions. Notably, the SGTLP subset exhibits the largest mean amplitudes (up to 288.71 ± 22.48 mm) and the smallest amplitude variability (7.6% on average compared with 35.0% for WB-NUL markers). Therefore, the SGTLP subset can be considered more relevant for the wave-propagation analysis. This suggests that its corresponding phase estimates are more robust than those obtained from WB-NUL, which may be more sensitive to persistent phase biases (such as those introduced by head markers). In fact, the largest individual marker deviations (see Supplementary Tables 2A and 2B) primarily occur in WB-NUL landmarks not included in the SGTLP subset.

## S12. Supplementary details on the resonator fit for sequence 3

Amplitude and phase are distinct observables with different scales and ranges. The fitting procedure was thus applied only to the amplitude, which contains the three model parameters (see. Eq. 3). The phase is displayed separately as an independent verification, ensuring that possible phase deviations do not affect the amplitude fit.

For illustrative purposes and simplicity, the data are grouped and plotted with error bars (see Fig. 2a-b and Extended Data Fig. 4a-f). However, the fitting procedure uses all individual data points. The uncertainties of the fitted parameters are obtained from the covariance matrix returned by the *nlinfit* function in MATLAB, while those of derived quantities are propagated accordingly (see Supplementary Table 1C). The additional equations employed for expressing the quantities and propagating the fit uncertainties are the following:

$$A(0) = \frac{\zeta}{\omega_0^2}, \; A_{max} = \frac{\zeta}{\gamma\sqrt{\omega_0^2 - \gamma^2/4}} \text{ and } \frac{A_{max}}{A(0)} = \frac{\omega_0^2}{\gamma\sqrt{\omega_0^2 - \gamma^2/4}}. \quad \text{(S45)}$$

For $\gamma < \sqrt{2}\,\omega_0$ , a resonance peak emerges, yielding $\frac{A_{max}}{A(0)} > 1$.

## S13. Supplementary details on audible waveforms from dance

To illustrate the musical *dyad* (i.e., a reduced form of a *chord*, typically containing three or more notes) we selected, for dance partners 1-3, the RSHO and RPSI experimental traces of sequence 5 corresponding to the least occluded markers in the recordings. From each sequence, we extracted a single period, concatenated it, and resampled all traces by a common scaling factor such that the lowest-frequency normal mode corresponds to 440 Hz, yielding an audible signal. Only one period was used to prevent additional spectral structure that would arise from concatenating a discrete number of periods. Gaps in the curves due to occlusion were filled by linear interpolation. To avoid discontinuities at the junctions between signal segments, the extracted period was baseline corrected by dividing by a linear function connecting its first and last points. For perceptual clarity, the signals were shaped with a 0.3 s exponential attack (5τ, reaching ~99.3% of full scale) and a symmetric exponential fade-out of the same duration. Supplementary Video 7 presents audio signals and corresponding graph segments of the normalized experimental signal, together with the extracted modes rendered as individual tones at their respective amplitudes and in superposition. The aim is to facilitate direct aural and visual comparison between the model and the experimental *dyads*.

## S14. Software employed

All video and data processing for the analysis were performed in MATLAB. This software was used for selective pixelation of still images and for the creation of illustrative videos (Supplementary Videos 1-6). For illustrative purposes only, Adobe Premiere was used to apply selective pixelation to video sequences, add additional text, adjust the output file size, and compose Supplementary Video 7. For illustrative purposes only, Adobe Photoshop was used to apply additional pixelation and adjust brightness and contrast in still images, including darkening the hands.

# SUPPLEMENTARY FIGURES

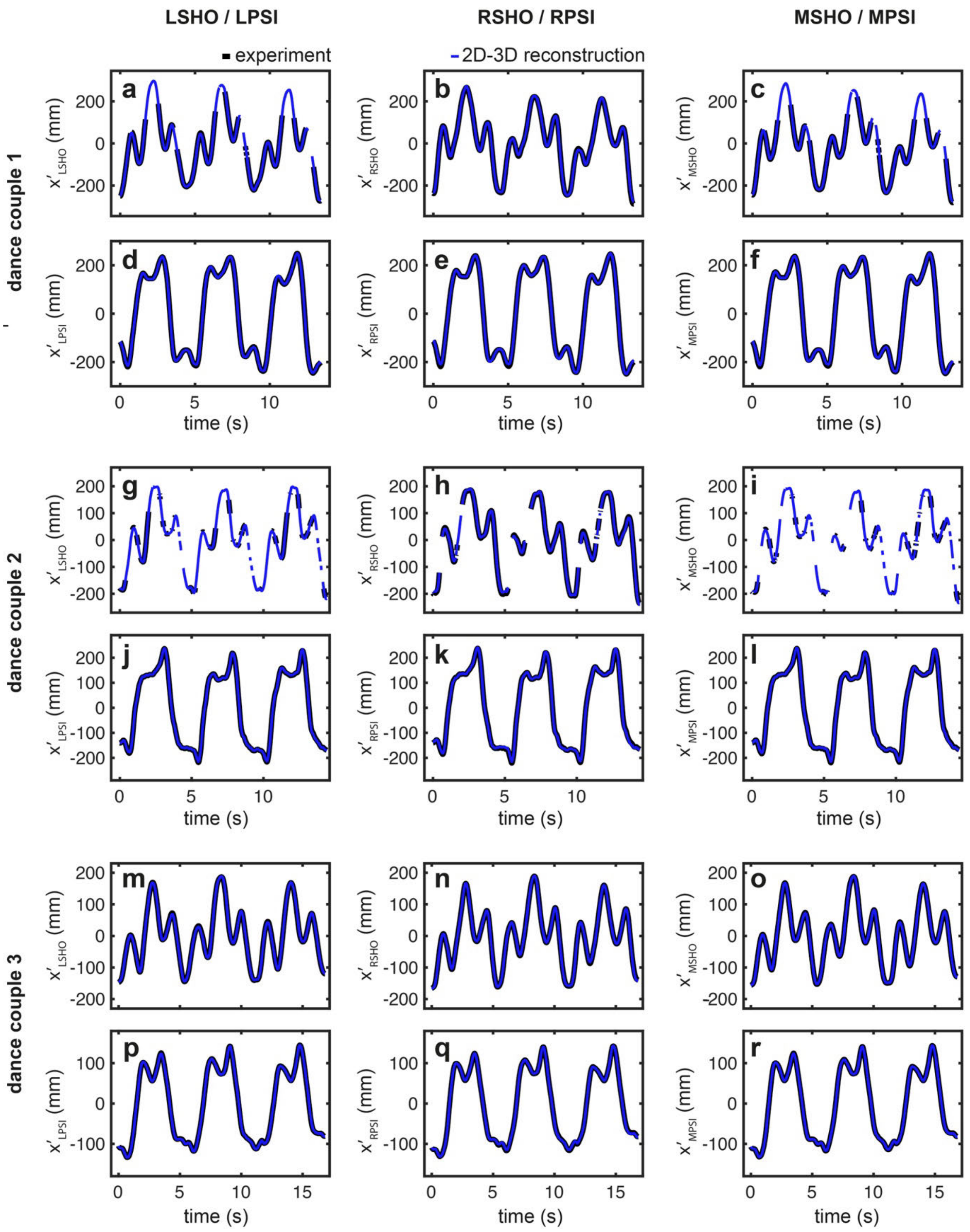

**Supplementary Fig. 1. 2D-3D reconstruction for sequence 5.** Panels **a-f**, **g-l** and **m-r** correspond to dance couples 1, 2 and 3, respectively. Left (**a**, **d**, **g**, **j**, **m** and **p**), centre (**b**, **e**, **h**, **k**, **n** and **q**) and right (**c**, **f**, **i**, **l**, **o** and **r**) panel columns show left, right and mid-position marker motions, respectively. Black curves denote the original projected 3D data and blue curves the fitted 2D-3D reconstructions.

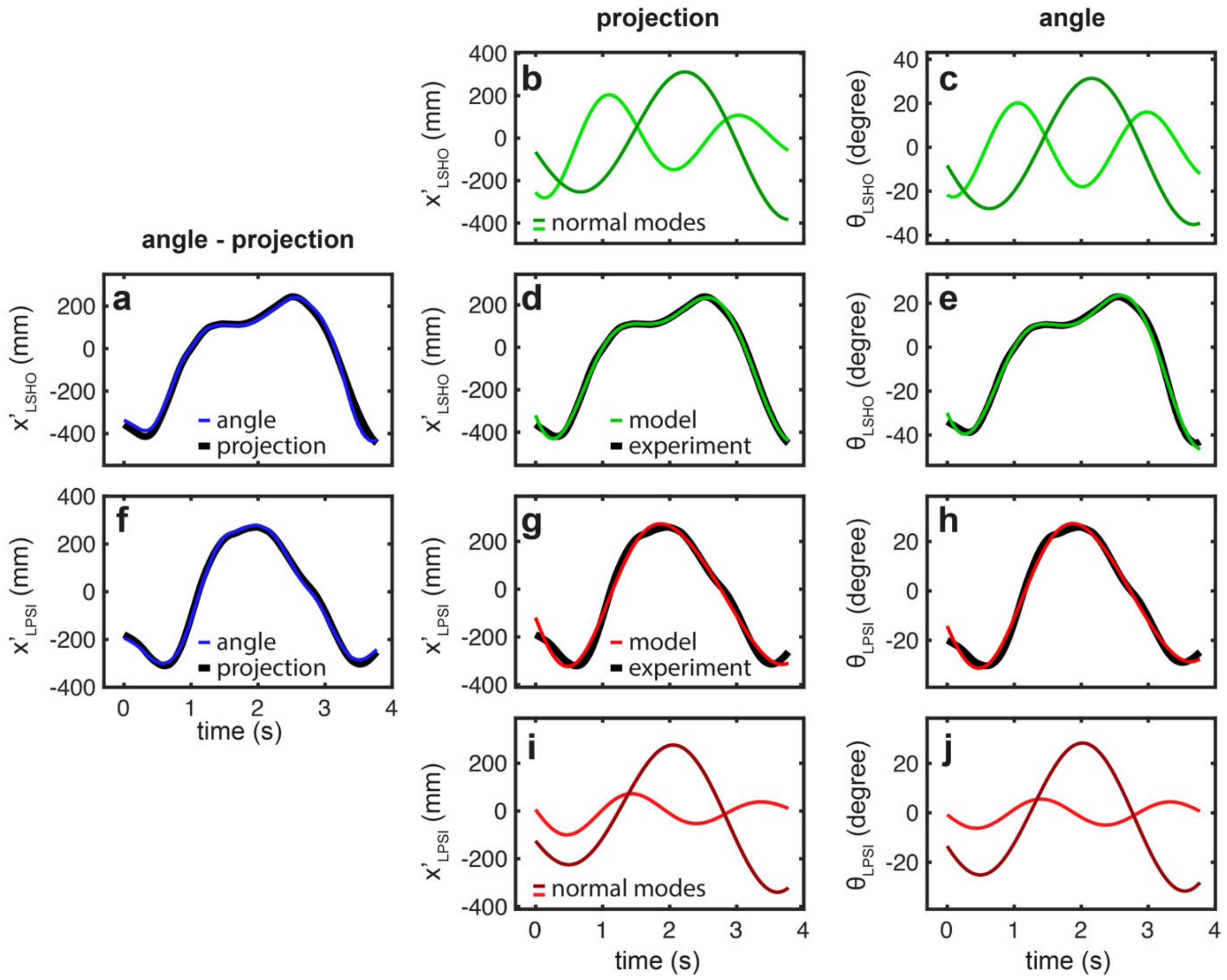

**Supplementary Fig. 2. Coupled oscillators in dance sequence 6a, dance couple 1.** The left panels **a** and **f** compare the azimuth-angle curves with the corresponding Cartesian projections after fitting with multiplicative and additive biases for LSHO and LPSI, respectively. Panels **d** and **e** show the model fitted to the projection and angle curves, respectively, for LSHO, while panels **g** and **h** show the same procedure for LPSI. Panels **b**, **c**, **i**, and **j** display the normal modes corresponding to **d**, **e**, **g**, and **h**, respectively.

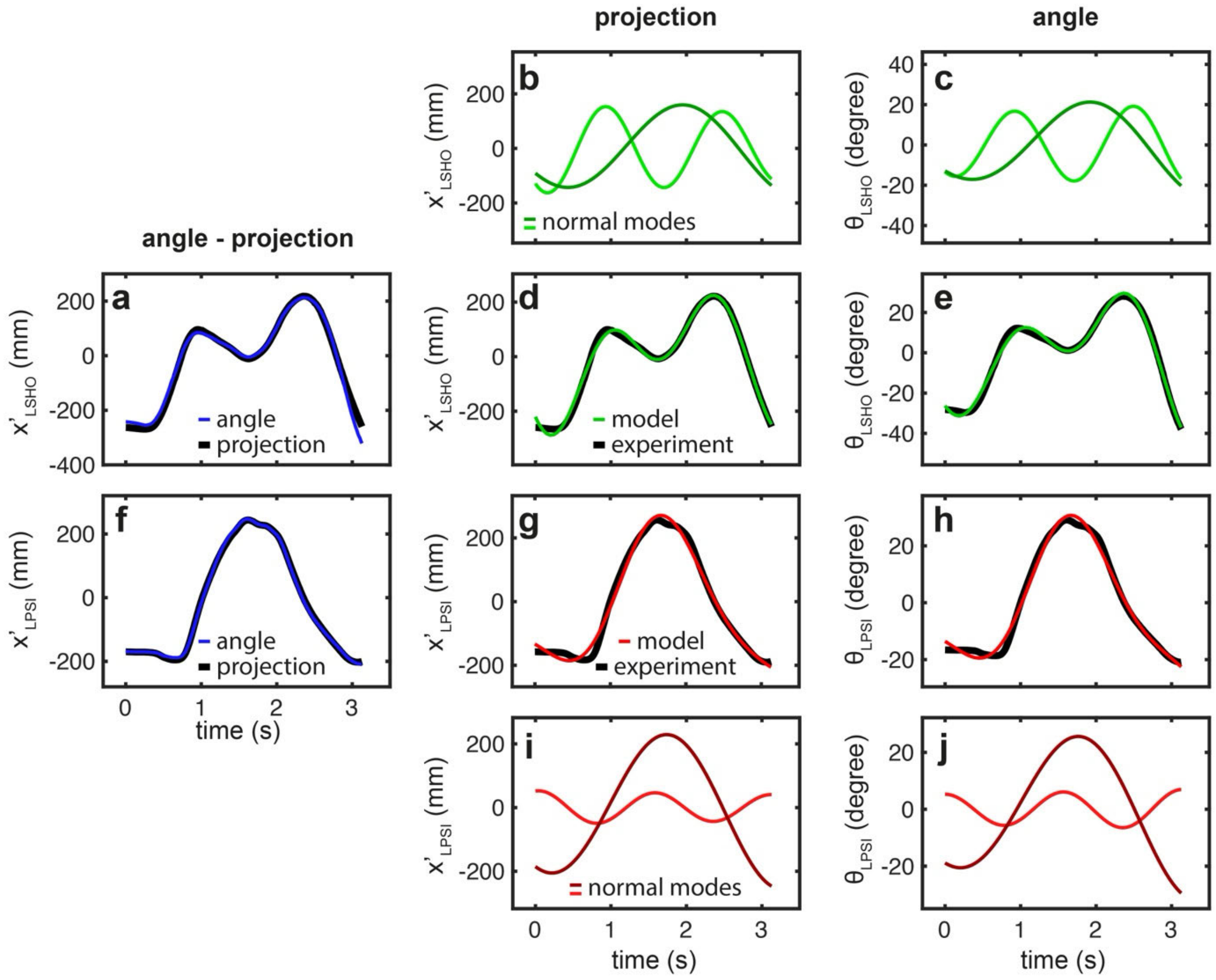

**Supplementary Fig. 3. Coupled oscillators in sequence 6a, dance couple 2.** The panels follow the same description as Supplementary Fig. 2.

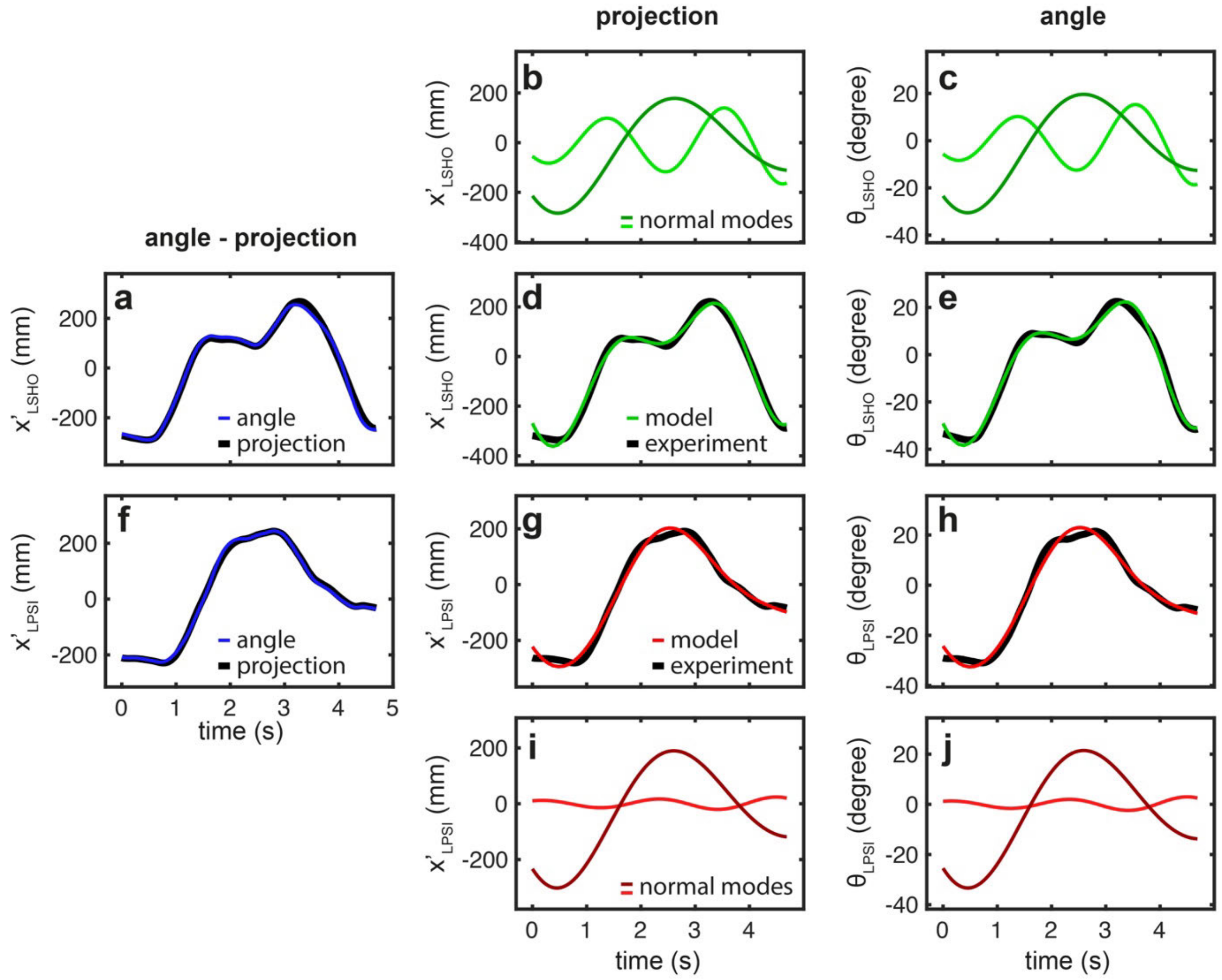

**Supplementary Fig. 4. Coupled oscillators in sequence 6a, dance couple 3.** The panels follow the same description as Supplementary Fig. 2.

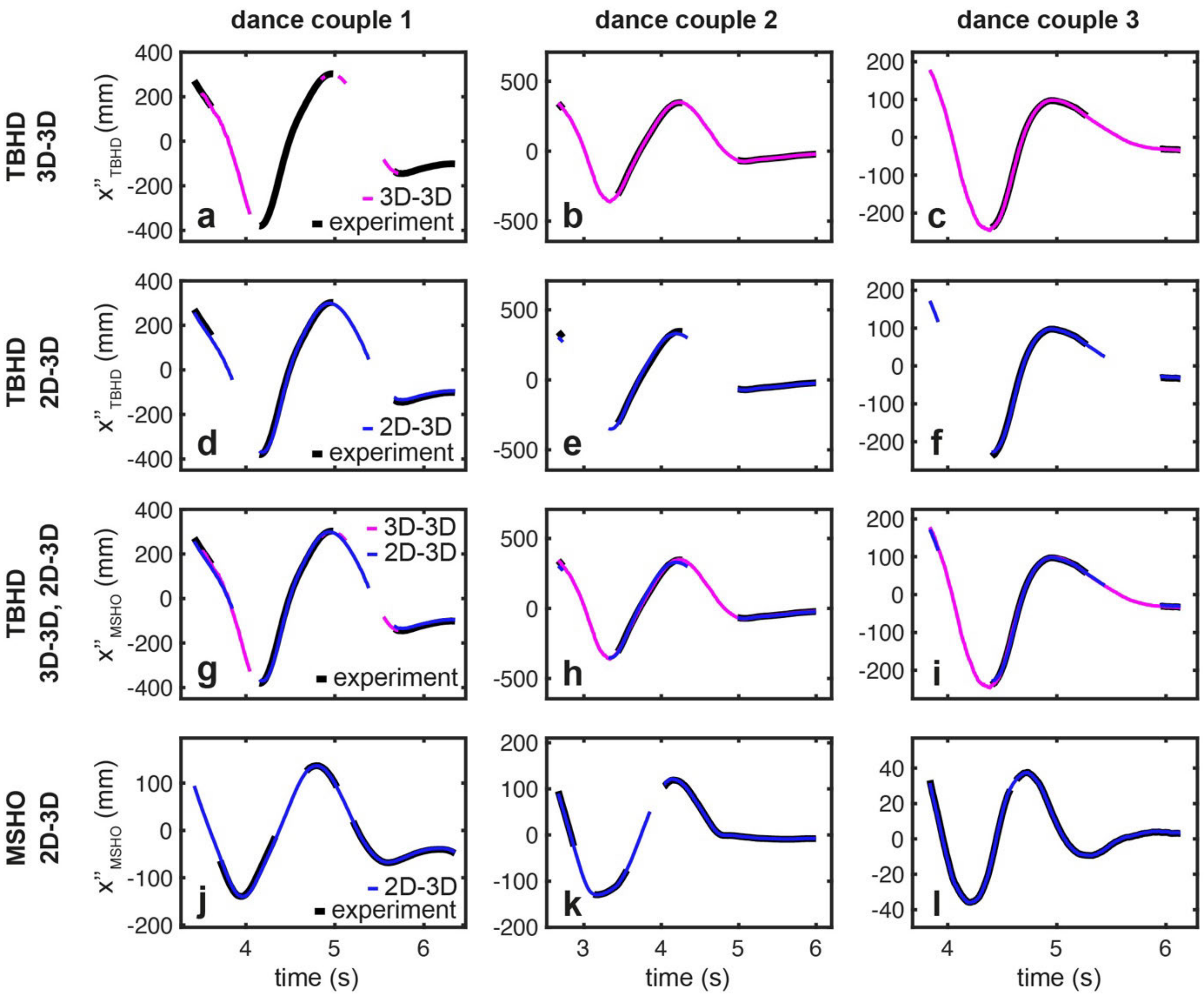

**Supplementary Fig. 5. Data reconstruction of sequence 6b.** Panels **a-i** show projection trajectories for the TBHD marker, and panels **j-l** for the MSHO marker. Columns correspond to dance couples 1-3. Black curves indicate the original projected 3D data (without reconstruction). For TBHD, pink curves (**a-c**) show the 3D-3D reconstruction used for model fitting, blue curves (**d-f**) show the 2D-3D reconstruction, and panels **g-i** show their superposition. For MSHO, blue curves (**j-l**) show the 2D-3D reconstruction.

# SUPPLEMENTARY TABLES

**Supplementary Table 1.**

**A**

| **Sequence 1** Parameter | Participant 1 | Participant 2 | Participant 3 | Units |
|---|---|---|---|---|
| $\omega^{D}/2\pi$ | $0.424 \pm 8.9\cdot10^{-5}$ | $0.422 \pm 7.4\cdot10^{-5}$ | $0.417 \pm 8.0\cdot10^{-5}$ | Hz |
| $\Delta\delta^{D}_{WB-NUL}$ | $69.56 \pm 0.50$ | $58.63 \pm 0.28$ | $68.20 \pm 0.30$ | degree |
| $\overleftrightarrow{\lambda}^{D}_{WB-NUL}$ | $0.19 \pm 1.4\cdot10^{-3}$ | $0.16 \pm 7.8\cdot10^{-4}$ | $0.19 \pm 8.3\cdot10^{-4}$ | a.u. |
| $\Delta\delta^{D}_{SGTLP}$ | $36.64 \pm 0.24$ | $44.55 \pm 0.28$ | $57.98 \pm 0.28$ | degree |
| $\overleftrightarrow{\lambda}^{D}_{SGTLP}$ | $0.10 \pm 6.7\cdot10^{-4}$ | $0.12 \pm 7.8\cdot10^{-4}$ | $0.16 \pm 7.8\cdot10^{-4}$ | a.u. |
| $\bar{A}^{D}_{WB-NUL}$ | $240.87 \pm 82.06$ | $214.12 \pm 73.80$ | $187.18 \pm 55.72$ | mm |
| $\bar{A}^{D}_{SGTLP}$ | $288.71 \pm 22.48$ | $253.23 \pm 22.70$ | $215.34 \pm 18.85$ | mm |
| $\omega^{U}/2\pi$ | $0.445\pm 1.6\cdot10^{-4}$ | $0.432 \pm 1.5\cdot10^{-4}$ | $0.432 \pm 1.3\cdot10^{-4}$ | Hz |
| $\Delta\delta^{U}_{WB-NUL}$ | $55.64 \pm 0.53$ | $95.25 \pm 0.67$ | $76.68 \pm 0.53$ | degree |
| $\overleftrightarrow{\lambda}^{U}_{WB-NUL}$ | $0.15 \pm 1.5\cdot10^{-3}$ | $0.26 \pm 1.9\cdot10^{-3}$ | $0.21 \pm 1.5\cdot10^{-3}$ | a.u. |
| $\Delta\delta^{U}_{SGTLP}$ | $44.96 \pm 0.51$ | $61.76 \pm 0.52$ | $66.46 \pm 0.43$ | degree |
| $\overleftrightarrow{\lambda}^{U}_{SGTLP}$ | $0.12 \pm 1.4\cdot10^{-3}$ | $0.17 \pm 1.4\cdot10^{-3}$ | $0.18 \pm 1.2\cdot10^{-3}$ | a.u. |
| $\bar{A}^{U}_{WB-NUL}$ | $192.45 \pm 59.27$ | $170.63 \pm 61.08$ | $166.71 \pm 43.95$ | mm |
| $\bar{A}^{U}_{SGTLP}$ | $228.15 \pm 11.22$ | $205.56 \pm 26.64$ | $189.34 \pm 7.95$ | mm |
| $A^{RD}_{RUCH}$ | $271.65 \pm 6.84$ | $234.16 \pm 7.71$ | $227.90 \pm 10.30$ | mm |
| $g^{RD}_{RUCH}$ | $2.73 \pm 9.8\cdot10^{-2}$ | $4.71 \pm 0.30$ | $2.99 \pm 0.20$ | $s^{-1}$ |
| $t^{RD}_{RUCH}$ | $13.60 \pm 3.4\cdot10^{-2}$ | $12.92 \pm 1.1\cdot10^{-2}$ | $13.64 \pm 1.9\cdot10^{-2}$ | s |
| $A^{RU}_{LUCH}$ | $230.67 \pm 9.63$ | $226.65 \pm 9.39$ | $204.15 \pm 8.51$ | mm |
| $g^{RU}_{LUCH}$ | $2.62 \pm 0.12$ | $3.05 \pm 0.12$ | $2.18 \pm 0.13$ | $s^{-1}$ |
| $t^{RU}_{LUCH}$ | $13.98 \pm 3.7\cdot10^{-2}$ | $13.15 \pm 2.5\cdot10^{-2}$ | $14.00 \pm 2.1\cdot10^{-2}$ | s |

**B**

| **Sequence 2** | | | | |
|---|---|---|---|---|
| Parameter | Dance couple 1 | Dance couple 2 | Dance couple 3 | Units |
| $\omega^D/2\pi$ | 0.427 ± 1.3·10$^{-4}$ | 0.414 ± 8.7·10$^{-5}$ | 0.407 ± 7.7·10$^{-5}$ | Hz |
| $\Delta\delta^D_{WB-NUL}$ | 88.01 ± 0.99 | 59.82 ± 0.39 | 54.01 ± 0.28 | degree |
| $\overleftrightarrow{\lambda}^D_{WB-NUL}$ | 0.24 ± 2.8·10$^{-3}$ | 0.17 ± 1.1·10$^{-3}$ | 0.15 ± 7.8·10$^{-4}$ | a.u. |
| $\Delta\delta^D_{SGTLP}$ | 42.16 ± 0.41 | 48.05 ± 0.32 | 36.84 ± 0.27 | degree |
| $\overleftrightarrow{\lambda}^D_{SGTLP}$ | 0.12 ± 1.1·10$^{-3}$ | 0.13 ± 8.9·10$^{-4}$ | 0.10 ± 7.5·10$^{-4}$ | a.u. |
| $\bar{A}^D_{WB-NUL}$ | 229.89 ± 82.58 | 171.56 ± 69.27 | 130.00 ± 46.62 | mm |
| $\bar{A}^D_{SGTLP}$ | 281.82 ± 11.76 | 214.92 ± 32.72 | 156.93 ± 6.18 | mm |
| $\omega^U/2\pi$ | 0.428 ± 1.2·10$^{-4}$ | 0.407 ± 6.9·10$^{-5}$ | 0.405 ± 1.1·10$^{-4}$ | Hz |
| $\Delta\delta^U_{WB-NUL}$ | 116.25 ± 0.48 | 73.45 ± 0.43 | 75.68 ± 0.81 | degree |
| $\overleftrightarrow{\lambda}^U_{WB-NUL}$ | 0.32 ± 1.3·10$^{-3}$ | 0.20 ± 1.2·10$^{-3}$ | 0.21 ± 2.3·10$^{-3}$ | a.u. |
| $\Delta\delta^U_{SGTLP}$ | 56.26 ± 0.45 | 53.69 ± 0.28 | 30.49 ± 0.37 | degree |
| $\overleftrightarrow{\lambda}^U_{SGTLP}$ | 0.16 ± 1.3·10$^{-3}$ | 0.15 ± 7.8·10$^{-4}$ | 0.08 ± 1.0·10$^{-3}$ | a.u. |
| $\bar{A}^U_{WB-NUL}$ | 220.03 ± 81.57 | 184.88 ± 63.06 | 146.91 ± 66.48 | mm |
| $\bar{A}^U_{SGTLP}$ | 257.63 ± 7.85 | 231.16 ± 10.19 | 188.22 ± 22.87 | mm |
| $A^{RD}_{LUCH}$ | 285.67 ± 2.08 | 241.90 ± 11.77 | 157.08 ± 0.46 | mm |
| $g^{RD}_{LUCH}$ | 10.41 ± 0.31 | 2.71 ± 0.12 | 7.08 ± 0.21 | $s^{-1}$ |
| $t^{RD}_{LUCH}$ | 10.28 ± 4.0·10$^{-3}$ | 12.85 ± 3.6·10$^{-2}$ | 13.36 ± 6.2·10$^{-3}$ | s |
| $A^{RU}_{RUCH}$ | 292.76 ± 4.14 | 241.90 ± 6.86 | 201.81 ± 10.60 | mm |
| $g^{RU}_{RUCH}$ | 2.24 ± 3.1·10$^{-2}$ | 5.02 ± 0.60 | 7.30 ± 0.63 | $s^{-1}$ |
| $t^{RU}_{RUCH}$ | 10.56 ± 7.8·10$^{-3}$ | 12.93 ± 6.3·10$^{-2}$ | 13.96 ± 3.1·10$^{-2}$ | s |

**C**

| **Sequence 3** | | | | |
|---|---|---|---|---|
| Parameter | Dance couple 1 | Dance couple 2 | Dance couple 3 | Units |
| $\omega_0/2\pi$ | 1.07 ± 7.9·10$^{-3}$ | 1.20 ± 5.3·10$^{-3}$ | 1.14 ± 5.2·10$^{-3}$ | Hz |
| $\omega_{A_{max}}/2\pi$ | 0.90 ± 5.1·10$^{-3}$ | 1.09 ± 3.8·10$^{-3}$ | 1.10 ± 4.7·10$^{-3}$ | Hz |
| $\gamma/2\pi$ | 0.83 ± 2.0·10$^{-2}$ | 0.72 ± 9.1·10$^{-3}$ | 0.43 ± 1.2·10$^{-2}$ | Hz |
| $\zeta/4\pi^2$ | 0.79 ± 1.7·10$^{-2}$ | 1.53 ± 1.9·10$^{-2}$ | 1.29 ± 2.2·10$^{-2}$ | $Hz^2$ |
| $A_{max}$ | 0.97 ± 8.4·10$^{-3}$ | 1.86 ± 8.3·10$^{-3}$ | 2.67 ± 4.7·10$^{-2}$ | a.u. |
| $A(0)$ | 0.69 ± 6.1·10$^{-3}$ | 1.06 ± 5.9·10$^{-3}$ | 0.98 ± 1.3·10$^{-2}$ | a.u. |
| $A_{max}/A(0)$ | 1.40 ± 2.2·10$^{-2}$ | 1.76 ± 1.5·10$^{-2}$ | 2.71 ± 6.8·10$^{-2}$ | a.u. |

**D**

| **Sequence 4** | | | | |
|---|---|---|---|---|
| Parameter | Dance couple 1 | Dance couple 2 | Dance couple 3 | Units |
| $\Delta\delta_{1,5}$ | 83.15 ± 17.90 | 98.93 ± 24.05 | 90.28 ± 9.78 | degree |
| $\Delta\delta_{2,6}$ | 106.98 ± 13.47 | 154.08 ± 26.41 | 111.15 ± 10.09 | degree |
| $\Delta\delta_{3,7}$ | 66.39 ± 12.90 | 58.00 ± 13.07 | 47.81 ± 8.92 | degree |
| $\Delta\delta_{\hat{4},\hat{8}}$ | 14.83 ± 4.08 | -9.22 ± 13.48 | -20.23 ± 8.84 | degree |
| $\Delta\delta_{4,8}$ | 153.06 ± 17.31 | 149.53 ± 18.75 | 149.82 ± 7.24 | degree |

**E**

| **Sequence 5. Markers LSHO/LPSI** | | | | |
|---|---|---|---|---|
| Parameter | Dance couple 1 | Dance couple 2 | Dance couple 3 | Units |
| $\mathcal{J}$ | $3.019 \pm 2.2\cdot10^{-3}$ | $2.992 \pm 2.6\cdot10^{-3}$ | $2.991 \pm 1.6\cdot10^{-3}$ | a.u. |
| $\omega_1$ | $4.166 \pm 2.0\cdot10^{-3}$ | $3.924 \pm 2.4\cdot10^{-3}$ | $3.341 \pm 1.4\cdot10^{-3}$ | rad $s^{-1}$ |
| $\omega_2$ | $1.3799 \pm 10.0\cdot10^{-4}$ | $1.311 \pm 1.1\cdot10^{-3}$ | $1.1171 \pm 9.0\cdot10^{-4}$ | rad $s^{-1}$ |
| $x'_{a0}$ | -245.83 ± 1.52 | -213.02 ± 1.49 | -165.32 ± 0.93 | mm |
| $x'_{b0}$ | -147.37 ± 1.45 | -131.37 ± 1.45 | -82.43 ± 0.77 | mm |
| $v'_{a0}$ | 207.48 ± 6.06 | -13.07 ± 7.06 | 117.30 ± 3.40 | $mm\cdot s^{-1}$ |
| $v'_{b0}$ | -126.14 ± 5.53 | -31.69 ± 5.09 | -66.81 ± 1.88 | $mm\cdot s^{-1}$ |
| $\omega_a$ | 3.45 ± 0.25 | 3.37 ± 0.26 | 2.94 ± 0.18 | rad $s^{-1}$ |
| $\beta_{ab}$ | $-6.81 \pm 5.2\cdot10^{-2}$ | $-6.15 \pm 5.8\cdot10^{-2}$ | $-5.79 \pm 4.6\cdot10^{-2}$ | $rad^2$ $s^{-2}$ |
| $\beta_{ba}$ | $-8.03 \pm 8.4\cdot10^{-2}$ | $-6.33 \pm 9.4\cdot10^{-2}$ | $-3.20 \pm 3.9\cdot10^{-2}$ | $rad^2$ $s^{-2}$ |
| $\omega_b$ | 2.72 ± 0.24 | 2.40 ± 0.26 | 1.94 ± 0.17 | rad $s^{-1}$ |
| $A_{11}$ | 104.15 ± 1.03 | 91.24 ± 1.07 | 84.58 ± 0.61 | mm |
| $A_{12}$ | 152.69 ± 1.04 | 122.56 ± 1.09 | 90.12 ± 0.62 | mm |
| $A_{21}$ | 83.85 ± 1.03 | 59.80 ± 1.02 | 36.49 ± 0.61 | mm |
| $A_{22}$ | 223.62 ± 1.04 | 192.39 ± 1.03 | 115.56 ± 0.62 | mm |
| $\delta_{11}$ | -154.11 ± 0.94 | -179.20 ± 1.22 | -153.35 ± 0.78 | degree |
| $\delta_{12}$ | -175.09 ± 0.54 | 173.54 ± 0.70 | 174.61 ± 0.59 | degree |
| $\delta_{21}$ | 25.89 ± 1.02 | 0.80 ± 1.38 | 26.65 ± 1.16 | degree |
| $\delta_{22}$ | -175.09 ± 0.47 | 173.54 ± 0.57 | 174.61 ± 0.53 | degree |

**F**

| **Sequence 5. Markers RSHO/RPSI** | | | | |
|---|---|---|---|---|
| Parameter | Dance couple 1 | Dance couple 2 | Dance couple 3 | Units |
| $\mathcal{J}$ | $3.008 \pm 2.3 \cdot 10^{-3}$ | $3.005 \pm 2.7 \cdot 10^{-3}$ | $2.991 \pm 1.7 \cdot 10^{-3}$ | a.u. |
| $\omega_1$ | $4.178 \pm 2.0 \cdot 10^{-3}$ | $3.934 \pm 2.5 \cdot 10^{-3}$ | $3.340 \pm 1.4 \cdot 10^{-3}$ | rad $s^{-1}$ |
| $\omega_2$ | $1.389 \pm 1.0 \cdot 10^{-3}$ | $1.309 \pm 1.1 \cdot 10^{-3}$ | $1.1167 \pm 9.3 \cdot 10^{-4}$ | rad $s^{-1}$ |
| $x'_{a0}$ | -269.72 ± 1.58 | -212.90 ± 1.45 | -161.49 ± 0.95 | mm |
| $x'_{b0}$ | -143.07 ± 1.53 | -133.45 ± 1.32 | -81.69 ± 0.87 | mm |
| $v'_{a0}$ | 171.40 ± 6.94 | -34.89 ± 4.96 | 115.29 ± 3.36 | mm·$s^{-1}$ |
| $v'_{b0}$ | -128.49 ± 5.76 | -13.45 ± 4.46 | -70.81 ± 2.05 | mm·$s^{-1}$ |
| $\omega_a$ | 3.41 ± 0.26 | 3.34 ± 0.24 | 2.93 ± 0.20 | rad $s^{-1}$ |
| $\beta_{ab}$ | -7.47 ± 5.9·10-2 | $-6.39 \pm 5.9 \cdot 10^{-2}$ | $-5.63 \pm 4.6 \cdot 10^{-2}$ | $rad^2$ $s^{-2}$ |
| $\beta_{ba}$ | -7.56 ± 8.2·10-2 | $-6.40 \pm 8.3 \cdot 10^{-2}$ | $-3.36 \pm 4.6 \cdot 10^{-2}$ | $rad^2$ $s^{-2}$ |
| $\omega_b$ | 2.78 ± 0.26 | 2.46 ± 0.23 | 1.96 ± 0.20 | rad $s^{-1}$ |
| $A_{11}$ | 107.71 ± 1.11 | 83.80 ± 1.02 | 81.96 ± 0.63 | mm |
| $A_{12}$ | 169.90 ± 1.09 | 129.89 ± 1.04 | 89.09 ± 0.63 | mm |
| $A_{21}$ | 83.89 ± 1.07 | 57.01 ± 1.00 | 37.68 ± 0.63 | mm |
| $A_{22}$ | 220.83 ± 1.08 | 191.37 ± 1.01 | 115.78 ± 0.63 | mm |
| $\delta_{11}$ | -157.94 ± 0.96 | 176.94 ± 1.24 | -152.73 ± 0.82 | degree |
| $\delta_{12}$ | -179.42 ± 0.54 | 174.17 ± 0.65 | 174.16 ± 0.61 | degree |
| $\delta_{21}$ | 22.06 ± 1.06 | -3.06 ± 1.44 | 27.27 ± 1.18 | degree |
| $\delta_{22}$ | -179.42 ± 0.49 | 174.17 ± 0.55 | 174.16 ± 0.55 | degree |

**G**

| **Sequence 5. Markers MSHO/MPSI** | | | | |
|---|---|---|---|---|
| Parameter | Dance couple 1 | Dance couple 2 | Dance couple 3 | Units |
| $\mathcal{I}$ | $3.013 \pm 2.0 \cdot 10^{-3}$ | $2.998 \pm 2.5 \cdot 10^{-3}$ | $2.991 \pm 1.6 \cdot 10^{-3}$ | a.u. |
| $\omega_1$ | $4.172 \pm 1.8 \cdot 10^{-3}$ | $3.927 \pm 2.3 \cdot 10^{-3}$ | $3.340 \pm 1.4 \cdot 10^{-3}$ | rad $s^{-1}$ |
| $\omega_2$ | $1.3846 \pm 9.1 \cdot 10^{-4}$ | $1.310 \pm 1.1 \cdot 10^{-3}$ | $1.1169 \pm 8.8 \cdot 10^{-4}$ | rad $s^{-1}$ |
| $x'_{a0}$ | -258.51 ± 1.37 | -214.57 ± 1.40 | -163.39 ± 0.88 | mm |
| $x'_{b0}$ | -145.20 ± 1.33 | -132.17 ± 1.36 | -82.00 ± 0.77 | mm |
| $v'_{a0}$ | 187.65 ± 4.51 | -7.79 ± 1.97 | 116.92 ± 2.88 | mm·$s^{-1}$ |
| $v'_{b0}$ | -126.39 ± 4.13 | -30.17 ± 3.33 | -69.18 ± 1.73 | mm·$s^{-1}$ |
| $\omega_a$ | 3.42 ± 0.24 | 3.35 ± 0.26 | 2.93 ± 0.18 | rad $s^{-1}$ |
| $\beta_{ab}$ | -7.15 ± 5.0·10-2 | $-6.28 \pm 5.6 \cdot 10^{-2}$ | $-5.71 \pm 4.4 \cdot 10^{-2}$ | $rad^2$ $s^{-2}$ |
| $\beta_{ba}$ | -7.81 ± 7.3·10-2 | $-6.34 \pm 8.9 \cdot 10^{-2}$ | $-3.29 \pm 4.0 \cdot 10^{-2}$ | $rad^2$ $s^{-2}$ |
| $\omega_b$ | 2.76 ± 0.24 | 2.43 ± 0.25 | 1.95 ± 0.18 | rad $s^{-1}$ |
| $A_{11}$ | 104.90 ± 0.97 | 88.69 ± 1.02 | 83.31 ± 0.60 | mm |
| $A_{12}$ | 162.74 ± 0.95 | 126.51 ± 1.05 | 89.64 ± 0.61 | mm |
| $A_{21}$ | 83.92 ± 0.93 | 58.96 ± 0.95 | 37.17 ± 0.60 | mm |
| $A_{22}$ | 222.24 ± 0.94 | 192.03 ± 0.96 | 115.65 ± 0.60 | mm |
| $\delta_{11}$ | -156.19 ± 0.85 | -178.62 ± 1.19 | -152.91 ± 0.77 | degree |
| $\delta_{12}$ | -177.21 ± 0.48 | 174.39 ± 0.65 | 174.38 ± 0.58 | degree |
| $\delta_{21}$ | 23.81 ± 0.93 | 1.38 ± 1.33 | 27.09 ± 1.13 | degree |
| $\delta_{22}$ | -177.21 ± 0.42 | 174.39 ± 0.53 | 174.38 ± 0.52 | degree |

**H**

| **Sequence 6a. Ellipsoids** | | | | |
|---|---|---|---|---|
| Parameter | Dance couple 1 | Dance couple 2 | Dance couple 3 | Units |
| $s_1$ | 658.3 ± 33.48 | 555.19 ± 30.66 | 573.6 ± 106.4 | mm |
| $s_2$ | 670.7 ± 58.47 | 535.94 ± 23.02 | 579.0 ± 110.1 | mm |
| $s_3$ | 506.9 ± 13.16 | 663.93 ± 19.56 | 536.6 ± 12.61 | mm |

**I**

| **Sequence 6a.** $x'$ **projection** | | | | |
|---|---|---|---|---|
| Parameter | Dance couple 1 | Dance couple 2 | Dance couple 3 | Units |
| $\mathcal{J}$ | $1.61 \pm 2.5 \cdot 10^{-2}$ | $1.97 \pm 2.3 \cdot 10^{-2}$ | $2.00 \pm 1.2 \cdot 10^{-2}$ | a.u. |
| $\omega_1$ | $3.25 \pm 2.0 \cdot 10^{-2}$ | $4.07 \pm 1.3 \cdot 10^{-2}$ | $2.91 \pm 9.2 \cdot 10^{-3}$ | rad $s^{-1}$ |
| $\omega_2$ | $2.02 \pm 1.4 \cdot 10^{-2}$ | $2.07 \pm 1.8 \cdot 10^{-2}$ | $1.46 \pm 7.5 \cdot 10^{-3}$ | rad $s^{-1}$ |
| $\kappa_1$ | $0.33 \pm 1.3 \cdot 10^{-2}$ | $8.16 \cdot 10^{-2} \pm 9.4 \cdot 10^{-3}$ | $-0.16 \pm 6.8 \cdot 10^{-3}$ | rad $s^{-1}$ |
| $\kappa_2$ | $-0.13 \pm 6.7 \cdot 10^{-3}$ | $-7.18 \cdot 10^{-2} \pm 8.1 \cdot 10^{-3}$ | $0.22 \pm 6.4 \cdot 10^{-3}$ | rad $s^{-1}$ |
| $x'_{a0}$ | -320.27 ± 3.32 | -220.36 ± 2.72 | -268.90 ± 2.39 | mm |
| $x'_{b0}$ | -119.57 ± 3.16 | -133.36 ± 2.50 | -222.69 ± 2.43 | mm |
| $v'_{a0}$ | -856.86 ± 15.82 | -629.62 ± 15.36 | -470.49 ± 8.86 | mm·$s^{-1}$ |
| $v'_{b0}$ | -741.92 ± 20.56 | -146.86 ± 10.42 | -286.23 ± 11.01 | mm·$s^{-1}$ |
| $\omega_a$ | 10.26 ± 0.11 | 3.75 ± 0.33 | 2.77 ± 0.24 | rad $s^{-1}$ |
| $\beta_{ab}$ | $-6.50 \pm 7.8 \cdot 10^{-2}$ | $-6.40 \pm 5.1 \cdot 10^{-2}$ | $-5.16 \pm 5.2 \cdot 10^{-2}$ | $rad^2$ $s^{-2}$ |
| $\beta_{ba}$ | 0.99 ± 0.12 | $-2.71 \pm 8.8 \cdot 10^{-2}$ | $-0.86 \pm 6.6 \cdot 10^{-2}$ | $rad^2$ $s^{-2}$ |
| $\omega_b$ | $3.64 \pm 8.2 \cdot 10^{-2}$ | 2.51 ± 0.20 | 1.73 ± 0.21 | rad $s^{-1}$ |
| $\gamma_{aa}$ | $-0.30 \pm 2.4 \cdot 10^{-2}$ | $0.25 \pm 2.6 \cdot 10^{-2}$ | $-0.30 \pm 1.7 \cdot 10^{-2}$ | rad $s^{-1}$ |
| $\gamma_{ab}$ | $1.20 \pm 2.6 \cdot 10^{-2}$ | $1.14 \pm 2.4 \cdot 10^{-2}$ | $0.78 \pm 2.5 \cdot 10^{-2}$ | rad $s^{-1}$ |
| $\gamma_{ba}$ | $-0.77 \pm 2.7 \cdot 10^{-2}$ | $-0.49 \pm 2.7 \cdot 10^{-2}$ | $1.03 \cdot 10^{-2} \pm 1.72 \cdot 10^{-2}$ | rad $s^{-1}$ |
| $\gamma_{bb}$ | $0.70 \pm 2.2 \cdot 10^{-2}$ | $-0.23 \pm 1.5 \cdot 10^{-2}$ | $0.41 \pm 1.5 \cdot 10^{-2}$ | rad $s^{-1}$ |
| $A_{11}$ | 295.10 ± 9.78 | 164.93 ± 3.13 | 79.03 ± 1.66 | mm |
| $A_{12}$ | 232.45 ± 3.20 | 138.75 ± 2.13 | 317.53 ± 3.90 | mm |
| $A_{21}$ | 117.18 ± 6.15 | 52.78 ± 1.92 | 11.67 ± 0.97 | mm |
| $A_{22}$ | 210.94 ± 3.76 | 202.44 ± 2.26 | 336.29 ± 3.62 | mm |
| $\delta_{11}$ | 149.84 ± 1.14 | 141.58 ± 1.17 | 133.81 ± 1.67 | degree |
| $\delta_{12}$ | 106.26 ± 2.29 | 131.06 ± 2.03 | 132.42 ± 0.90 | degree |
| $\delta_{21}$ | 87.22 ± 3.23 | -10.57 ± 1.96 | -26.73 ± 3.59 | degree |
| $\delta_{22}$ | 126.43 ± 2.07 | 156.22 ± 1.80 | 133.88 ± 0.96 | degree |

**J**

| **Sequence 6a. Angle $\theta$** | | | | |
|---|---|---|---|---|
| Parameter | Dance couple 1 | Dance couple 2 | Dance couple 3 | Units |
| $\mathcal{J}$ | $1.58 \pm 2.3\cdot10^{-2}$ | $1.98 \pm 2.5\cdot10^{-2}$ | $1.97 \pm 1.3\cdot10^{-2}$ | a.u. |
| $\omega_1$ | $3.26 \pm 1.8\cdot10^{-2}$ | $3.99 \pm 1.4\cdot10^{-2}$ | $2.89 \pm 1.0\cdot10^{-2}$ | rad $s^{-1}$ |
| $\omega_2$ | $2.06 \pm 1.4\cdot10^{-2}$ | $2.01 \pm 2.0\cdot10^{-2}$ | $1.47 \pm 8.1\cdot10^{-3}$ | rad $s^{-1}$ |
| $\kappa_1$ | $0.12 \pm 1.1\cdot10^{-2}$ | $-8.67\cdot10^{-2} \pm 8.4\cdot10^{-3}$ | $-0.19 \pm 7.6\cdot10^{-3}$ | rad $s^{-1}$ |
| $\kappa_2$ | $-7.56\cdot10^{-2} \pm 5.4\cdot10^{-3}$ | $-0.14 \pm 9.5\cdot10^{-3}$ | $0.21 \pm 6.7\cdot10^{-3}$ | rad $s^{-1}$ |
| $x'_{a0}$ | -29.92 ± 0.29 | -26.29 ± 0.28 | -29.08 ± 0.27 | degree |
| $x'_{b0}$ | -14.25 ± 0.26 | -13.59 ± 0.27 | -24.28 ± 0.29 | degree |
| $v'_{a0}$ | -74.18 ± 1.28 | -53.66 ± 1.69 | -48.21 ± 0.99 | degree$\cdot s^{-1}$ |
| $v'_{b0}$ | -63.75 ± 1.51 | -16.60 ± 1.22 | -32.60 ± 1.29 | degree$\cdot s^{-1}$ |
| $\omega_a$ | $10.38 \pm 9.4\cdot10^{-2}$ | 3.62 ± 0.33 | 2.75 ± 0.25 | rad $s^{-1}$ |
| $\beta_{ab}$ | $-6.38 \pm 6.7\cdot10^{-2}$ | $-7.20 \pm 5.3\cdot10^{-2}$ | $-4.91 \pm 5.3\cdot10^{-2}$ | $rad^2$ $s^{-2}$ |
| $\beta_{ba}$ | 0.52 ± 0.10 | $-2.66 \pm 9.1\cdot10^{-2}$ | $-0.92 \pm 7.2\cdot10^{-2}$ | $rad^2$ $s^{-2}$ |
| $\omega_b$ | $4.05 \pm 6.5\cdot10^{-2}$ | 2.53 ± 0.20 | 1.75 ± 0.21 | rad $s^{-1}$ |
| $\gamma_{aa}$ | $-0.43 \pm 2.1\cdot10^{-2}$ | $-0.19 \pm 2.4\cdot10^{-2}$ | $-0.37 \pm 1.8\cdot10^{-2}$ | rad $s^{-1}$ |
| $\gamma_{ab}$ | $1.18 \pm 2.3\cdot10^{-2}$ | $1.07 \pm 2.5\cdot10^{-2}$ | $0.69 \pm 2.4\cdot10^{-2}$ | rad $s^{-1}$ |
| $\gamma_{ba}$ | $-0.56 \pm 2.2\cdot10^{-2}$ | $-0.47 \pm 2.7\cdot10^{-2}$ | $8.18\cdot10^{-3} \pm 1.87\cdot10^{-2}$ | rad $s^{-1}$ |
| $\gamma_{bb}$ | $0.51 \pm 1.8\cdot10^{-2}$ | $-0.26 \pm 1.5\cdot10^{-2}$ | $0.41 \pm 1.5\cdot10^{-2}$ | rad $s^{-1}$ |
| $A_{11}$ | 22.87 ± 0.78 | 15.43 ± 0.30 | 7.93 ± 0.19 | degree |
| $A_{12}$ | 26.65 ± 0.39 | 16.28 ± 0.26 | 33.84 ± 0.43 | degree |
| $A_{21}$ | 6.58 ± 0.39 | 5.29 ± 0.18 | 1.29 ± 0.11 | degree |
| $A_{22}$ | 24.25 ± 0.32 | 20.02 ± 0.25 | 37.02 ± 0.42 | degree |
| $\delta_{11}$ | 161.28 ± 1.46 | 150.13 ± 1.31 | 135.55 ± 1.87 | degree |
| $\delta_{12}$ | 108.05 ± 1.93 | 142.45 ± 2.18 | 133.80 ± 0.94 | degree |
| $\delta_{21}$ | 96.58 ± 4.42 | 2.56 ± 2.04 | -22.51 ± 3.73 | degree |
| $\delta_{22}$ | 123.83 ± 1.85 | 160.50 ± 2.04 | 133.48 ± 1.02 | degree |

**K**

| **Sequence 6b. Ellipsoids** | | | | |
|---|---|---|---|---|
| Parameter | Dance couple 1 | Dance couple 2 | Dance couple 3 | Units |
| $s_1$ | 343.2 ± 6.38 | 360.8 ± 78.30 | 272.9 ± 6.55 | mm |
| $s_2$ | 389.2 ± 5.40 | 376.0 ± 86.70 | 277.2 ± 6.23 | mm |
| $s_3$ | 419.3 ± 25.54 | 399.7 ± 37.41 | 311.5 ± 17.96 | mm |

**L**

| **Sequence 6b. $x''$ projection** | | | | |
|---|---|---|---|---|
| Parameter | Dance couple 1 | Dance couple 2 | Dance couple 3 | Units |
| $\mathcal{J}$ | 1.72 ± 0.14 | 2.24 ± 7.5·10$^{-2}$ | 1.87 ± 8.3·10$^{-2}$ | a.u. |
| $\omega_1$ | 4.69 ± 7.0·10$^{-2}$ | 3.95 ± 2.4·10$^{-2}$ | 6.77 ± 7.2·10$^{-2}$ | rad s$^{-1}$ |
| $\omega_2$ | 2.73 ± 0.12 | 1.76 ± 7.1·10$^{-2}$ | 3.63 ± 4.1·10$^{-2}$ | rad s$^{-1}$ |
| $\kappa_1$ | 0.56 ± 5.3·10$^{-2}$ | 0.96 ± 2.2·10$^{-2}$ | 1.70 ± 4.4·10$^{-2}$ | rad s$^{-1}$ |
| $\kappa_2$ | -4.29·10$^{-2}$ ± 4.09·10$^{-2}$ | 0.64 ± 4.6·10$^{-2}$ | 0.94 ± 2.1·10$^{-2}$ | rad s$^{-1}$ |
| $x'_{a0}$ | 79.97 ± 3.11 | 68.96 ± 2.86 | 31.26 ± 1.65 | mm |
| $x'_{b0}$ | 214.16 ± 5.17 | 322.61 ± 5.57 | 151.49 ± 1.69 | mm |
| $v'_{a0}$ | -242.84 ± 17.24 | -261.56 ± 12.16 | -174.31 ± 17.07 | mm·s$^{-1}$ |
| $v'_{b0}$ | 487.19 ± 70.83 | 225.02 ± 26.30 | 2.12 ± 0.92 | mm·s$^{-1}$ |
| $\omega_a$ | 5.70 ± 2.42 | 1.52 ± 0.82 | 11.58 ± 3.41 | rad s$^{-1}$ |
| $\beta_{ab}$ | -14.48 ± 2.30 | 1.34 ± 0.24 | -2.84 ± 0.79 | rad$^2$ s$^{-2}$ |
| $\beta_{ba}$ | -196.21 ± 38.43 | -20.84 ± 2.13 | 240.93 ± 36.91 | rad$^2$ s$^{-2}$ |
| $\omega_b$ | 9.63 ± 4.48 | 3.61 ± 0.91 | 2.51·10$^{-3}$ ± 1.30 | rad s$^{-1}$ |
| $\gamma_{aa}$ | -8.43 ± 0.27 | -5.44 ± 0.25 | 10.34 ± 1.33 | rad s$^{-1}$ |
| $\gamma_{ab}$ | 1.81 ± 0.31 | 2.21 ± 8.9·10$^{-2}$ | -3.44 ± 0.45 | rad s$^{-1}$ |
| $\gamma_{ba}$ | 8.65 ± 7.22 | -24.50 ± 0.83 | -3.65 ± 3.95 | rad s$^{-1}$ |
| $\gamma_{bb}$ | 9.47 ± 0.33 | 8.63 ± 0.28 | -5.06 ± 1.32 | rad s$^{-1}$ |
| $A_{11}$ | 131.02 ± 13.07 | 225.64 ± 9.21 | 61.03 ± 2.07 | mm |
| $A_{12}$ | 70.06 ± 3.96 | 153.52 ± 11.73 | 30.59 ± 1.73 | mm |
| $A_{21}$ | 351.37 ± 35.14 | 771.32 ± 26.77 | 226.89 ± 10.91 | mm |
| $A_{22}$ | 154.33 ± 9.23 | 482.72 ± 42.15 | 284.81 ± 8.04 | mm |
| $\delta_{11}$ | 3.33 ± 2.30 | 12.48 ± 1.37 | 8.94 ± 2.99 | degree |
| $\delta_{12}$ | 136.51 ± 12.12 | -170.38 ± 6.13 | 161.64 ± 3.62 | degree |
| $\delta_{21}$ | -38.82 ± 2.54 | -6.44 ± 1.61 | -55.81 ± 2.42 | degree |
| $\delta_{22}$ | 112.72 ± 12.91 | -156.85 ± 5.51 | 85.17 ± 2.56 | degree |

**Supplementary Table 1. Main fitted and derived parameters.** A, sequence 1. B, sequence 2. C, sequence 3. D, sequence 4. E, sequence 5 for markers LSHO and LPSI. F, sequence 5 for markers RSHO and RPSI. G, sequence 5 for markers MSHO and MPSI. H, sequence 6a, ellipsoids. I, sequence 6a, $x'$ projection. J, sequence 6a, angle $\theta$. K, sequence 6b, ellipsoids. L, sequence 6b, $x''$ projection. For simplicity, in sequence 6b (Table L), all data start at $t = 0$ in the fitting procedures.

**Supplementary Table 2.**

**A**

| **Sequence 1** Participant | Curve | Root Mean Square Deviation (RMSD) | Experimental Data Range (EDR) | Units | RMSD/ EDR (%) |
|---|---|---|---|---|---|
| 1 | $y'^{D}_{REOB}$ | 14.33 | 616.91 | mm | 2.32 (min) |
| 1 | $y'^{D}_{LFHD}$ | 47.65 | 621.51 | mm | 7.6 (max) |
| 1 | $y'^{U}_{RASI}$ | 28.48 | 477.91 | mm | 6.00 (min) |
| 1 | $y'^{U}_{LFHD}$ | 34.44 | 400.93 | mm | 8.60 (max) |
| 1 | $y'^{R}_{RUCH}$ | 5.86 | 263.35 | mm | 2.23 |
| 2 | $y'^{D}_{REOB}$ | 16.26 | 537.14 | mm | 3.03 (min) |
| 2 | $y'^{D}_{RANK}$ | 5.02 | 82.50 | mm | 6.09 (max) |
| 2 | $y'^{U}_{RUCH}$ | 24.58 | 499.40 | mm | 4.92 (min) |
| 2 | $y'^{U}_{LFHD}$ | 33.27 | 428.38 | mm | 7.78 (max) |
| 2 | $y'^{R}_{RUCH}$ | 8.99 | 249.65 | mm | 3.59 |
| 3 | $y'^{D}_{RICH}$ | 11.56 | 433.93 | mm | 2.66 (min) |
| 3 | $y'^{D}_{RANK}$ | 6.43 | 112.39 | mm | 5.72 (max) |
| 3 | $y'^{U}_{REOB}$ | 20.62 | 431.18 | mm | 4.78 (min) |
| 3 | $y'^{U}_{LFHD}$ | 25.46 | 347.22 | mm | 7.33 (max) |
| 3 | $y'^{R}_{RUCH}$ | 5.12 | 185.16 | mm | 2.76 |

**B**

| **Sequence 2** Dance couple | Curve | Root Mean Square Deviation (RMSD) | Experimental Data Range (EDR) | Units | RMSD/ EDR (%) |
|---|---|---|---|---|---|
| 1 | $x'^{D}_{MSHO}$ | 29.81 | 629.17 | mm | 4.74 (min) |
| 1 | $x'^{D}_{MANK}$ | 10.40 | 105.03 | mm | 9.90 (max) |
| 1 | $x'^{U}_{MSHO}$ | 21.66 | 576.85 | mm | 3.75 (min) |
| 1 | $x'^{U}_{MANK}$ | 8.02 | 84.27 | mm | 9.51 (max) |
| 1 | $x'^{R}_{MBAK}$ | 6.80 | 284.30 | mm | 2.39 |
| 2 | $x'^{D}_{MIDO}$ | 14.58 | 467.25 | mm | 3.12 (min) |
| 2 | $x'^{D}_{MANK}$ | 7.72 | 113.39 | mm | 6.81 (max) |
| 2 | $x'^{U}_{MLUM}$ | 9.92 | 488.87 | mm | 2.03 (min) |
| 2 | $x'^{U}_{TBHD}$ | 22.95 | 361.62 | mm | 6.35 (max) |
| 2 | $x'^{R}_{MBAK}$ | 8.71 | 231.39 | mm | 3.76 |
| 3 | $x'^{D}_{MPSI}$ | 11.22 | 336.64 | mm | 3.33 (min) |
| 3 | $x'^{D}_{MANK}$ | 4.36 | 65.88 | mm | 6.62 (max) |
| 3 | $x'^{U}_{TBHD}$ | 14.49 | 406.95 | mm | 3.56 (min) |
| 3 | $x'^{U}_{MANK}$ | 5.74 | 54.68 | mm | 10.50 (max) |
| 3 | $x'^{R}_{MBAK}$ | 2.54 | 215.55 | mm | 1.18 |

**C**

| **Sequence 3**<br>Dance couple | Curve | Root Mean Square Deviation (RMSD) | Experimental Data Range (EDR) | Units | RMSD/ EDR (%) |
|---|---|---|---|---|---|
| 1 | $A_R$ | 0.07 | 1.39 | a.u. | 5.18 |
| 1 | $\Delta\varphi_R$ | 18.13 | 135.05 | degree | 13.42 |
| 2 | $A_R$ | 0.14 | 2.42 | a.u. | 5.92 |
| 2 | $\Delta\varphi_R$ | 31.32 | 106.17 | degree | 29.50 |
| 3 | $A_R$ | 0.74 | 6.47 | a.u. | 11.39 |
| 3 | $\Delta\varphi_R$ | 31.01 | 127.48 | degree | 24.32 |

**D**

| **Sequence 5**<br>Dance couple | Curve | Root Mean Square Deviation (RMSD) | Experimental Data Range (EDR) | Units | RMSD/ EDR (%) |
|---|---|---|---|---|---|
| 1 | $x'^{2D-3D}_{LSHO}$ | 3.12 | 546.96 | mm | 0.57 |
| 1 | $x'^{2D-3D}_{LPSI}$ | 3.46 | 497.48 | mm | 0.69 |
| 1 | $x'_{LSHO}$ | 31.33 | 546.96 | mm | 5.73 |
| 1 | $x'_{LPSI}$ | 26.18 | 497.48 | mm | 5.26 |
| 1 | $x'^{2D-3D}_{RSHO}$ | 2.38 | 567.47 | mm | 0.42 |
| 1 | $x'^{2D-3D}_{RPSI}$ | 2.24 | 492.36 | mm | 0.46 |
| 1 | $x'_{RSHO}$ | 34.01 | 567.47 | mm | 5.99 |
| 1 | $x'_{RPSI}$ | 25.93 | 492.36 | mm | 5.27 |
| 1 | $x'^{2D-3D}_{MSHO}$ | 1.95 | 559.70 | mm | 0.35 |
| 1 | $x'^{2D-3D}_{MPSI}$ | 2.60 | 494.91 | mm | 0.53 |
| 1 | $x'_{MSHO}$ | 26.79 | 559.70 | mm | 4.79 |
| 1 | $x'_{MPSI}$ | 25.75 | 494.91 | mm | 5.20 |
| 2 | $x'^{2D-3D}_{LSHO}$ | 3.39 | 429.69 | mm | 0.79 |
| 2 | $x'^{2D-3D}_{LPSI}$ | 2.96 | 455.75 | mm | 0.65 |
| 2 | $x'_{LSHO}$ | 29.42 | 429.69 | mm | 6.85 |
| 2 | $x'_{LPSI}$ | 30.46 | 455.75 | mm | 6.68 |
| 2 | $x'^{2D-3D}_{RSHO}$ | 1.41 | 417.43 | mm | 0.34 |
| 2 | $x'^{2D-3D}_{RPSI}$ | 1.72 | 451.71 | mm | 0.38 |
| 2 | $x'_{RSHO}$ | 28.32 | 417.43 | mm | 6.79 |
| 2 | $x'_{RPSI}$ | 30.44 | 451.71 | mm | 6.74 |
| 2 | $x'^{2D-3D}_{MSHO}$ | 1.95 | 426.98 | mm | 0.46 |
| 2 | $x'^{2D-3D}_{MPSI}$ | 2.22 | 453.56 | mm | 0.49 |
| 2 | $x'_{MSHO}$ | 25.85 | 426.98 | mm | 6.05 |
| 2 | $x'_{MPSI}$ | 29.58 | 453.56 | mm | 6.52 |
| 3 | $x'^{2D-3D}_{LSHO}$ | 1.60 | 357.89 | mm | 0.45 |
| 3 | $x'^{2D-3D}_{LPSI}$ | 1.18 | 274.47 | mm | 0.43 |
| 3 | $x'_{LSHO}$ | 21.67 | 357.89 | mm | 6.06 |
| 3 | $x'_{LPSI}$ | 16.85 | 274.47 | mm | 6.14 |
| 3 | $x'^{2D-3D}_{RSHO}$ | 1.74 | 334.60 | mm | 0.52 |
| 3 | $x'^{2D-3D}_{RPSI}$ | 1.21 | 278.14 | mm | 0.44 |
| 3 | $x'_{RSHO}$ | 22.85 | 334.60 | mm | 6.83 |
| 3 | $x'_{RPSI}$ | 16.49 | 278.14 | mm | 5.93 |
| 3 | $x'^{2D-3D}_{MSHO}$ | 0.98 | 346.22 | mm | 0.28 |
| 3 | $x'^{2D-3D}_{MPSI}$ | 0.68 | 276.01 | mm | 0.25 |
| 3 | $x'_{MSHO}$ | 21.20 | 346.22 | mm | 6.12 |
| 3 | $x'_{MPSI}$ | 16.60 | 276.01 | mm | 6.01 |

**E**

| **Sequence 6a** Dance couple | Curve | Root Mean Square Deviation (RMSD) | Experimental Data Range (EDR) | Units | RMSD/ EDR (%) |
|---|---|---|---|---|---|
| 1 | $r_{LSHO}$ | 24.11 | 699.62 | mm | 3.45 |
| 1 | $r_{LPSI}$ | 37.16 | 720.23 | mm | 5.16 |
| 2 | $r_{LSHO}$ | 12.64 | 586.10 | mm | 2.16 |
| 2 | $r_{LPSI}$ | 42.34 | 667.41 | mm | 6.34 |
| 3 | $r_{LSHO}$ | 14.11 | 586.66 | mm | 2.41 |
| 3 | $r_{LPSI}$ | 33.72 | 622.50 | mm | 5.42 |
| 1 | $x'^{\theta-3D}_{LSHO}$ | 19.70 | 688.95 | mm | 2.86 |
| 1 | $x'^{\theta-3D}_{LPSI}$ | 10.48 | 580.59 | mm | 1.80 |
| 1 | $x'_{LSHO}$ | 7.47 | 688.95 | mm | 1.08 |
| 1 | $x'_{LPSI}$ | 21.55 | 580.59 | mm | 3.71 |
| 1 | $\theta_{LSHO}$ | 1.08 | 66.57 | degree | 1.62 |
| 1 | $\theta_{LPSI}$ | 1.82 | 56.11 | degree | 3.24 |
| 2 | $x'^{\theta-3D}_{LSHO}$ | 16.78 | 487.26 | mm | 3.44 |
| 2 | $x'^{\theta-3D}_{LPSI}$ | 4.42 | 446.62 | mm | 0.99 |
| 2 | $x'_{LSHO}$ | 10.19 | 487.26 | mm | 2.09 |
| 2 | $x'_{LPSI}$ | 14.99 | 446.62 | mm | 3.36 |
| 2 | $\theta_{LSHO}$ | 1.30 | 66.01 | degree | 1.97 |
| 2 | $\theta_{LPSI}$ | 1.58 | 49.98 | degree | 3.17 |
| 3 | $x'^{\theta-3D}_{LSHO}$ | 10.19 | 558.29 | mm | 1.83 |
| 3 | $x'^{\theta-3D}_{LPSI}$ | 5.63 | 471.91 | mm | 1.19 |
| 3 | $x'_{LSHO}$ | 13.12 | 558.29 | mm | 2.35 |
| 3 | $x'_{LPSI}$ | 15.36 | 471.91 | mm | 3.26 |
| 3 | $\theta_{LSHO}$ | 1.51 | 58.69 | degree | 2.57 |
| 3 | $\theta_{LPSI}$ | 1.84 | 52.51 | degree | 3.51 |

**F**

| **Sequence 6b**<br>Dance couple | Curve | Root Mean Square Deviation (RMSD) | Experimental Data Range (EDR) | Units | RMSD/ EDR (%) |
|---|---|---|---|---|---|
| 1 | $r_{TBHD}$ | 7.42 | 428.77 | mm | 1.73 |
| 2 | $r_{TBHD}$ | 12.25 | 408.95 | mm | 3.00 |
| 3 | $r_{TBHD}$ | 3.61 | 315.51 | mm | 1.14 |
| 1 | $x''^{2D-3D}_{MSHO}$ | 3.46 | 277.06 | mm | 1.25 |
| 1 | $x''^{3D-3D}_{TBHD}$ | 1.00 | 683.77 | mm | 0.15 |
| 1 | $x''^{2D-3D}_{TBHD}$ | 11.38 | 683.77 | mm | 1.67 |
| 2 | $x''^{2D-3D}_{MSHO}$ | 1.32 | 250.90 | mm | 0.53 |
| 2 | $x''^{3D-3D}_{TBHD}$ | 3.13 | 711.34 | mm | 0.44 |
| 2 | $x''^{2D-3D}_{TBHD}$ | 11.34 | 698.41 | mm | 1.62 |
| 3 | $x''^{2D-3D}_{MSHO}$ | 0.63 | 66.63 | mm | 0.94 |
| 3 | $x''^{3D-3D}_{TBHD}$ | 1.21 | 424.14 | mm | 0.29 |
| 3 | $x''^{2D-3D}_{TBHD}$ | 3.01 | 410.33 | mm | 0.73 |
| 1 | $x''_{MSHO}$ | 6.85 | 276.71 | mm | 2.48 |
| 1 | $x''_{TBHD}$ | 23.35 | 683.77 | mm | 3.41 |
| 2 | $x''_{MSHO}$ | 8.39 | 250.90 | mm | 3.34 |
| 2 | $x''_{TBHD}$ | 15.23 | 711.34 | mm | 2.14 |
| 3 | $x''_{MSHO}$ | 3.38 | 66.21 | mm | 5.10 |
| 3 | $x''_{TBHD}$ | 6.55 | 424.14 | mm | 1.55 |

**Supplementary Table 2.** Goodness of fit using the Root Mean Square Deviation (RMSD), the Experimental Data Range (EDR), and the normalized goodness-of-fit metric (RMSD/EDR). A-F correspond to sequences 1, 2, 3, 5, 6a and 6b, respectively.

# SUPPLEMENTARY VIDEO CAPTIONS

## Wave physics as a choreographic notation for partner dance

Fernando Ramiro-Manzano[1*]

[1]*Nanomaterials for Optoelectronics, Photonics and Energy, Instituto de Tecnología Química, Universitat Politècnica de València - Consejo Superior de Investigaciones Científicas (UPV-CSIC), Avd. de los Naranjos s/n, 46022, Valencia, Spain.*

*e-mail: ferraman@fis.upv.es

Supplementary Videos 1-7 can be accessed at the following URL:

https://osf.io/2vkqg/overview?view_only=e3693b66672d4c2786e2bea143fc5edf

**Supplementary Video 1. Sequence 1: downward-propagating waves reflected into upward-propagating waves for anteroposterior body displacements (participant 1).** **a,c,** Camera recordings. **b,d,** Corresponding whole-body skeleton-like animations with markers labelled. **e,** Three-dimensional view of the skeleton-like motion. **f,** Waterfall plots of the anterior projection (*y'*) of selected marker trajectories (visible for most of the sequence); the vertical offset corresponds to each marker's mean z position, given that the data are mean-subtracted (see axis directions in panels **a** and **c**). From top to bottom, markers are LFHD, RSHO, RUCH, REOB, RICH, RASI, RTHI, RTIB and RANK (see Extended Data Table 1). **g,** Explanatory animation of wave propagation and reflection. Blue and green skeleton-like figures show the motion at the movie timeline and a phase-shifted version, respectively. The video shows the data acquisition for Fig. 1 and Extended Data Figs. 1a,b and 3a,d.

**Supplementary Video 2. Sequence 2,: downward-propagating waves reflected into upward-propagating waves for mediolateral body displacements (dance couple 1).** **a,c,** Camera recordings. **b,d,** Corresponding skeleton-like animations with markers labelled. **e,** Waterfall plots of the lateral projection (*x'*) of mid-position marker trajectories; the vertical offset corresponds to their mean z position, given that the data are mean-subtracted (see axis directions in panel **a**). From top to bottom, the traces correspond to TBHD, MSHO, MBAK, MIDO, MLUM, MPSI, MTHI, MKNE and MANK. M denotes the midpoint between the corresponding left and right markers (see Extended Data Table 1). The video shows the data acquisition for Extended Data Figs. 2a,b and 3g,j.

**Supplementary Video 3. Sequence 3: resonator spectra (dance couple 2).** **a,b,** Camera recordings with markers LBAK, RBAK, LPSI and RPSI labelled (see Extended Data Table 1). **c,** Tracked transverse-plane motion (*x'*-*y'*); see axis directions in panel **a**): LBAK, RBAK and their midpoint (virtual marker MBAK) plotted against the left axis, and LPSI, RPSI and their midpoint (virtual marker MPSI) plotted against the right axis. **d,e,** Normalized amplitude (MPSI/MBACK) and phase difference spectra, respectively; black points show the experimental data and red lines show the model fits. The video

illustrates the data acquisition and analysis outputs for Fig. 2a,b and Extended Data Fig. 4b,e.

**Supplementary Video 4. Sequence 4: phase shifts resembling a resonant response (dance couple 1). a,b,** Camera recordings with markers LSHO, RSHO, LPSI and RPSI labelled (see Extended Data Table 1). **c,** Tracked frontal-plane motion: LPSI, RPSI and their $z'$ difference (2·DPSI) are plotted in centred $x'$-$z'$ coordinates, and LSHO, RSHO and their midpoint (virtual marker MSHO) are plotted in centred $x'$-$z''$ coordinates (see axis directions in panel **b**). **d,** MSHO lateral drive (top; labelled with counts 1-8) and DPSI response (bottom; response labelled with the same count numbers, including decorative $\hat{4}$ and $\hat{8}$). Offsets between the top and bottom colour-band boundaries highlight the phase shifts. The video illustrates the data acquisition for Fig. 2c and Extended Data Fig. 4g.

**Supplementary Video 5. Sequence 5: coupled undamped oscillators (dance couple 1). a,b,** Camera recordings with markers LSHO, RSHO, LPSI and RPSI labelled (see Extended Data Table 1). **c,** Tracked frontal-plane motion: LPSI, RPSI and their midpoint (virtual marker MPSI) plotted in centred $x'$-$z'$ coordinates, and LSHO, RSHO and their midpoint (virtual marker MSHO) plotted in centred $x'$-$z''$ coordinates (see axis directions in panel **b**). **d,** Black curves show the experimental trajectories (with 2D-3D reconstruction) of MSHO and MPSI; green and red curves show the fitted coupled undamped-oscillator models for MSHO and MPSI, respectively. The video illustrates the data acquisition and analysis outputs for Fig. 3b,c and Extended Data Fig. 5f,i.

**Supplementary Video 6. Sequence 6: combined coupled oscillators (dance couple 1). a,b,** Camera recordings with markers TBHD, LSHO, RSHO and LPSI labelled (see Extended Data Table 1). **c,d** and **e,f** correspond to sequences 6a and 6b, respectively. **c,** Tracked motion of LSHO and LPSI on an ellipsoidal surface. **d,** Projections of these trajectories onto the $x'$ coordinate (chosen to maximize the projection range; see $x'$ axis direction in **c**). Black curves show the experimental trajectories of LSHO and LPSI; green and red curves show the fitted coupled-oscillator models for MSHO and MPSI, respectively. **e,** Tracked motion of a TBHD trajectory (3D–3D reconstructed) on an ellipsoidal surface. **f, middle,** $x''$ projections (see $x''$ direction in **e**) of TBHD (black; from 3D-3D reconstruction) and the mid-position virtual marker MSHO (black; from 2D-3D reconstruction); the grey trace, derived from TBHD 2D–3D reconstruction, is illustrative and was not used in the fitting process. Red and green curves show the fitted coupled-oscillator models for TBHD and MPSI, respectively. **f, top and bottom,** modal decompositions of the fitted TBHD and MPSI trajectories, respectively. The video illustrates the data acquisition and analysis outputs for Fig. 4 and Extended Data Figs. 8a,g,j and 9a,d,g,j,m.

**Supplementary Video 7. Audible waveforms from dance.** The video presents audio signals and corresponding graph excerpts derived from the experimental data and fitted models of Sequence 5 for dance couples 1-3. Markers RSHO and RPSI were selected because they show minimal occlusion across participants (see Extended Data Figs. 5e,h, 6e,h and 7e,h). Audio segments were generated from concatenated single-period traces after background removal; occluded intervals were bridged by straight-line joins. The pitch is set by mapping the lowest-frequency mode from the modal analysis to 440 Hz, with all other components scaled accordingly. For perceptual clarity, the signals were shaped with a 0.3 s exponential attack and a symmetric 0.3 s exponential fade-out. The sound sequence is: single tones 1 and 2 from the modal analysis, the fitted *dyad* model, and the experimental *dyad*. The musical *dyad* forms a *perfect fifth*, extended by an *octave* (frequency ratio $\approx$ 3:1).